\documentclass[paper]{JHEP3} 
\usepackage{epsfig,multicol}
\usepackage{amsmath}
\usepackage{amscd}
\usepackage{amsfonts}
\usepackage{amsthm}

\def\IC{\mathbb{C}}
\def\IN{\mathbb{N}}
\def\IZ{\mathbb{Z}}
\def\IR{\mathbb{R}}
\def\ID{\mathbb{D}}

\def\IX{\mathbb{X}}
\def\IP{\mathbb{P}}
\def\IY{\mathbb{Y}}
\def\IQ{\mathbb{Q}}

\newcommand{\rref}[1]{(\ref{#1})}

\newcommand{\Ann}{\operatorname{Ann}}

\newcommand{\pp}{{=\!\!\!|}}
\newcommand\fverb{\setbox\pippobox=\hbox\bgroup\verb}
\newcommand\fverbdo{\egroup\medskip\noindent%
            \fbox{\unhbox\pippobox}\ }
\newcommand\fverbit{\egroup\item[\fbox{\unhbox\pippobox}]}
\newbox\pippobox

\title{An $N=2$ worldsheet approach to D-branes in bihermitian geometries: II. The general case}

\author{Alexander Sevrin, Wieland Staessens\thanks{Aspirant FWO.}\\
Theoretische Natuurkunde, Vrije Universiteit Brussel\\ and\\The International Solvay Institutes\\
Pleinlaan 2, B-1050 Brussels, Belgium \\
E-mail:  \email{Alexandre.Sevrin@vub.ac.be},
\email{Wieland.Staessens@vub.ac.be}}
\author{Alexander Wijns\\
NORDITA\\
Roslagstullsbacken 23, SE-106 91 Stockholm, Sweden\\
and\\
Science Institute, University of Iceland,\\
Dunhaga 3, 107 Reykjav\'\i k, Iceland\\
E-mail: \email{awijns@nordita.org}}

\abstract{We complete the investigation of $N=(2,2)$ supersymmetric nonlinear $\sigma$-models in the presence of a
boundary. We study the full bihermitian geometry parameterized by chiral, twisted chiral and semi-chiral
superfields and identify the D-brane configurations
preserving an $N=2$ worldsheet supersymmetry. Combining twisted with semi-chiral superfields leads
to a clearly defined notion of lagrangian and coisotropic branes generalizing lagrangian and coisotropic A-branes on K{\"a}hler
manifolds to manifolds which are not necessarily K{\"a}hler (but still bihermitian).
Adding chiral fields complicates the picture and results in hybrid configurations interpolating
between lagrangian/coisotropic branes and branes wrapping around a holomorphic cycle. Even here the branes can be viewed as coisotropic
submanifolds albeit in a generalized sense. All supersymmetric D-brane configurations are characterized in the context of generalized complex geometry.
Duality transformations interchanging the various types
of superfields while preserving all supersymmetries are explicitly
constructed and provide for a powerful technique to construct various highly non-trivial
D-brane configurations. Several explicit examples are given.}


\begin{document}
\setcounter{equation}{0}

%
%

\section{Introduction} \label{introduction}
The discovery of moduli stabilization \cite{Kachru:2003aw} led to the
recognition that there is a landscape of (metastable) string vacua. This
resulted in a highly increased interest in string theory in backgrounds
with fluxes. While many results were obtained within the framework of
effective field theories (gauged supergravity), a direct stringy approach
is desirable. The pure spinor formalism \cite{Berkovits:2000fe} succeeds
in doing so, however here developing work remains to be done. On the
other hand we do have an alternative description for a subclass of these
backgrounds. Indeed non-linear $\sigma$-models in two dimensions with an
$N=(2,2)$ supersymmetry -- the so-called RNS models -- provide a
worldsheet description of type II superstrings in backgrounds including
general NSNS-fluxes (no RR-fluxes and a constant dilaton however)
\cite{Alvarez-Gaume:1981hm}-\cite{MS}. The requirement of $N=(2,2)$
supersymmetry imposes severe restrictions on the allowed geometries.
Imposing conformal invariance at the quantum level (the vanishing of the
$\beta$-functions) gives further conditions allowing an analysis which
potentially surpasses a supergravity one as higher order $\alpha '$
corrections are -- in principle -- calculable.

A manifestly supersymmetric formulation of these models clarifies the
geometric structure and greatly facilitates (quantum) calculations. Such
a formulation is now known: any $N=(2,2)$ non-linear $\sigma $-model can
be parameterized in $N=(2,2)$ superspace in terms of chiral, twisted
chiral and semi-chiral superfields \cite{Lindstrom:2005zr}.

When dealing with backgrounds which contain D-branes one has to consider
non-linear $\sigma $-models with boundaries. The presence of boundaries
breaks the $N=(2,2)$ worldsheet supersymmetry to an $N=2$ supersymmetry
and further enriches the geometric structure. The present paper concludes
the study of the classical geometry of these models in a manifestly
supersymmetric formulation ($N=2$ boundary superspace).

While a lot of pioneering work was done on supersymmetric D-brane
configurations \cite{Ooguri:1996ck}-\cite{Lindstrom:2002mc}, the study of
a manifestly $N=2$ supersymmetric worldsheet formulation of D-branes
started only in \cite{Koerber:2003ef} where $N=1$ and $N=2$ boundary
superspace was set up. This was subsequently applied to the study of A-
and B-branes on K{\"a}hler manifolds \cite{Sevrin:2007yn}. Contrary to
expectations, A-type boundary conditions were indeed possible in $N=2$
superspace. This was then extended to models which include NSNS-fluxes
where in first instance the simplest case -- mutually commuting complex
structures (or put differently models described in terms of twisted
chiral and chiral superfields) -- was studied \cite{Sevrin:2008tp}. An
interesting observation was that very involved brane configurations, {\em
e.g.} coisotropic branes, could easily be constructed from simple brane
configurations through supersymmetry preserving T-duality
transformations.

In the present case we turn our attention to the most general $N=(2,2)$
non-linear $\sigma $-model. In such models the complex structures do not
necessarily commute and a complete description needs, besides twisted
chiral and chiral superfields, semi-chiral superfields as well. We
identify the brane configurations compatible with worldsheet supersymmetry. The most
transparant case is where only semi-chiral and twisted chiral superfields
are present. Here one finds a very clear and explicit generalization of
lagrangian and coisotropic branes on K{\"a}hler manifolds to the non-K{\"a}hler
case. Having models described solely by chiral fields results in B-branes
wrapping around holomorphic cycles of K{\"a}hler manifolds. The general case
-- where all three types of superfields are present -- interpolates
between the two previous cases. Even here the branes can be interpreted
as generalized coisotropic submanifolds however in the context of a
foliation by symplectic leaves of a Poisson manifold.

In the next section we review $N=(2,2)$ non-linear $\sigma $-models in
$N=(2,2)$ superspace. We also introduce boundaries, reducing $N=(2,2)$
superspace to $N=2$ boundary superspace. We identify the three types of
superfields in boundary superspace.

Section 3 classifies the boundary conditions compatible with $N=2$
supersymmetry and leads to the identification of the various D-brane
configurations. Some of these results were already announced in
\cite{Sevrin:2008qx}. The various configurations are interpreted in terms
of generalized complex submanifolds of a generalized K{\"a}hler manifold.

Section 4 turns to duality transformations which interchange the various
types of superfields. After briefly reviewing the duality transformations
which do not need isometries we make a thorough study of duality
transformations in the presence of isometries.

In section 5 we illustrate our results through several examples. In
particular we focus on the non-linear $\sigma $-model with the Hopf
surface $S^3\times S^1$ as target manifold (also known as the
Wess-Zumino-Witten model on $SU(2)\times U(1)$) where we explicitly
construct langrangian D2-branes and coisotropic D4-branes. In order
to achieve this we start from the much simpler D1- and D3-brane
configurations on $D\times T^2$ which we then dualize to the above
mentioned D2- and D4-branes on $S^3\times S^1$.

We end with our conclusions and an outlook on future developments. The
first appendix summarizes our conventions. In appendix B we briefly
review $N=(1,1)$ and $N=1$ supersymmetric non-linear $\sigma $-models in
superspace. Appendix C summarizes some useful notions of generalized
complex geometry. In the last appendix we digress on the role of
auxiliary fields in T-duality transformations.

\section{N=2 superspace}
\subsection{$N=(2,2)$ supersymmetry in the absence of boundaries}
An $N=(2,2)$ non-linear $\sigma $-model is determined by the following
data:
\begin{itemize}
  \item An even dimensional (target) manifold ${\cal M}$. We denote the
  local coordinates by $X^a$, $a\in\{1,\cdots ,2n\}$.
  \item A metric $g_{ab}(X)$ on the manifold.
  \item A closed three-form $H_{abc}(X)$ on the manifold. Locally we
  introduce a two-form potential $b_{ab}(X)$ and we write $H_{abc}=-
  (3/2)\,  \partial _{[a}b_{bc]}$. Obviously the two-form potential
  is only defined modulo a gauge transformation, $b_{ab}\simeq
  b_{ab}+\partial _ak_b-\partial _bk_a$.
  \item Two (integrable) complex structures $J^a_{\pm}{}_b(X)$, $J^a_{\pm}{}_cJ^c_{\pm}{}_b
  =-\delta ^a_b$, which are such that the metric is hermitian with
  respect to both of them: $J_\pm^c{}_a\,J_\pm^d{}_b\,g_{cd}=g_{ab}$.
  \item The complex structures are covariantly constant though with
  different connections:
  \begin{eqnarray} 0=\nabla_c^\pm
\,J_\pm^a{}_b\equiv
\partial _c\,J_\pm^a{}_b+\Gamma^a_{\pm dc}J_\pm^d{}_{b}-
\Gamma^d_{\pm bc}J_\pm^a{}_d\,,\label{covconst}
\end{eqnarray}
with the connections $\Gamma_\pm$ given by,
\begin{eqnarray}
\Gamma^a_{\pm bc}\equiv  \left\{ {}^{\, a}_{bc} \right\} \pm
H^a{}_{bc}\,.\label{cons}
\end{eqnarray}
\end{itemize}
For obvious reasons this type of target manifold geometry is called a
bihermitian geometry. Note that if $\{{\cal M}, g, H, J_+,J_-\}$ defines
a bihermitian geometry then so does $\{{\cal M}, g, H, J_+, -J_-\}$. This
is a local realization of mirror symmetry.

The hermiticity of the metric with respect to the two complex structures
implies the existence of two two-forms,
\begin{eqnarray}
 \omega^\pm _{ab}=-\omega^\pm _{ba}\equiv
-g_{ac}J^c_\pm{}_b.
\end{eqnarray}
In general they are not closed. Using eq.~(\ref{covconst}), one shows
that,
\begin{eqnarray}
 \omega ^\pm_{[ab,c]}=\pm 2 J_\pm^d{}_{[a}H_{bc]d}=\pm (2/3)J^d_\pm{}_a
J^e_\pm{}_bJ^f_\pm{}_cH_{def},
\end{eqnarray}
where for the last step one uses the fact that the Nijenhuis
tensors\footnote{ Out of two $(1,1)$ tensors $R^a{}_b$ and $ S^a{}_b$,
one constructs a $(1,2)$ tensor ${\cal N}[R,S]^a{}_{bc}$, the Nijenhuis
tensor, as ${\cal N}[R,S]^a{}_{bc}=
R^a{}_dS^d{}_{[b,c]}+R^d{}_{[b}S^a{}_{c],d}+R\leftrightarrow S$. In the
present context, the integrability of $J_+$ and $J_-$ is equivalent to
${\cal N}[J_+,J_+]={\cal N}[J_-,J_-]=0$.} vanish. When the torsion
vanishes, the two-forms are closed and this reduces to the usual K{\"a}hler
geometry. Later in this section we will show that even when the torsion
does not vanish one might have -- under special circumstances -- closed
two-forms defined out of the metric $g$ and the complex structures
$J_\pm$.

From a local point of view, the equations above might be viewed as a set
of differential and algebraic equations which should be solved. For a
single complex structure, say $J_+$, this is indeed easily done. Going to
complex coordinates $Z^A$ and $Z^{\bar A}$, $a\in\{1,\cdots , n\}$, where
$J_+$ assumes its canonical form, $J^A_+{}_B=i\delta ^A_B$, $J^{\bar
A}_+{}_{\bar B}=-i\delta ^A_B$, $J^A_+{}_{\bar B}=J^{\bar A}_+{}_B=0$,
one immediately finds using eq.~(\ref{covconst}) that all conditions are
solved provided metric and torsion potential are parameterized in terms
of a (locally defined) one form $m_A$:
\begin{eqnarray}
&& g_{A\bar B}=\frac 1 2\, \big(\partial _Am_{\bar B}+
 \partial _{\bar B}m_A\big),\nonumber\\
&& b_{AB}=-\frac 1 2\,\big(\partial _Am_B-\partial _Bm_A\big),\qquad
b_{\bar A\bar B}=-\frac 1 2\,\big(\partial _{\bar A}m_{\bar B}-\partial _{\bar B}m_{\bar A}\big),
\label{sol1cs}
\end{eqnarray}
and all other components zero. There is a residual freedom in defining
the one-form $m_A$: $m_A\simeq m_A +n_A+i\,\partial _Af$, where $n_A$ is
holomorphic -- $\partial _{\bar B}n_A=0$ -- and $f$ is an arbitrary real
function. The precise form of $b$ is obviously gauge dependent, only the
torsion 3-form $H_{AB\bar C}=\partial _{\bar C}(
\partial _Am_B-\partial _Bm_A)/4$ has an invariant meaning.

Solving the conditions for both complex structures $J_+$ and $J_-$
simultaneously is more involved. Nonetheless -- as the off-shell
description of these models in $N=(2,2)$ superspace is known
\cite{Lindstrom:2005zr} (building on earlier work in
\cite{Buscher:1987uw}-\cite{Bogaerts:1999jc}) -- it can be done in terms
of a single real potential. The construction starts from the observation
that the terms in the algebra which do not close off-shell are all
proportional to the commutator of the two complex structures $[J_+,J_-]$.
As a consequence one expects that additional auxiliary fields will be
needed in the direction of $\mbox{coker}{[}J_+,J_-{]}$ while this will
not be the case for $\ker[J_+,J_-]=\ker(J_+-J_-)\oplus\ker(J_++J_-)$.

Decomposing the tangent space as
$\ker(J_+-J_-)\oplus\ker(J_++J_-)\oplus\mbox{coker}{[}J_+,J_-{]}$ one
shows that the first subspace gets parameterized by chiral, the second by
twisted chiral and the last one by semi-chiral $N=(2,2)$ superfields
\cite{Lindstrom:2005zr}. The three types of superfields are defined by
the following constraints\footnote{We refer to the appendix for our
conventions. We make a distinction between letters from the beginning
($\alpha ,\,\beta ,\,\gamma,\,$...) and letters from the middle of the
Greek alphabet ($\mu ,\,\nu ,\,\rho ,\,$...)}:
\begin{description}
  \item[Semi-chiral superfields:] $l^{\tilde \alpha }$, $l^{\bar{ \tilde
  \alpha}  }$, $r^{\tilde \mu  }$, $r^{\bar{ \tilde \mu }  }$,
  $\qquad\tilde \alpha,\,\bar{ \tilde \alpha} ,\,\tilde \mu ,\,\bar{
  \tilde \mu }\in \{1,\cdots n_s\}$,
\begin{eqnarray}
\bar \ID_+ l^{\tilde \alpha }=\ID_+l^{\bar{ \tilde
  \alpha}  }= \bar \ID_-r^{\tilde \mu  }=\ID_-r^{\bar{ \tilde \mu }  }=0.
\end{eqnarray}
  \item[Twisted chiral superfields:] $w^\mu $, $w^{\bar \mu }$, $\qquad\mu,\,\bar \mu
  \in\{1,\cdots n_t\}$,
\begin{eqnarray}
 \bar \ID_+w^\mu=\ID_-w^\mu=\ID_+w^{\bar \mu }=\bar \ID_-w^{\bar \mu }=0.
\end{eqnarray}
  \item[Chiral superfields:] $z^\alpha $, $z^{\bar \alpha }$, $\qquad\alpha,\, \bar \alpha
  \in\{1,\cdots n_c\}$,
\begin{eqnarray}
 \bar \ID_\pm z^\alpha =\ID_\pm z^{\bar \alpha }=0.
\end{eqnarray}
\end{description}
It is clear that chiral and twisted chiral $N=(2,2)$ superfields have the
same number of components as $N=(1,1)$ superfields while semi-chiral
$N=(2,2)$ superfields have twice as many, half of which are -- from
$N=(1,1)$ superspace point of view -- auxiliary.

Note that given $\{{\cal M},g,H\}$, the choice for $J_+$ and $J_-$ is not
necessarily unique. Consider {\em e.g.} a hyper-K{\"a}hler manifold (so
$H=0$, this discussion was given in \cite{Sevrin:1996jr}) where one has
three complex structures $J_i$, $i\in\{1,2,3\}$, satisfying
$J_iJ_j=-\delta _{ij}+\varepsilon_{ijk}J_k$. If one chooses
$J_+=J_-=\sin\theta\cos\phi J_1+\sin\theta \sin\phi J_2+\cos\theta J_3$
with $\phi \in[0,2\pi ]$, $\theta \in[0,\pi] $, one gets a description in
terms of chiral fields only. Choosing $J_+=-J_-=\sin\theta\cos\phi
J_1+\sin\theta \sin\phi J_2+\cos\theta J_3$, gives a description in terms
of twisted chiral fields. Finally, one could also put $J_+=J_1$ and
$J_-=\cos\phi J_2+\sin \phi J_3$ in which case $\{J_+,J_-\}=0$ which
implies $\ker [J_+,J_-]=\emptyset$. As a consequence the model is now
formulated in terms of semi-chiral superfields.

The most general action involving these superfields and consistent with
dimensions is given by,
\begin{eqnarray}
{\cal S}=4\,\int\,d^2 \sigma \,d^2\theta \,d^2 \hat \theta \, V(l,\bar l,r,\bar r,w,\bar w, z, \bar z),
\label{actionN22}
\end{eqnarray}
where the Lagrange density $V(l,\bar l,r,\bar r,w,\bar w, z, \bar z)$ is
an arbitrary real function of the semi-chiral, the twisted chiral and the
chiral superfields. It is defined modulo a generalized K{\"a}hler
transformation,
\begin{eqnarray}
V\rightarrow V+F(l,w,z)+ \bar F(\bar l,\bar w,\bar z)+ G(\bar r,w,\bar z)+ \bar G(r,\bar w,z).\label{genKahltrsf1}
\end{eqnarray}
These generalized K{\"a}hler transformations are essential for the global
consistency of the model, see {\em e.g.} \cite{Hull:2008vw}. Reducing the
action in eq.~(\ref{actionN22}) to $N=(1,1)$ superspace one finds that
$\hat{D}_-l^{\tilde \alpha }$ and $\hat{D}_+r^{\tilde \mu }$ (and their
complex conjugates) are auxiliary fields. Before doing so, let us
introduce some notation. We write,
\begin{eqnarray}
 M_{AB}=\left( \begin{array}{cc}
   V_{ab} & V_{a\bar b} \\
   V_{\bar a b} & V_{\bar a\bar b}
 \end{array}\right),
\end{eqnarray}
where, $(A,a)\in\{(l,\tilde \alpha ),\,(r,\tilde \mu ),\,(w,\mu
),\,(z,\alpha )\}$ and $(B,b)\in\{(l,\tilde \beta ),\,(r,\tilde \nu
),\,(w,\nu ),\,(z,\beta )\}$. In this way {\em e.g.} we get that $M_{zr}$
is the $2n_c\times 2n_s$ matrix given by,
\begin{eqnarray}
 M_{zr}=\left( \begin{array}{cc}
   V_{\alpha \tilde \nu } & V_{\alpha \bar{\tilde \nu }} \\
   V_{\bar \alpha \tilde \nu } & V_{\bar \alpha \bar{\tilde \nu }}
 \end{array}\right).
\end{eqnarray}
Note that $M_{AB}^T=M_{BA}$. We also introduce the matrix $\IP$,
\begin{eqnarray}
 \IP\equiv\left(
\begin{array}{cc}
  {\bf 1} & 0 \\
  0 & -{\bf 1}
\end{array}
 \right),
\end{eqnarray}
with $\bf 1$ the unit matrix and using this we write,
\begin{eqnarray}
 C_{AB}\equiv \IP M_{AB}-M_{AB}\IP,\qquad
A_{AB}\equiv \IP M_{AB}+M_{AB}\IP. \label{canda}
\end{eqnarray}

Using this notation one obtains -- after elimination of the auxiliary
fields -- the complex structures,
\begin{eqnarray}
 J_+&=&\left(\begin{array}{cccc}
   i\,\IP & 0 & 0 & 0 \\
   i\,M_{lr}^{-1}C_{ll} &  i\,M_{lr}^{-1}\IP M_{lr} &
   i\,M_{lr}^{-1}C_{lw}&i\,M_{lr}^{-1} C_{lz}   \\
   0 & 0 & i\,\IP & 0 \\
   0 & 0 & 0 & i\,\IP
 \end{array}   \right),\nonumber\\
J_-&=&\left(\begin{array}{cccc}
   i\,M_{rl}^{-1}\IP M_{rl} &  i\,M_{rl}^{-1} C_{rr} &
   i\,M_{rl}^{-1}A_{rw}&i\,M_{rl}^{-1} C_{rz}    \\
0&i\,\IP  & 0 & 0 \\
   0 & 0 & -i\,\IP & 0 \\
   0 & 0 & 0 & i\,\IP
 \end{array}   \right),\nonumber\\
\end{eqnarray}
where we labeled rows and columns in the order $l$, $\bar l$, $r$, $\bar
r$, $w$, $\bar w$, $z$, $\bar z$. Note that neither of them is in the
canonical (diagonal) form. One easily shows \cite{Sevrin:1996jr} that
making a coordinate transformation which replaces $r^{\tilde \mu }$ and
$r^{\bar {\tilde \mu} }$ by $V_{\tilde \alpha }$ and $V_{\bar{\tilde
\alpha} }$ resp.~while keeping the other coordinates as they are,
diagonalizes $J_+$. Similarly, a coordinate transformation which goes
from $l^{\tilde \alpha }$ and $l^{\bar {\tilde \alpha} }$ to $V_{\tilde
\mu }$ and $V_{\bar{\tilde \mu} }$ and keeping the other coordinates fixed
diagonalizes $J_-$. This allows one to reinterpret the generalized K{\"a}hler
potential as the generating functional for a canonical transformation
bringing one from a coordinate system where $J_+$ assumes its standard
diagonal form to another coordinate system where $J_-$ has its canonical
form (and vice-versa) \cite{Lindstrom:2005zr}.

{}From the second order action one reads off the metric $g$ and the torsion
potential $b$. We write both of them together $e=g+b$ with $e$ given by,
\begin{eqnarray}
 e=\frac 1 2 J_+^T\,\left(
\begin{array}{cccc}
  0 & M_{lr} & M_{lw}&M_{lz}   \\
  -M_{rl} & 0 & 0 & 0 \\
  0 & M_{wr} & M_{ww}&M_{wz}  \\
  0 & M_{zr} &  M_{zw}&M_{zz}
\end{array}
 \right)\,J_-+\frac 1 4 \,
 \left(\begin{array}{cccc}
   0 & 0 & -M_{lw}&M_{lz}  \\
   0 & 0 & -M_{rw}&M_{rz} \\
   -M_{wl} & -M_{wr} &  -2M_{ww}&0\\
   M_{zl} & M_{zr} & 0& 2M_{zz}
 \end{array}   \right).\label{eexplicit}
\end{eqnarray}
We will give a more elegant expression for the metric and torsion
potential later in this section. However, when only semi-chiral fields
are present, the expressions for the metric and torsion potential
following from eq.~(\ref{eexplicit}) greatly simplify,
\begin{eqnarray}
 g&=&\frac 1 4 \,\left(\begin{array}{cc}
   0 & M_{lr}\\
   - M_{rl} &0
 \end{array}   \right)\big[J_+,J_-\big],\nonumber\\
 b&=&\frac 1 4 \,\left(\begin{array}{cc}
   0 & M_{lr}\\
   - M_{rl} &0
 \end{array}   \right)\big\{J_+,J_-\big\}. \label{gbs}
\end{eqnarray}
Similarly, if only twisted chiral and chiral fields are present, the
metric and torsion potential following from eq.~(\ref{eexplicit}) are
given by,
\begin{eqnarray}
 g_{\alpha \bar \beta }=+V_{\alpha \bar \beta },\quad
 g_{\mu \bar \nu }=-V_{\mu \bar \nu },\quad
b_{\alpha  \nu }=+\frac 1 2 V_{\alpha  \nu },\quad
 b_{\alpha \bar \nu }=-\frac 1 2 V_{\alpha \bar \nu },
\end{eqnarray}
and complex conjugate. Note that as we are not yet considering
boundaries, $b$ is only defined modulo a gauge transformation. The
relevant gauge invariant object is the torsion 3-form $H\sim db$ whose
explicit form is unfortunately in general rather involved.

We already noted the existence of the ``local mirror transform'',
$\{{\cal M}, g, H, J_+,J_-\}$ $\rightarrow$ $\{{\cal M}, g, H,
J_+,-J_-\}$. In superspace this is simply realized by,
\begin{eqnarray}
 V(l,\bar l,r,\bar r,w,\bar w, z, \bar z)\rightarrow
-V(l,\bar l,\bar r, r,z,\bar z, w, \bar w).\label{mirrorpot}
\end{eqnarray}

Moreover, let us remark that depending on the field content one can have
several two-forms which -- using the conditions which guarantee the
existence of an $N=(2,2)$ bulk supersymmetry -- can be shown to be {\em
closed} and which are {\em linear} in the generalized K{\"a}hler potential.
\begin{itemize}
  \item There are no chiral fields, so $\ker (J_+-J_-)=\emptyset$.
  Then,
\begin{eqnarray}
 \Omega ^{(-)} _{ab}\equiv 2\,g_{ac}\big( (J_+-J_-)^{-1}\big)^c{}_b,\label{Omega}
\end{eqnarray}
is a closed form. It is {\em linear} in the generalized K{\"a}hler
potential and it is explicitely given by,
\begin{eqnarray}
\Omega ^{(-)}= - \frac{i}{2} \,
 \left(\begin{array}{ccc}
    C_{ll} &  A_{lr} & C_{lw} \\
    -A_{rl} & -C_{rr} & -A_{rw} \\
   C_{wl} & A_{wr} & C_{ww}
 \end{array}   \right), \label{2form}
\end{eqnarray}
where the matrices $C$ and $A$ were defined in eq.~(\ref{canda}). We
used a basis $(l, \bar l, r, \bar r, w, \bar w)$. When only twisted
chiral fields are present we have that $J\equiv J_+=-J_-$, the
geometry becomes K{\"a}hler and $\Omega ^{(-)}$ reduces to the usual
K{\"a}hler two-form, $\Omega ^{(-)}_{ab}=-g_{ac}J^c{}_b$. The two-form
$\Omega ^{(-)}$ has generically no well defined holomorphicity
properties with respect to either $J_+$ or $J_-$. One finds,
\begin{eqnarray}
 \Omega ^{(-)}_{ac}\,J^c_{\pm b}=
 -\Omega ^{(-)}_{bc}\,J^c_{\mp a}.
\end{eqnarray}
Having the two-form $\Omega ^{(-)}$ and the complex structures
$J_\pm$ allows one to neatly characterize the remainder of the
geometry. One finds,
\begin{eqnarray}
 g_{ab}&=&+\frac 1 2 \,\Omega ^{(-)}_{ac}\big(J_+-J_-)^c{}_b,\nonumber\\
 b_{ab}&=&-\frac 1 2 \,\Omega ^{(-)}_{ac}\big(J_++J_-)^c{}_b,\label{spform1}
\end{eqnarray}
where $b$ is equivalent -- modulo a gauge transformation -- to the
previously given expression ({\em i.e.} we still have $H=db$).
  \item There are no twisted chiral fields, so $\ker (J_++J_-)=\emptyset$.
  Then,
\begin{eqnarray}
 \Omega ^{(+)} _{ab}\equiv 2\,g_{ac}\big( (J_++J_-)^{-1}\big)^c{}_b,  \label{Omega2}
\end{eqnarray}
is a closed form. Again it is linear in the generalized K{\"a}hler
potential,
\begin{eqnarray}
\Omega ^{(+)}= \frac{i}{2} \,
 \left(\begin{array}{ccc}
    C_{ll} &  C_{lr} & C_{lz} \\
    C_{rl} & C_{rr} & C_{rz} \\
   C_{zl} & C_{zr} & C_{zz}
 \end{array}   \right),
 \end{eqnarray}
 where we used a basis $(l, \bar l, r, \bar r, z, \bar z)$. When no
semi-chiral fields are present we get $J\equiv J_+=J_-$ and the
geometry becomes K{\"a}hler with $\Omega ^{(+)}$ being precisely the K{\"a}hler
two-form.  Here as well one finds that $\Omega ^{(+)}$ has no
particular properties with respect to either $J_+$ or $J_-$,
\begin{eqnarray}
 \Omega ^{(+)}_{ac}\,J^c_{\pm b}=
 \Omega ^{(+)}_{bc}\,J^c_{\mp a}.
\end{eqnarray}
As before we can express the metric and the torsion potential in
terms of the closed two-form and the complex structures,
\begin{eqnarray}
 g_{ab}&=&+\frac 1 2 \,\Omega ^{(+)}_{ac}\big(J_++J_-)^c{}_b,\nonumber\\
 b_{ab}&=&-\frac 1 2 \,\Omega ^{(+)}_{ac}\big(J_+-J_-)^c{}_b,\label{spform2}
\end{eqnarray}
where again it should be noted that $b$ is only defined modulo a
gauge transformation.
  \item There are only semi-chiral fields, so $\ker [J_+,J_-]=\emptyset$.
  Then both $ \Omega ^{(-)}$ and $ \Omega ^{(+)}$ exist. On top of
  that we have that,
\begin{eqnarray}
 \Omega ^{(\pm)} _{ab}\equiv 2\,g_{ac}\big( [J_+,J_-]^{-1}\big)^c{}_b,\label{omplmin}
\end{eqnarray}
is a closed two-form as well. In terms of the generalized K{\"a}hler
potential it is given by,
\begin{eqnarray}
\Omega ^{(\pm)}= \frac{1}{2} \,
 \left(\begin{array}{cc}
   0 & M_{lr}\\
   - M_{rl} &0
 \end{array}   \right),
\end{eqnarray}
where we used a basis $(l, \bar l, r, \bar r)$. In this case we find
that $\Omega ^{(\pm)}$ is a $(2,0)+(0,2)$ two-form with respect to
both $J_+$ and $J_-$,
\begin{eqnarray}
 \Omega ^{(\pm)}_{ac}\,J^c_{+ b}=
- \Omega ^{(\pm)}_{bc}\,J^c_{+ a},\qquad
\Omega ^{(\pm)}_{ac}\,J^c_{- b}=
- \Omega ^{(\pm)}_{bc}\,J^c_{- a}.
\end{eqnarray}
The relation with $ \Omega ^{(-)}$ and $ \Omega ^{(+)}$ is explicitly
given by,
\begin{eqnarray}
  \Omega ^{(-)}=- \Omega ^{(\pm)}\,(J_++J_-)\,,\qquad
  \Omega ^{(+)}=+ \Omega ^{(\pm)}\,(J_+-J_-)\,.
\end{eqnarray}
\end{itemize}

Finally, let us return to the general case where semi-chiral, twisted
chiral and chiral superfields are simultaneously present. The expressions
in eqs.~(\ref{spform1}) and (\ref{spform2}) suggest the following
parameterization for $g$ and $b$,
\begin{eqnarray}
 g_{ab}&=&+\frac 1 2 \,\Omega_{+ac}J_+^c{}_b+\frac 1 2 \,\Omega_{-ac}J_-^c{}_b
 ,\nonumber\\
 b_{ab}&=&-\frac 1 2 \,\Omega_{+ac}J_+^c{}_b+\frac 1 2 \,\Omega_{-ac}J_-^c{}_b
 ,\label{spform3}
\end{eqnarray}
where $\Omega _\pm$ are two-tensors with a priory no particular
(symmetry) properties. Through a suitable gauge choice for $b$ one can
always turn either $\Omega _+$ or $\Omega _-$ into a closed two-form as
can be verified for {\em e.g.} $\Omega _+$ using the expressions for $g$
and $b$ given in eq.~(\ref{sol1cs}). Using those one finds that $\Omega
_+$ can be written as $\Omega _{+ab}=\partial _ak_b-\partial _bk_a$ with
$k_A=-(i/2)m_A$ and $k_{\bar A}=(i/2)m_{\bar A}$. However the other
$\Omega $ will in general not be a two-form. An explicit example of this
is the case where only twisted chiral and chiral superfields are present.
The one-form used in eq.~(\ref{sol1cs}) is then explicitly given by
$m_\alpha =V_\alpha $ and $m_\mu =-V_\mu $ (and complex conjugate). Using
this one easily verifies that $\Omega _+$ is a closed two-form while
$\Omega _-$ is neither anti-symmetric nor symmetric.

Choosing a gauge for $b$ such that $\Omega_+$ is a closed two-form, given
by,
 \begin{eqnarray}
\Omega_+ = -\frac i 2 \left(\begin{array}{cccc} C_{ll} & A_{lr} &
C_{lw} & A_{lz}\\
-A_{rl} & -C_{rr} & -A_{rw} & -C_{rz} \\
C_{wl}& A_{wr} & C_{ww} & A_{wz} \\
-A_{zl} & -C_{zr} & -A_{zw} & -C_{zz}
\end{array}
 \right), \label{omplus}
\end{eqnarray}
with respect to the basis $(l,\bar l, r, \bar r, w, \bar w, z, \bar z)$,
one finds that $\Omega_-$ is generically neither antisymmetric nor
symmetric and it can not be expressed in terms of linear derivatives of
the potential. It is explicitly given by,
\begin{eqnarray}
\Omega_- = \frac i 2 \left(\begin{array}{cccc} C_{ll} & A_{lr} &
C_{lw} & A_{lz}\\
-A_{rl} & -C_{rr} & -A_{rw} &- C_{rz} \\
C_{wl}& A_{wr} & C_{ww} & A_{wz} \\
C_{zl} + Z_{zl} & C_{zr} + Z_{zr} & A_{zw} + Z_{zw} & C_{zz} +
Z_{zz}
\end{array}
 \right),\label{omplus2}
\end{eqnarray}
where we have,
\begin{eqnarray}
Z_{zl} &=& -2\, M_{zl}M_{rl}^{-1}\IP M_{rl},\\
Z_{zr} &=&  + 2\, M_{zl} M_{rl}^{-1} \IP C_{rr} \IP, \\
Z_{zw} &=& - 2\, M_{zl}M_{rl}^{-1} \IP A_{rw} \IP, \\
Z_{zz} &=& + 2\, M_{zl} M_{rl}^{-1} \IP C_{rz} \IP.
\end{eqnarray}
Locally we can write $2\,\Omega_{+ab}=\partial _aB_b-\partial _bB_a$
where
 $B_a =  i
\big(V_l,-V_{\bar l}, - V_r, V_{\bar r},$ $V_w, - V_{\bar w}, -V_z,
V_{\bar z}  \big) $. When there are no chiral fields present,
$\Omega_\pm$ reduces to $\pm \Omega^{(-)}$. We thus reproduce the
situation defined in eq.~(\ref{Omega}) and subsequent relations.

However, using a different gauge choice for $b$ one makes $\Omega_+$
non-linear in $V$ and $\Omega_-$ a closed two-form,
\begin{eqnarray}
\Omega_- = \frac i 2 \left(\begin{array}{cccc} C_{ll} & C_{lr} &
C_{lw} & C_{lz}\\
C_{rl} & C_{rr} & C_{rw} & C_{rz} \\
C_{wl}& C_{wr} & C_{ww} & C_{wz} \\
C_{zl} & C_{zr} & C_{zw} & C_{zz}
\end{array}
 \right),
\end{eqnarray}
w.r.t. the basis $(l,\bar l, r, \bar r, w, \bar w, z, \bar z)$. We get
for $\Omega _+$ now,
\begin{eqnarray}
\Omega_+ = \frac i 2 \left(\begin{array}{cccc} C_{ll} & C_{lr} &
C_{lw} & C_{lz}\\
C_{rl} & C_{rr} & C_{rw} & C_{rz} \\
-C_{wl} + W_{wl} & -A_{wr} + W_{wr} & -C_{ww} + W_{ww} & -C_{wz} + W_{wz} \\
C_{zl} & C_{zr} & C_{zw} & C_{zz}
\end{array}
 \right),
\end{eqnarray}
with,
\begin{eqnarray}
W_{wl} &=& - 2\, M_{wr} M_{lr}^{-1}\IP C_{ll} \IP, \\
W_{wr} &=& + 2\,  M_{wr} M_{lr}^{-1}\IP M_{lr}, \\
W_{ww} &=& - 2\, M_{wr} M_{lr}^{-1} \IP C_{lw} \IP, \\
W_{wz} &=& - 2\, M_{wr} M_{lr}^{-1} \IP C_{lz} \IP.
\end{eqnarray}
In absence of twisted chiral fields, $\Omega_\pm$ reduces to
$\Omega^{(+)}$, which yields the same relations as in eq.~(\ref{Omega2})
and subsequent expressions.

We stress once more that while the introduction of the two-form $\Omega
_+$ will turn out to be most useful, it is not globally well defined as
its precise form explicitly depends on the gauge choice for $b$.

\subsection{Boundaries and $N=2$ superspace} \label{boundaries}
We now introduce a boundary in $N=(2,2)$ superspace which breaks half of
the supersymmetries, reducing $N=(2,2)$ to $N=2$. The
boundary\footnote{This is a so called B-type boundary. Alternatively we
could have introduced an A-type boundary defined by $\sigma =0$,  $
\theta '\equiv (\theta ^+- \theta ^-)/2=0$ and $ \hat \theta '\equiv(\hat
\theta ^++\hat \theta ^-)/2=0$. Throughout this paper we will always use
B-type boundary conditions as switching to A-type boundary conditions
amounts to performing the local version of the mirror transform as
defined in eq.~(\ref{mirrorpot}) \cite{Sevrin:2007yn}.} is defined by
$\sigma =0$, $ \theta '\equiv (\theta ^+- \theta ^-)/2=0$ and $ \hat
\theta '\equiv(\hat \theta ^+-\hat \theta ^-)/2=0$.

When passing to $N=2$ superspace, we get the following structure for the
superfields:
\begin{description}
  \item[Semi-chiral superfields:] $l^{\tilde \alpha }$, $l^{\bar{ \tilde
  \alpha}  }$, $r^{\tilde \mu  }$, $r^{\bar{ \tilde \mu }  }$,
  $\ID'l^{\tilde \alpha }$, $\bar \ID 'l^{\bar{ \tilde \alpha}  }$,
  $\ID'r^{\tilde \mu }$, $\bar \ID'r^{\bar{ \tilde \mu }  }$ are {\em
  unconstrained} $N=2$ superfields. The remaining components are
  determined by
\begin{eqnarray}
\bar \ID' l^{\tilde \alpha }=-\bar \ID l^{\tilde \alpha },\quad
\ID'l^{\bar{ \tilde \alpha}  }= -\ID l^{\bar{ \tilde
  \alpha}  },\quad \bar \ID'r^{\tilde \mu  }=+\bar \ID r^{\tilde \mu  },\quad
  \ID'r^{\bar{ \tilde \mu }  }=+\ID r^{\bar{ \tilde \mu }  }.\label{semicon}
\end{eqnarray}
Reducing the action to $N=1$ superspace, one finds that $\ID'
l^{\tilde \alpha }$, $\bar \ID'l^{\bar{ \tilde \alpha} }$, $
\ID'r^{\tilde \mu  }$ and $\bar \ID'r^{\bar{ \tilde \mu }  }$ are all
auxiliary.
 \item[Twisted chiral superfields:] $w^\mu $, $w^{\bar \mu }$ are
  {\em unconstrained} $N=2$ superfields. The other components are
  determined by,
\begin{eqnarray}
\ID'w^\mu =+\ID w^\mu ,\quad \bar \ID'w^{\mu} =-\bar \ID w^{\mu},\quad
\ID'w^{\bar\mu} =-\ID w^{\bar\mu} ,\quad \bar \ID'w^{\bar \mu} =+\bar\ID w^{\bar \mu}.\label{twichcon}
\end{eqnarray}
  \item[Chiral superfields:] $z^\alpha $, $z^{\bar \alpha }$, $\ID' z^\alpha $, $\bar \ID 'z^{\bar \alpha }$
  are {\em constrained} $N=2$ superfields. They satisfy,
\begin{eqnarray}
&& \bar \ID z^\alpha =\ID z^{\bar \alpha }=0,\nonumber\\
&& \bar \ID\ID' z^\alpha =-2i\partial _\sigma z^\alpha ,\quad
\ID\bar \ID' z^{\bar \alpha }=-2i\partial _\sigma z^{\bar \alpha }.\label{chicon}
\end{eqnarray}
The other components are fixed by,
\begin{eqnarray}
\bar \ID' z^{ \alpha }=\ID' z^{\bar \alpha }=0.
\end{eqnarray}
 \end{description}
Concluding: viewed from the boundary, both semi-chiral and twisted chiral
superfields are very similar as they both give rise to unconstrained
superfields. Chiral fields on the other hand remain constrained (chiral)
on the boundary.

One verifies that the difference between the two measures $\int d^2\sigma
\,D_+D_- \hat D_+\hat D_-$ and $\int d^2\sigma \,D\hat DD'\hat D '$ is
just a boundary term. So the most general $N=2$ invariant action which
reduces to the usual action when boundaries are absent is,
\begin{eqnarray}
{\cal S}&=&-\int d^2 \sigma\, d^2\theta \,d^2 \theta '\, V(X, \bar X)+
i\,\int d \tau \,d^2 \theta \,W( X, \bar X) ,\label{bsfa}
\end{eqnarray}
with $V(X, \bar X)$ and $W( X , \bar X)$ real functions of the
semi-chiral, the twisted chiral and the chiral superfields. The
generalized K{\"a}hler potential $V$ is arbitrary but the dependence of the
boundary potential on the semi-chiral and twisted chiral fields will be
determined by the boundary conditions as we will show later on. The
action is still invariant under the generalized K{\"a}hler transformations
eq.~(\ref{genKahltrsf1}) provided the boundary potential $W$ transforms
as well,
\begin{eqnarray}
W\rightarrow W-i\,\big(F(l,w,z)- \bar F(\bar l,\bar w,\bar z)\big)-
i\,\big(G(\bar r,w,\bar z)- \bar G(r,\bar w,z)\big).\label{genKahltrsf2}
\end{eqnarray}
This implies that $\big(V+i\,W\big)_{\bar \mu }$, $\big(V+i\,W\big)_{\bar
{\tilde{\alpha }}}$ and $\big(V+i\,W\big)_{\tilde \mu }$ (and their
complex conjugates) are invariant expressions. Note that when dealing
with the global definition of the geometry, eqs.~(\ref{genKahltrsf1}) and
(\ref{genKahltrsf2}) play an important role. Indeed when going from one
coordinate system to another on the overlap of two neighbourhoods one
finds that the generalized K{\"a}hler potential is invariant modulo a
generalized K{\"a}hler transformation eq.~(\ref{genKahltrsf1}). The
requirement that the boundary potential should transform as in
eq.~(\ref{genKahltrsf2}) imposes then severe restrictions on the form of
$W$. An explicit example of this can be found in \cite{Sevrin:2008tp}.

Reducing the action, eq.~(\ref{bsfa}) to $N=1$ boundary superspace
yields,
\begin{eqnarray}
 {\cal S}={\cal S}_{bulk}+i\,\int d \tau \,d \theta \,\big(B_a+\partial _aW\big)\hat DX^ a, \label{trixto1}
\end{eqnarray}
where we denoted the superfields collectively by $X$. The locally defined
one-form $B$ satisfies $2\,\Omega _{+ab}=\partial _aB_b-\partial _bB_a$,
where $\Omega _+$ was given in eq.~(\ref{omplus}). Upon eliminating the
auxiliary fields one finds for ${\cal S}_{bulk}$,
\begin{eqnarray}
 {\cal S}_{bulk}=\int d^2 \sigma\, d\theta \,D' \Big(2\,DX^ aD' X^ b\,g_{ab}-
 DX^ aDX^ b\,b_{ab}+D' X^ aD' X^ b\,b_{ab}\Big),\label{trixto}
\end{eqnarray}
where $g$ and $b$ are given in eq.~(\ref{spform3}) and the gauge choice
for $b$ is such that $\Omega _+$ and $\Omega _-$ are given by
eqs.~(\ref{omplus}) and (\ref{omplus2}). One verifies that the bulk
action, eq.~(\ref{trixto}) is indeed equivalent to the expression given
in eq.~(\ref{an1}). Obviously a detailed comparison of the boundary term
obtained in eq.~(\ref{trixto1}) with the generic one in eq.~(\ref{an1})
requires a careful analysis of the boundary conditions imposed on the
superfields.

When varying the action eq.~(\ref{bsfa}), one needs to take into account
that the chiral fields are constrained, eq.~(\ref{chicon}). Introducing
unconstrained fields $\Lambda ^{\alpha }$,  $\Lambda ^{\bar\alpha }$, $M^
\alpha $ and $M^{\bar \alpha }$ we can solve the constraints,
\begin{eqnarray}
&& z^\alpha =\bar \ID\Lambda ^ \alpha,\qquad z^ {\bar \alpha }=\ID\Lambda ^{\bar \alpha },\nonumber\\
&&\ID'z^\alpha =\bar \ID M^ \alpha -2i\,\partial _\sigma \Lambda ^\alpha ,\qquad
\bar \ID' z^ {\bar \alpha }=\ID M^{\bar \alpha }-2i\,\partial _\sigma \Lambda ^{\bar \alpha }.
\end{eqnarray}
Upon varying the action we get the bulk equations of motion and a
boundary term,
\begin{eqnarray}
\delta {\cal S}\Big|_{boundary}&=&\int d \tau \,d^2 \theta  \,
\Big\{
\delta \Lambda ^{ \alpha } \big(
\bar \ID'V_ { \alpha }+i\, \bar \ID W_{ \alpha }
\big)-
\delta \Lambda ^{ \bar \alpha } \big(
 \ID'V_ { \bar \alpha }-i \, \ID W_{ \bar \alpha }\big) \nonumber\\
&&\qquad
- \delta w^{ \mu } \big(V_{ \mu }-i W_{ \mu }\big)+
 \delta w^{ \bar \mu } \big(V_{ \bar \mu }+i W_{ \bar \mu }\big)
-\delta l^{ \tilde \alpha  } \big(V_{ \tilde \alpha }-i W_{ \tilde \alpha}\big)\nonumber\\&&\qquad +
 \delta l^{ \bar{\tilde \alpha} } \big(V_{ \bar{\tilde \alpha} }+i W_{ \bar{\tilde \alpha} }\big)+
\delta r^{ \tilde \mu   } \big(V_{ \tilde \mu  }+i W_{ \tilde \mu }\big)-
 \delta r^{ \bar{\tilde \mu } } \big(V_{ \bar{\tilde \mu } }-i W_{ \bar{\tilde \mu } }\Big\}
,\label{bsfab}
\end{eqnarray}
which should vanish by imposing proper boundary conditions. The
expression above can also be rewritten as,
\begin{eqnarray}
 \delta {\cal S}\Big|_{boundary}&=&i\,\int d \tau \,d^2 \theta  \,
\Big\{\delta \Lambda ^{ \alpha } \big(
\bar \ID'-\bar \ID\big)B_ { \alpha }+
\delta \Lambda ^{\bar  \alpha } \big(
\ID'- \ID\big)B_ {\bar  \alpha }+B_a\,\delta X^ a+\delta W\Big\},\nonumber\\\label{varspecial}
\end{eqnarray}
where $X^a$ collectively denotes all superfields and $B_a$ is the locally
defined one-form such that $2\,\Omega _{+ab}=\partial _aB_b-\partial
_bB_a$ where $\Omega _+$ is defined in eq.~(\ref{omplus}).

Finally, when reducing the action eq.~(\ref{bsfa}) to $N=1$ superspace,
one finds that $\bar \ID' l^{\tilde \alpha }$, $\ID'l^{\bar{ \tilde
\alpha} }$, $\bar \ID'r^{\tilde \mu  }$ and $\ID'r^{\bar{ \tilde \mu } }$
are all auxiliary. It is interesting to note that upon their elimination
one recovers a (matrix) structure which has a very different appearance,
though it remains equivalent of course, from the one we get in the case
without boundaries.

\section{Boundary conditions}
\subsection{Unconstrained $N=2$ fields and lagrangian and coisotropic branes} \label{generala}
\subsubsection{Generalities}
In this section we will study the case where all fields are a priori
unconstrained from the $N=2$ boundary superspace point of view. Put
differently: the bulk $N=(2,2)$ superfields consist of a number ($n_t$,
corresponding to $2n_t$ real directions) of twisted chiral superfields
and a number ($n_s$, corresponding to $4n_s$ real directions) of
semi-chiral multiplets. No chiral $N=(2,2)$ superfields are present. We
denote the unconstrained superfields collectively as $X^a$,
$a\in\{1,\cdots ,2n_t+4n_s\}$. Having that $\ker (J_+ -J_-)=\emptyset$
implies the existence of the non-degenerate two-form
$\Omega^{(-)}=2\,g\,(J_+-J_-)^{-1} $ introduced in eq.~(\ref{Omega}). We
will use throughout section (\ref{generala}) the expression for $b$ given
in eq.~(\ref{spform1}), {\em i.e.} $b=-(1/2)\,\Omega ^{(-)}(J_++J_-)$.

Whenever $\ker (J_+-J_-)$ is non-degenerate, one finds that imposing a
Dirichlet boundary condition $Y(X)=0$ implies a Neumann boundary
condition as well. Indeed, using the general relation,
\begin{eqnarray}
 \hat D X ^a=\frac 1 2 \,\big(J_+-J_-\big)^a{}_bD'X^b+
\frac 1 2 \,\big(J_++J_-\big)^a{}_bDX^b,\label{ser1}
\end{eqnarray}
we get from the Dirichlet boundary condition that,
\begin{eqnarray}
 0=\hat D Y=\partial _aY\,\Big(\big(J_+-J_-\big)^a{}_bD'X^b+
\,\big(J_++J_-\big)^a{}_bDX^b\Big), \label{sdir}
\end{eqnarray}
which is a Neumann boundary condition as it relates $D'X$ to $DX$ (see
eq.~(\ref{nis1NEU})). So the number of Dirichlet boundary conditions one
can impose is bounded and can maximally be $n_t+2n_s$.

As $\ker (J_+ -J_-)=\emptyset$, we can rewrite eq.~(\ref{ser1}) as,
\begin{eqnarray}
 g_{ab}\,D'X^b =\Omega^{(-)}_{ab}\hat D X^b+b_{ab} D X^b,\label{ser2}
\end{eqnarray}
where we used eq.~(\ref{spform1}). This is very reminiscent of the
Neumann boundary conditions in eq.~(\ref{nis1NEU}). The boundary
conditions will allow for the identification of $\hat DX$ in terms of
$DX$.

In the present case -- only twisted chiral and semi-chiral superfields --
the boundary term in the variation of the action eq.~(\ref{varspecial})
reduces to,
\begin{eqnarray}
 \delta {\cal S}\Big|_{boundary}&=&i\,\int d \tau \,d^2 \theta  \,
\Big\{B_a(X)\,\delta X^a+\delta W(X)
\Big\},\label{bdytunc}
\end{eqnarray}
where $B_a$ is a locally defined one-form whose external derivative is
precisely the closed two-form $\Omega^{(-)} $, $2\,\Omega^{(-)} _{ab}=
\partial _aB_b-\partial _bB_a$, introduced in eq.~(\ref{Omega}).
The vanishing of eq.~(\ref{bdytunc}) requires appropriate boundary
conditions. In what follows we will show that this gives rise to lagrangian
and coisotropic D-branes which generalize lagrangian and coisotropic
A-branes on K{\"a}hler manifolds to manifolds which are bihermitian but not
necessarily K{\"a}hler. For the necessary background on lagrangian and coisotropic branes on symplectic manifolds, see appendix \ref{sympl}.

\subsubsection{Lagrangian branes}
We first consider the case where we impose the maximal number of
Dirichlet boundary conditions. We make a coordinate transformation such
that the Dirichlet conditions are expressed by $Y^{\hat A}(X)=0$ for
$\hat A\in\{1,\cdots, n_t+2n_s\}$, $Y^{\hat A}\in\IR$. The remainder of
the coordinates -- the world volume coordinates on the
D($n_t+2n_s$)-brane -- are written as $\sigma^A(X)\in\IR$ with $
A\in\{1,\cdots,n_t+2n_s\}$. The boundary term eq.~(\ref{bdytunc})
vanishes provided a boundary potential $W(\sigma )$ can be found which
satisfies,
\begin{eqnarray}
 \frac{\partial \,W}{\partial \sigma ^A}=-B_b\,\frac{\partial X^b}{\partial \sigma ^A}.
\end{eqnarray}
The integrability conditions for these equations state that the pullback
of $\Omega^{(-)} $ to the world volume of the brane vanishes. Put
differently: we are dealing with a brane which is {\em lagrangian} with
respect to the symplectic structure defined by $\Omega^{(-)} $.

The Neumann boundary conditions can be written as,
\begin{eqnarray}
 \frac{\partial X^c}{\partial \sigma ^A}\,g_{cb}D'X^b=
\frac{\partial X^c}{\partial \sigma ^A}\,
b _{cd}\,\frac{\partial X^d}{\partial \sigma ^B}\,D\sigma ^B,
\end{eqnarray}
where we used eq.~(\ref{ser2}) and the fact that the pullback of
$\Omega^{(-)} $ to the world volume of the brane vanishes. Comparing this
to the Neumann boundary conditions in eq.~(\ref{nis1NEU}), one finds that
the invariant field strength ${\cal F}$ is of the form,
\begin{eqnarray}
{\cal F}_{ab} =b_{ab} =- \frac 1 2 \,\Omega^{(-)}_{ac} \left(J_+ + J_-\right)^c{}_b.\label{flux1}
\end{eqnarray}

\subsubsection{Maximally coisotropic branes}
The other extremal case is when we have Neumann boundary conditions in
all directions. The only way to achieve this is to constrain the fields
such that they become chiral on the boundary,
\begin{eqnarray}
 \hat D X^a=K^a{}_b(X)\,DX^b. \label{comNeu}
\end{eqnarray}
{}From $\hat D^2=D^2=-i\partial /\partial \tau $ we obtain integrability
conditions which tell us that $K$ is a(n integrable) complex structure.
Going to complex coordinates adapted to the complex structure $K$, one
immediately finds that the boundary term in the action
eq.~(\ref{bdytunc}) vanishes provided the one-form $B_a+\partial _aW$ is
holomorphic with respect to $K$. This in its turn implies that
$\Omega^{(-)} $ is a closed holomorphic $(2,0)+(0,2)$ two-form with
respect to $K$. As $\Omega^{(-)} $ is non-degenerate, this requires that
$n_t\in 2\IN$. So we end up with a space filling brane which is {\em
maximally coisotropic} with respect to the symplectic structure
$\Omega^{(-)} $.

The Neumann boundary conditions follow from eqs.~(\ref{comNeu}) and
(\ref{ser2}) and are given by,
\begin{eqnarray}
 g_{ab}D'X^b=\Big( \Omega^{(-)} _{ac}K^c{}_b+b_{ab}\Big)DX^b,
\end{eqnarray}
where $b$ was given in eq.~(\ref{spform1}). Comparing this to
eq.~(\ref{nis1NEU}), we get that,
\begin{eqnarray}
 {\cal F}_{ab}= \Omega^{(-)} _{ac}K^c{}_b +b_{ab}. \label{flux2}
\end{eqnarray}
As $\Omega ^{(-)}$ is a $(2,0)+(0,2)$ form with respect to the complex
structure $K$, we get that,
\begin{eqnarray}
 \hat\Omega _{ab}=\Omega ^{(-)}_{ac}\,K^c{}_b,\label{fsX1}
\end{eqnarray}
is a globally defined non-degenerate two-form. Furthermore, using the
integrability of the complex structure $K$ (the vanishing of the
Nijenhuis tensor), one shows that it is {\em closed} as well.

Following a strategy very similar to the the discussion around and
following eq.~(4.41) in \cite{Sevrin:2007yn}, we rewrite the boundary
term in the variation of the action eq.~(\ref{bdytunc}) as,
\begin{eqnarray}
 \delta {\cal S}\Big|_{boundary}&=&2i\,\int d \tau \,d^2 \theta  \, \delta \Lambda^a
\Big\{
\partial _{[a}\big(M_{|c|}K^c{}_{b]}\big)-\partial _{[a}M_{c]}K^c{}_b
\Big\}\,DX^b,\label{bdytunc10}
\end{eqnarray}
where $\Lambda $ is an unconstrained anti-commuting superfield and
$M_a=B_a+\partial _aW$. This vanishes provided,
\begin{eqnarray}
 \hat\Omega _{ab}=\partial _a\Big(\frac 1 2 \, M_cK^c{}_b\Big)-
\partial _b\Big(\frac 1 2 \, M_cK^c{}_a\Big),
\end{eqnarray}
holds. This leads us to the $U(1)$ potential,
\begin{eqnarray}
A_a = \frac 1 2 (B_b + \partial_b W) K^b{}_a,
\label{fsX2}
\end{eqnarray}
which is fully consistent with eqs.~(\ref{trixto1}), (\ref{comNeu}) and
(\ref{frfr2}). From this it follows again that $d{\cal F} = H$, as
required. Comparing eq.~(\ref{flux2}) to eq.~(\ref{flux1}), we conclude
that we now have a $U(1)$ bundle with fieldstrength $\hat\Omega _{ab}$
given in eq.~(\ref{fsX1}) and potential $A_a$, eq.~(\ref{fsX2}).

\subsubsection{Coisotropic branes}
Finally we consider the intermediate case. We use adapted coordinates
$Y^{\hat A}(X)$, $\sigma ^A(X)$, $\sigma ^\alpha(X) $ and $\sigma ^{\bar
\alpha }(X)$, with $\hat A, A\in\{1,\cdots,k\}$ and $\alpha ,\bar \alpha
\in\{1,\cdots,n_t+2n_s-k\}$. We impose the Dirichlet boundary conditions,
\begin{eqnarray}
 Y^{\hat A}=0,
\end{eqnarray}
and we require that the worldvolume coordinates $\sigma ^\alpha $ are
boundary chiral,
\begin{eqnarray}
 \hat D\sigma ^\alpha =+i\,D\sigma ^\alpha ,\qquad
 \hat D\sigma ^{\bar \alpha }=-i\,D\sigma ^{\bar \alpha }.
\end{eqnarray}
The boundary term in the variation of the action -- taking into account
that we now have constrained fields on the boundary -- vanishes provided,
\begin{eqnarray}
 &&\frac{\partial \,W}{\partial \sigma ^A}=-B_b\,\frac{\partial X^b}{\partial \sigma ^A},\nonumber\\
 &&\frac{\partial \ }{\partial \sigma ^{\bar \alpha }}\Big(\frac{\partial X^c}{\partial \sigma ^{\beta }}\, B_c
 +\frac{\partial \,W}{\partial \sigma ^{\beta }}\Big)=
 \frac{\partial \ }{\partial \sigma ^{ \alpha }}\Big(\frac{\partial X^c}{\partial \sigma ^{\bar \beta }}\, B_c
 +\frac{\partial \,W}{\partial \sigma ^{\bar \beta }}\Big)=0,\nonumber\\
 &&\frac{\partial \ }{\partial \sigma ^{A }}\Big(\frac{\partial X^c}{\partial \sigma ^{\beta }}\, B_c
 +\frac{\partial \,W}{\partial \sigma ^{\beta }}\Big)=
 \frac{\partial \ }{\partial \sigma ^{ A }}\Big(\frac{\partial X^c}{\partial \sigma ^{\bar \beta }}\, B_c
 +\frac{\partial \,W}{\partial \sigma ^{\bar \beta }}\Big)=0.
\end{eqnarray}
The integrability conditions which follow from this imply that all
components of the pullback of $\Omega^{(-)} $ to the D-brane world volume
vanish except for $\Omega^{(-)} _{\alpha \beta }$ and $\Omega^{(-)}
_{\bar \alpha \bar \beta }$ and we end up with a D$(2n_t+4n_s-k)$-brane
which is coisotropic\footnote{When no semi-chiral fields are present,
this reduces to coisotropic A-branes on K{\"a}hler manifolds whose
existence was discovered in \cite{Kapustin:2001ij}.} with respect to the
symplectic structure $\Omega^{(-)} $. Note that $n_t+2n_s-k$ must be
even. We distinguish three different sets of Neumann boundary conditions,
\begin{eqnarray}
 &&\frac{\partial X^c}{\partial \sigma ^A}\,g_{cb}\,\left(\frac{\partial X ^b}{\partial \sigma ^B}
 D' \sigma ^B +\frac{\partial X ^b}{\partial \sigma ^\beta}D' \sigma ^\beta  +
 \frac{\partial X ^b}{\partial \sigma ^{\bar \beta}}D' \sigma ^{\bar \beta} \right) =  \nonumber\\
&&\qquad\frac{\partial X^c}{\partial \sigma ^A}\,
b_{cd}\, \left( \frac{\partial X^d}{\partial \sigma ^B}\,
D\sigma ^B +\frac{\partial X ^d}{\partial \sigma ^\beta}D \sigma ^\beta  +
\frac{\partial X ^d}{\partial \sigma ^{\bar \beta}}D \sigma ^{\bar \beta}  \right), \nonumber\\
&&\frac{\partial X^c}{\partial \sigma
^\alpha}\,g_{cb}\,\left(\frac{\partial X ^b}{\partial \sigma ^B} D'
\sigma ^B +\frac{\partial X ^b}{\partial \sigma ^\beta}D' \sigma ^\beta
+ \frac{\partial X ^b}{\partial \sigma ^{\bar \beta}}D' \sigma ^{\bar
\beta} \right) =
 i \, \frac{\partial X^c}{\partial \sigma ^\alpha}\, \Omega^{(-)} _{cd}
\frac{\partial X ^d}{\partial \sigma ^\beta}D \sigma ^\beta \nonumber\\
&&\qquad + \,\frac{\partial X^c}{\partial \sigma ^\alpha}\, b_{cd}\,
\left(\frac{\partial X^d}{\partial \sigma ^B}\,D\sigma ^B +
\frac{\partial X ^d}{\partial \sigma ^\beta}D \sigma ^\beta  +
\frac{\partial X ^d}{\partial \sigma ^{\bar \beta}}D \sigma ^{\bar \beta}  \right),\nonumber \\
&&\frac{\partial X^c}{\partial \sigma ^{\bar
\alpha}}\,g_{cb}\,\left(\frac{\partial X ^b}{\partial \sigma ^B}D' \sigma
^B +\frac{\partial X ^b}{\partial \sigma ^\beta}D' \sigma ^\beta  +
\frac{\partial X ^b}{\partial \sigma ^{\bar \beta}} D' \sigma ^{\bar
\beta} \right) =
 - i \, \frac{\partial X^c}{\partial \sigma ^{\bar\alpha}}\,
\Omega^{(-)} _{cd} \frac{\partial X ^d}{\partial \sigma ^{\bar\beta}}D
\sigma ^{\bar\beta} \nonumber\\
&&\qquad +\,\frac{\partial X^c}{\partial \sigma ^{\bar\alpha}}\, b_{cd}\,
\left(\frac{\partial X^d}{\partial \sigma ^B}\,D\sigma ^B +
\frac{\partial X ^d}{\partial \sigma ^\beta}D \sigma ^\beta  +
\frac{\partial X ^d}{\partial \sigma ^{\bar \beta}}D \sigma ^{\bar \beta}  \right).\nonumber \\
\end{eqnarray}
Comparing these boundary conditions with eq.~(\ref{nis1NEU}), we can read
off the flux ${\cal F}$, which is generically of the form,
\begin{eqnarray}
 {\cal F}_{ab} &=& b_{ab}+F_{ab},
\end{eqnarray}
where $b$ was given in eq.~(\ref{spform1}) and the only non-vanishing
components of $F$ -- the $U(1)$ field strength -- are given by,
\begin{eqnarray}
 F_{\alpha \beta} = i \, \Omega^{(-)}_{\alpha \beta},\qquad  F_{\bar\alpha \bar\beta}
= -i \, \Omega^{(-)}_{\bar\alpha \bar\beta}.
\end{eqnarray}

\subsection{Chiral $N=2$ fields}
We now turn to the case where only chiral fields, $z^\alpha $, $\alpha
\in\{1,\cdots,n_c\}$, are present. The bulk geometry is K{\"a}hler. This case
has been thoroughly studied in \cite{Sevrin:2007yn} where as a starting
point the Dirichlet boundary conditions on the unconstrained superfields
were taken (see eq.~(\ref{bsfab})). The result was that through a
holomorphic coordinate transformation one can always find coordinates
$z^{\tilde \alpha }$, $\tilde\alpha \in\{1,\cdots, k\}$ and $z^{\hat
\alpha }$, $\hat\alpha \in\{k+1,\cdots, n_c\}$, such that the Dirichlet
boundary conditions are simply the statement that $z^{\hat \alpha }$'s
are constant. The worldvolume coordinates are then given by $z^{\tilde
\alpha }$ and the worldvolume itself is also K{\"a}hler. Put differently, we
obtain a type B D$2k$-brane wrapping around a holomorphic cycle of the
target manifold.

In order that the boundary term in the variation eq.~(\ref{bsfab})
vanishes, we need to impose $2k$ Neumann boundary conditions as well,
\begin{eqnarray}
 V_{\tilde\alpha \bar \beta }\,\bar \ID'z^{\bar \beta }=-i\, W_{\tilde\alpha \bar{\tilde \beta }}\,
 \bar \ID z^{\bar{\tilde \beta }},
\end{eqnarray}
and complex conjugate. Comparing to eq.~(\ref{nis1NEU}), we find a $U(1)$
field strength with as non-vanishing elements ${\cal F}_{\tilde\alpha
\bar{\tilde \beta }}=-i\, W_{\tilde\alpha \bar{\tilde \beta }}$. Note
that here -- at least at the classical level -- we have no restrictions
on the form of the boundary potential $W$ \footnote{Superconformal
invariance at the quantum level does give additional conditions, see {\em
e.g.} \cite{Nevens:2006ht}.}.

\subsection{The general case}
We now turn to the most generic case where we have a model in terms of
$n_s$ semi-chiral multiplets, $n_t$ twisted chiral superfields and $n_c$
chiral superfields. This generic case is an -- at least in principle -- combination of
the two cases discussed below. Since expressions become more and more involved, we will
restrict ourselves to some important remarks which capture the essence of the ideas involved.
We can however already make some general remarks without going into more detail.

First of all, note that while the dependence of the boundary potential $W$ on the
semi-chiral and twisted chiral coordinates is fixed by the boundary
conditions, we are still free to add some function of the chiral fields
to the potential. This reflects the freedom to switch on an arbitrary
$U(1)$ holomorphic bundle in the chiral directions.

Finally we still have that $W\simeq W+f+\bar f$ where $f$ is an arbitrary
holomorphic function of all the {\em boundary} chiral fields. This
freedom can {\em e.g.} be used to make certain isometries manifest in the
boundary potential.

\subsubsection{Generalized maximally coisotropic branes}

Let us first assume that all twisted chiral and semi-chiral fields obey Neumann conditions.
A first thing to realize is that one can do parts of the analysis in section \ref{generala}
more generally. We again start from the generally valid eq.~\rref{ser1}. In this subsection,
we denote the collection of all twisted chiral and semi-chiral fields, {\em and} the chiral Neumann
fields by $X^a$. The Neumann conditions then take the usual form
\begin{eqnarray}
D' X^a = g^{ab} {\cal F}_{bc} DX^ c.
\end{eqnarray}
Plugging this into \rref{ser1} yields
\begin{eqnarray}
\hat D X^a = K^a{}_b DX^ b, \quad K^a{}_b =  \frac 1 2 \,\big(J_++J_-\big)^a{}_b +
\frac 1 2 \,\big(J_+-J_-\big)^a{}_c \, g^{cd} {\cal F}_{db}. \label{acs}
\end{eqnarray}
Note that when $X^a$ is a chiral field this simply reduces to the usual
chirality condition. The other components of eq.~\rref{acs} mix chiral
and non-chiral fields. The integrability of these equations requires $K$
to be a complex structure. If $\ker (J_+ -J_-) = 0$ -- i.e. in absence of
chiral fields -- we can solve for ${\cal F}$ as a function of $K$ and we
recover eq.~\rref{flux2}. When only chiral and twisted chiral fields are
present, we recover the expression for the complex structure we presented
in eq.~(4.47) of \cite{Sevrin:2008tp}.

The remainder of the analysis of the generic case is similar to the one in section 4.2.2 of
\cite{Sevrin:2008tp} ($\pi_+ = 1$ case). While $X^a$ still denotes any
superfield (in the Neumann directions), we write $X^{\tilde a}$ for the
chiral superfields and $X^{\hat a}$ for the semi-chiral and twisted
chiral superfields. The vanishing of eq.~(\ref{varspecial}) requires,
\begin{eqnarray}
 \partial _{\hat a}\big(M_cK^c{}_b\big)-\partial _{b}\big(M_cK^c{}_{\hat a}\big)=
 2\,\Omega _{+\hat a c}K^c{}_b,\label{alet1}
\end{eqnarray}
to hold where $\Omega _+$ was given in eq.~(\ref{omplus}). Using this we
find that ${\cal F}_{ab}=b_{ab}+F_{ab}$ where $b$ is in the gauge where
$\Omega _+$ is a closed two-form and $F$ is the $U(1)$ fieldstrength. The
explicit expressions for the fieldstrength follow from combining
eqs.~(\ref{trixto1}) and (\ref{alet1}) which results in,
\begin{eqnarray}
&& F_{\hat a\hat b}=\big(\Omega _+K)_{\hat a \hat b}\,,\nonumber\\
&& F_{\hat a\tilde b}=\big(\Omega _+K)_{\hat a \tilde b}\,,\nonumber\\
&& F_{\alpha \bar \beta }=-iW_{\alpha \bar \beta }+
\partial _\alpha \Big(\frac 1 2 \,M_{\hat c}K^{\hat c}{}_{\bar \beta }\Big) -
\partial _{\bar \beta }\Big(\frac 1 2 \,M_{\hat c}K^{\hat c}{}_{\alpha  }\Big),\nonumber\\
&&F_{\alpha \beta }=\partial _{[\alpha }\big(M_{|\hat c|}K^{\hat c}{}_{\beta ]}\big),\qquad
F_{\bar \alpha \bar \beta }=\partial _{[\bar \alpha }\big(M_{|\hat c|}K^{\hat c}{}_{\bar \beta ]}\big),
\end{eqnarray}
where we used the original (complex) notation for the chiral fields
again. Note that the expressions significantly simplify when $K$ has no
components which mix the chiral with the twisted and semi-chiral
superfields.

Denoting the number of chiral fields for which we choose Neumann
conditions by $\hat n_c$, the conditions of this subsection describe a
$(2 \hat n_c + 2 n_t + 4 n_s)$-dimensional brane. Although the target space is here no longer symplectic, there is a very natural way in which such a brane still wraps a coisotropic submanifold, namely in the sense of Poisson geometry. This is explained in appendix \ref{app:pois} and further discussed in section \ref{sec:GCG}. Hence the branes described above will be called generalized maximally coisotropic.

\subsubsection{Generalized lagrangian branes}

Let us now turn to the other extreme, namely impose the maximal number of
Dirichlet conditions on twisted chiral and semi-chiral fields as
possible. The discussion surrounding eq.~(\ref{sdir}) still holds, so
that this means that there are an equal number of Dirichlet and Neumann
conditions on these fields.

We denote the chiral fields (in the Neumann directions) by $z^\alpha $
and $z^{\bar \alpha }$ and we write the semi-chiral and twisted chiral
superfields collectively as $X^{\hat a}$. Through a coordinate
transformation we exchange $X^{\hat a}$ for adapted (real) coordinates
$Y^{\hat A}(X,z)$ and $\sigma ^ A(X,z)$, $\hat A,\,A\in\{1,\cdots,
n_t+2n_s\}$. The Dirichlet boundary conditions are given by $Y^{\hat
A}(X,z)=0$ and $\sigma ^A$, $z^\alpha $ and $z^{\bar \alpha }$ are the
worldvolume coordinates.

The vanishing of the boundary term in the variation of the action,
eq.~(\ref{varspecial}), requires the existence of a boundary potential
$W(\sigma ,z)$ such that,
\begin{eqnarray}
 \frac{\partial \,W}{\partial \sigma ^A}=-B_{\hat b}\,\frac{\partial X^{\hat b}}{\partial \sigma ^A},
\end{eqnarray}
holds. The integrability condition for this is given by,
\begin{eqnarray}
 \frac{\partial X^{\hat c}}{\partial \sigma ^A}\,
\Omega _{+\hat c\hat d}\,
  \frac{\partial X^{\hat d}}{\partial \sigma ^B}\,=0.
\end{eqnarray}
The Neumann boundary conditions assume their standard form,
eq.~(\ref{nis1NEU}), with ${\cal F}=b+F$. The torsion potential $b$ is in
the gauge where $\Omega _+=-(g-b)J_+$ is a closed two-form. The $U(1)$
fieldstrength follows from the gauge potentials,
\begin{eqnarray}
&& A_\alpha =V_\alpha +i\,W_\alpha +i\,B_{\hat b}\,\frac{\partial X^{\hat b}}{\partial z^\alpha },\nonumber\\
&& A_{\bar \alpha} =V_{\bar \alpha} -i\,W_{\bar \alpha}
-i\,B_{\hat b}\,\frac{\partial X^{\hat b}}{\partial z^{\bar \alpha} },\nonumber\\
&&A_A=0.
\end{eqnarray}

Denoting the number of chiral fields for which we choose Neumann
conditions again by $\hat n_c$, the conditions of this subsection
describe a $(2 \hat n_c + n_t + 2n_s)$-dimensional brane. Such a brane wraps a minimally\footnote{Here we mean minimal in the non-chiral directions. Any number of chiral fields can be chosen to obey Neumann boundary conditions without affecting the minimality we refer to here.} coisotropic submanifold -- again in the sense of Poisson geometry -- as will be explained in the next section. We therefore refer to it as a generalized lagrangian brane. Notice that it however need no longer be half-dimensional because of the chiral directions.

\subsection{Embedding in Generalized Complex Geometry} \label{sec:GCG}

In flux compactification scenarios, the presence of non-trivial fluxes
along cycles of the internal manifold forces the internal manifold to no
longer be Calabi-Yau. A good language for capturing some essential features
of the required internal geometry was proposed by Hitchin
\cite{Hitchin:2004ut} and subsequently developed by Gualtieri
\cite{Gualtieri:2003dx}. Generalized complex geometry or GCG, as it is
called, contains both complex and symplectic geometry as special cases.
As such it turns out to be the right setting for the formulation of what
the Calabi-Yau condition generalizes to in the presence of fluxes.
Perhaps not surprisingly, this is called the (weak) generalized
Calabi-Yau condition \cite{Hitchin:2004ut}. Since in this paper we are
not yet concerned with conformal invariance on the worldsheet, we have
no need to discuss all conditions that go into the generalized Calabi-Yau
requirement. Demanding $N=(2,2)$ supersymmetry on the worldsheet
nevertheless has a very nice interpretation in the language of GCG. Ever
since the work of \cite{Gates:1984nk} we know that in the presence of
NSNS-flux (but in the absence of RR-flux and for constant dilaton)
the relevant target space geometry is a bihermitian geometry. It was
however shown in \cite{Gualtieri:2003dx} that this is equivalent to what
is called generalized K{\"a}hler geometry in the GCG approach. In appendix
\ref{app:geo} some basic constructions in GCG -- as well its limiting
cases of complex and symplectic geometry -- are discussed, with special
emphasis on certain natural classes of submanifolds, i.e. generalized
complex, complex and coisotropic submanifolds, respectively. In
\cite{Gualtieri:2003dx} (see also \cite{Zabzine:2004dp}) it was shown
that both A branes (on symplectic manifolds) and B branes (on complex
manifolds) can be understood as being generalized complex submanifolds.
In this section and appendix \ref{app:sub} we rederive some of these
results, flesh them out a bit and find more clues for the relevance of
generalized complex submanifolds in describing D-branes on  generic
generalized K{\"a}hler manifolds by comparing our findings with the
$\sigma$-model results of the previous section.

\subsubsection{Generalized complex submanifols of bihermitian manifolds}

In appendix \ref{app:geo}, eq.~\rref{gcs2}, we present the pair of
commuting $H$-twisted generalized complex structures $({\cal J}_+, {\cal
J}_-)$ comprising the generalized K{\"a}hler structure associated with the
data $(g,H,J_+, J_-)$ of a bihermitian geometry. As we discussed
before, sending $J_-$ to $-J_-$ interchanges chiral and twisted chiral
fields in the local parameterization of the manifold. Since this also
interchanges ${\cal J}_+$ and ${\cal J}_-$ it is sufficient to focus on
one of them when analyzing the conditions for a generalized complex
submanifold. In our conventions it turns out that the natural choice is
${\cal J}_+$ which from now on we simply call ${\cal J}$.

Before we proceed, it will be useful to introduce some new notation. We
combine the complex structures $J_\pm$ into the combinations\footnote{In
this section, it is more appropriate to use a slightly more abstract
notation, as is explained in footnote \ref{ftn:ind} of appendix
\ref{app:geo}.}
\begin{eqnarray}
J_{(\pm)} = \frac 1 2 (J_+ \pm J_-).
\end{eqnarray}
From the non-degenerate two-forms,
$\omega_{\pm} = -g J_{\pm}$, and more precisely
their inverses $\omega^{-1}_{\pm} = J_{\pm}\, g^{-1}$, we can then define two Poisson structures \cite{Lyakhovich:2002kc}
\begin{eqnarray}
\Pi_{(\pm)} = J_{(\pm)} g^{-1}= \frac 1 2 (\omega^{-1}_+ \pm \omega^{-1}_-).
\end{eqnarray}
For a brief discussion of some relevant facts about Poisson structures, see appendix \ref{app:pois}. When one of these Poisson bi-vectors is invertible, the inverse is a symplectic structure,
\begin{eqnarray}
\Omega^{(\pm)} \equiv \Pi_{(\pm)}^{-1} =g J_{(\pm)}^{-1}.
\end{eqnarray}
As we mentioned before, from these symplectic structures we can define a symmetric 2-covector and a 2-form in a natural way,
\begin{eqnarray}
g^{(\pm)} &=& \Omega^{(\pm)} J_{(\pm)}, \\
b^{(\pm)} &=& -\Omega^{(\pm)} J_{(\mp)}.
\end{eqnarray}
For backgrounds for which $\Pi_{(\pm)}$ is invertible, $g^{(\pm)}$ and $b^{(\pm)}$ are precisely the metric and b-field of the bihermitian geometry respectively. Notice that using this notation, the additional complex structure $K$ in eq.~(\ref{acs}) can also be written as,\footnote{To be precise, the objects in this equation should be pulled back in the proper way to the world-volume, as will be discussed below.}
\begin{eqnarray}
K = J_{(+)} + \Pi_{(-)} {\cal F}  = \Pi_{(+)} g + \Pi_{(-)} {\cal F} . \label{genK}
\end{eqnarray}

Now consider a (generalized) submanifold $({\cal N}, {\cal F})$ of a generalized K{\"a}hler manifold $({\cal M},{\cal J}_\pm, H)$ in the sense discussed in appendix \ref{app:geo}, i.e in particular $d{\cal F} = H\vert_{\cal N}$. Such a submanifold is called generalized complex if its generalized tangent bundle, $\tau_{\cal N}^{\;\cal F}$ defined in eq.~\rref{gtb}, is stable under the following $H$-twisted generalized complex structure
\begin{eqnarray}
{\cal J}               =
                                        \left( \begin{array}{cc}
                                        J_{(+)} & \Pi_{(-)} \\
                                        g J_{(-)}  & - J^t_{(+)}
                                        \end{array}\right), \label{gcsm}
\end{eqnarray}
which is simply a rewriting of ${\cal J}_+$ in eq.~\rref{gcs2}.
Requiring ${\cal J}$ to stabilize $\tau_{\cal N}^{\;\cal F}$, we get the following conditions,
\begin{eqnarray}
\Pi_{(-)} (\Ann T_{\cal N}) &\subset& T_{\cal N}, \label{st2} \\
(J_{(+)} + \Pi_{(-)} {\cal F}) (T_{\cal N}) &\subset& T_{\cal N}, \label{st1} \\
\left(g J_{(-)} - J^t_{(+)} {\cal F} - {\cal F} J_{(+)} - {\cal F} \Pi_{(-)} {\cal F}\right) (T_{\cal N})  &\subset& \Ann T_{\cal N}, \label{st3}
\end{eqnarray}
where $\Ann T_{\cal N}$ is defined in eq.~\rref{ann}.
We now distinguish the following, gradually more complicated cases:
\begin{enumerate}
  \item \underline{$J_{(-)} = 0$}\\
  When $J_{(-)} = 0$ -- so that $J_{(+)} = J_+ = J_-$ gives rise to a K{\"a}hler structure -- condition (\ref{st2}) becomes empty, while the other two reduce to the conditions for a B brane in a K{\"a}hler manifold with complex structure $J_+$, as is reviewed in appendix \ref{app:geo}.
  \item \underline{$\Pi_{(-)} $ is invertible}\\
  As explained before, this implies that $\Pi_{(-)}^{-1} = \Omega^{(-)}$ is symplectic. Condition (\ref{st2}) then reduces to the requirement that ${\cal N}$ be coisotropic with respect to $\Omega^{(-)}$. Indeed, in this case we have that $\Pi_{(-)} (\Ann T_{\cal N}) = T_{\cal N}^\bot$, the symplectic complement introduced in eq.~(\ref{sc}).

  Condition (\ref{st1}) is most straightforwardly analyzed by first
  introducing $F = {\cal F} - b^{(-)} = {\cal F} + \Omega^{(-)}
  J_{(+)}$. In terms of $F$, the condition becomes $\Pi_{(-)}
  (\iota_X F) = \Pi_{(-)} F X \in T_{\cal N}$. This condition was
  analyzed in subsection \ref{sympl} and the conclusion is that $F$
  is zero on $T_{\cal N}^\bot$ and descends to a two-form on $T_{\cal
  N}/T_{\cal N}^\bot$. This implies that on $T_{\cal N}^\bot$, ${\cal
  F} = b^{(-)}$. In particular, on a lagrangian submanifold ${\cal F}
  = b^{(-)}$ on the whole of ${\cal N}$, which agrees with
  (\ref{flux1}).

  Multiplying (\ref{st3}) by $\Pi_{(-)}$ from the left and using simple identities like $J_{(+)}^2 + J_{(-)}^2 = -1$ and $\Pi_{(-)} J_{(+)}^t = J_{(+)} \Pi_{(-)}$, we see that it implies
\begin{eqnarray}
(J_{(+)} + \Pi_{(-)} {\cal F})^2  = -1 \quad\mbox{on}\quad T_{\cal N}/T_{\cal N}^\bot.
\end{eqnarray}
It follows that $K = J_{(+)} + \Pi_{(-)} {\cal F}$ is an almost complex structure on $T_{\cal N}/T_{\cal N}^\bot$. This is precisely the complex structure $K$ arising from the $\sigma$-model, as follows from eq.~(\ref{flux2}). Indeed, since $\Pi_{(-)}$ is invertible, we can solve for ${\cal F}$,
\begin{eqnarray}
{\cal F} &=& \Omega^{(-)} (K - J_{(+)}) \\
&=& \Omega^{(-)} K + b^{(-)},
\end{eqnarray}
which is precisely eq.~(\ref{flux2}).

These results are actually nothing but the already known conditions for a coisotropic brane on a symplectic manifold with three-form flux $H = db^{(-)}$, albeit stated more explicitly than is usually done. In fact, the above conclusions could have been obtained more straightforwardly by first performing a b-transform of (\ref{gcsm}) with $b = -b^{(-)}$. As discussed in appendix \ref{app:geo}, the resulting generalized complex structure ${\cal J}_b = e^b {\cal J}e^{-b}$ is untwisted since $H +db = H - db^{(-)} = 0$. The resulting ${\cal J}_b$ actually turns out to be of canonical symplectic form, eq.~(\ref{cands}) with $\Omega = \Omega^{(-)}$. All the above results then follow from the results for a canonical generalized complex structure for a symplectic manifold, reviewed in subsection \ref{sympl}. For instance, the extra complex structure is $K = \Pi_{(-)} F = \Pi_{(-)}({\cal F} - b^{(-)}) = J_{(+)} + \Pi_{(-)} {\cal F}$ as before.

  \item \underline{No semi-chiral superfields}\\
  Even when $\Pi_{(-)}$ is not invertible, condition eq.~(\ref{st2}) is a coisotropy condition in the sense discussed in appendix \ref{app:pois} in the context of Poisson geometry. Indeed, while isotropic submanifolds have no natural generalization for (non-symplectic) Poisson structures, coisotropic submanifolds do. While mathematicians would call such submanifolds coisotropic in the generic case, in order to make the distinction clear, we speak of generalized coisotropic once the Poisson structure in question is not invertible.

  The simplest non-symplectic case is the one where no semi-chiral fields are present. Since in this case, we can compute things quite explicitly, let us try to get some intuition for the general case by first considering this one. We write the tangent space of ${\cal M}$ at some point $x$ as a sum of a chiral and a twisted chiral part, $T_{\cal M} = T_{\cal C} \oplus T_{\cal T}$. Denoting the canonical (diagonal) complex structure by $J$, and $\omega_{c,t} = -g_{c,t}J$, then we get the Poisson structures (we also use that the metric has a block diagonal form with blocks $g_c$ and $g_t$)
  \begin{eqnarray}
  \Pi_{(+)}  =
                                        \left( \begin{array}{cc}
                                        \omega_c^{-1} & 0 \\
                                        0  & 0
                                        \end{array}\right), \quad
   \Pi_{(-)}  =
                                        \left( \begin{array}{cc}
                                        0 & 0 \\
                                        0  & \omega_t^{-1}
                                        \end{array}\right). \quad
  \end{eqnarray}
  This implies that a lot of the analysis splits up in conditions on $T_{\cal C}$ and $T_{\cal T}$ seperately. Using the language of symplectic foliations introduced in appendix \ref{app:pois}, a symplectic leave associated to $\Pi_{(\pm)}$ is denoted by $S^\pm$. At a point $x$, this implies that $S^+_x = T_{\cal C}$ and $S^-_x = T_{\cal T}$. Now according to eq.~\rref{pcois2}, $\Pi_{(-)} (\Ann T_{\cal N}) = T_{\cal N,T} ^\bot$, the symplectic complement of $T_{\cal N,T} \equiv T_{\cal N} \cap S^-_x$ in $S^-_x$, where the symplectic structure is the inverse of the restriction of $\Pi_{(-)}$ to $S^-_x = T_{\cal T}$, namely $\omega_t$.

  Eq.~(\ref{st2}) then implies that $T_{\cal N,T}^\bot \subset T_{\cal N}$. In fact, because of the block diagonal structure of $\Pi_{(-)}$, $T_{\cal N,T}$ should be a coisotropic subspace of $T_{\cal T}$.

  Note however that ${\cal F}$ can a priori still have mixed indices. Condition (\ref{st1}) on one hand says $J (T_{\cal N,C}) \subset T_{\cal N,C}$, where $T_{\cal N,C} = T_{\cal N} \cap T_{\cal C}$, so that the chiral directions of ${\cal N}$ are `holomorphic'. The term involving ${\cal F}$ however reduces to a condition on $T_{\cal N,T}$. It implies that $\iota_X {\cal F} = 0$ for $X \in T_{\cal N,T}^\bot$, the symplectic complement of $T_{\cal N,T}$ for $\omega_t$ restricted to $T_{\cal T}$. Note that, since this condition follows from restricting $\Pi_{(-)}$ to $T_{\cal T}$, this says nothing about components of ${\cal F}$ with one leg in $T_{\cal N,T}^\bot$ and one along $T_{\cal N,C}$. Indeed such components were shown to be non-zero in \cite{Sevrin:2007yn}.

  Finally, eq.~(\ref{st3}) requires more care. First of all, multiplying it by $\Pi_{(-)}$, we get as before that $K = J_{(+)} + \Pi_{(-)} {\cal F}$ is a complex structure on $T_{\cal N} / T_{\cal N,T}^\bot$. This is indeed the object we called ${\cal K}$ in \cite{Sevrin:2007yn}. However, since $\Pi_{(-)}$ is not invertible, multiplying eq.~(\ref{st3}) by $\Pi_{(-)}$ yields only part of the necessary conditions. The remaining conditions are obtained by multiplying eq.~(\ref{st3}) by $\Pi_{(+)}$. This yields a $\Pi_{(+)}(\Ann T_{\cal N})$ on the right hand side of the inclusion. This equals $(T_{\cal N,C})^\bot$, where now $\omega_c$ on $T_{\cal C}$ has to be used. Since $T_{\cal N,C}$ is a symplectic subspace of $T_{\cal C}$ (see appendix \ref{app:sub}), its symplectic complement is zero when all chiral fields are taken to be Neumann. Restricting to this case for simplicity (and using the fact that $J_{(+)}  J_{(-)} = 0$ in absence of semi-chiral fields), we find
\begin{eqnarray}
J_{(+)} \Pi_{(+)} {\cal F} + \Pi_{(+)} {\cal F} J_{(+)} + \Pi_{(+)} {\cal F} \Pi_{(-)} {\cal F} = 0 \quad\mbox{on}\quad T_{\cal N}.  \label{noholo}
\end{eqnarray}
This equation generalizes the holomorphicity condition for the $U(1)$ flux on a B brane, showing that for instance the field strengths along the chiral directions are generically no longer holomorphic.
Indeed, letting $\alpha$ and $\beta$ run over chiral, and $\mu$ and $\nu$ over twisted chiral fields, one of the equations implied by eq.~(\ref{noholo}) is the following condition on ${\cal F}_{\alpha\beta}$,
\begin{eqnarray}
2 {\cal F}_{\alpha\beta} + {\cal F}_{\alpha\mu} g^{\mu\bar\nu} {\cal
F}_{\bar\nu\beta} - {\cal F}_{\alpha\bar\mu} g^{\bar\mu\nu} {\cal
F}_{\nu\beta} = 0.
\end{eqnarray}

  \item \underline{General case}\\
   Here again, we use the notation $S^\pm$ for the symplectic leaves associated to $\Pi_{(\pm)}$. Writing locally $T_{\cal M} = T_{\cal C} \oplus T_{\cal T} \oplus T_{\cal S}$ where the last term now adds the semi-chiral fields, we find that $S^+_x = T_{\cal C} \oplus T_{\cal S}$ and $S^-_x = T_{\cal T} \oplus T_{\cal S}$. Let us denote $S^-_x \cap T_{\cal N}$ by $S^-_{\cal N}$ in the following. Condition \rref{st2} then states that
   \begin{eqnarray}
   \left( S^-_{\cal N} \right)^\bot \subset T_{\cal N},
   \end{eqnarray}
   where the symplectic complement is with respect to the inverse of the restriction of $\Pi_{(-)}$ to $S^-$. This is indeed essentially the structure that was found by analyzing boundary conditions in the $\sigma$-model. Of course much more remains to be analyzed, especially concerning the invariant field strength ${\cal F}$ on the brane.\footnote{As far as the authors are aware, the most general analysis has so far not appeared in the literature in the amount of detail required for comparison with $\sigma$-model results.
   An equation similar and related to eq.~(\ref{st3}) has been studied in \cite{Gualtieri:2007,Kapustin:2005vs} for slightly different, but ultimately related reasons.}
   Let us simply note here that one can of course still multiply eq.~(\ref{st3}) by $\Pi_{(-)}$ from the right to obtain
   \begin{eqnarray}
   K^2 = -1 \quad \mbox{on} \quad T_{\cal N}/ \left( S^-_{\cal N} \right)^\bot,
   \end{eqnarray}
  where $K$ is still of the form (\ref{genK}). This indeed agrees with eq.~(\ref{acs}).
\end{enumerate}

\section{Duality transformations}
In $N=(2,2)$ supersymmetric models there exists a variety of duality
transformations which allows one to change the nature of the superfields.
These duality transformations fall into two categories: those which need
an isometry and those which do not. The former are what is usually
understood as a T-duality transformation while the latter are a
consequence of the constraints which are imposed on $N=(2,2)$
superfields. A complete catalogue of duality transformations in $N=(2,2)$
superspace was obtained in \cite{Grisaru:1997ep}. Here we generalize this
to the situation where boundaries are present. The main subtlety consists
in finding the proper boundary terms in the first order action which
guarantee that the boundary conditions consistently pass through the
duality transformation.

\subsection{Dualities without an isometry}
The basic idea of dualities without an isometry is to impose the
constraints on the superfields through Lagrange multipliers
(unconstrained superfields). In a first order formulation one takes the
original fields as unconstrained superfields. Integrating over the
Lagrange multipliers brings us back to the original model. However, if we
integrate over the original unconstrained fields we get the dual
formulation. In this way one has the following dual combinations:
\begin{itemize}
\item Four dual semi-chiral formulations.
  \item Twisted chiral field $\leftrightarrow$ twisted complex linear superfield.
  \item Chiral field $\leftrightarrow$ complex linear superfield.
\end{itemize}
In the present paper we briefly introduce these duality transformations
and postpone a detailed analysis of them -- which requires a careful
treatment of the boundary conditions -- to a forthcoming paper.

\subsubsection{The four dual semi-chiral formulations}
The starting point is the first order action,
\begin{eqnarray}
{\cal S}&=&-\int\,d^2 \sigma \,d^2\theta \,d^2  \theta' \, \Big(V(l,\bar l,r,\bar r,\cdots)-
\Lambda ^ +\,\bar \ID_+ l-\bar \Lambda ^+\,\ID_+ \bar l- \Lambda ^-\,\bar \ID_- r-
\bar \Lambda ^-\,\ID_- \bar r\Big)\nonumber\\
&&+i\int d\tau \,d^2\theta \Big(W(l,\bar l,r,\bar r,\cdots)+i
\Lambda ^ +\,\bar \ID_+ l-i\bar \Lambda ^+\,\ID_+ \bar l- i\Lambda
^-\,\bar \ID_- r+i \bar \Lambda ^-\,\ID_- \bar r\Big),
\end{eqnarray}
where $l$, $\bar l$, $r$ and $\bar r$ are unconstrained bosonic complex
superfields and $\Lambda ^\pm$ and $\bar \Lambda ^\pm$ are unconstrained
complex fermionic superfields. Integrating over the Lagrange multipliers
constrains $l$ and $r$ to form a semi-chiral multiplet. Upon partial
integration we can rewrite the action in three ways,
\begin{eqnarray}
{\cal S}&=&-\int\,d^2 \sigma \,d^2\theta \,d^2  \theta' \, \Big(V(l,\bar l,r,\bar r,\cdots)-
l\, l'-\bar l\, \bar l'- \Lambda ^-\,\bar \ID_- r- \bar \Lambda ^-\,\ID_- \bar r\Big)\nonumber\\
&&+i\int d\tau \,d^2\theta \Big(W(l,\bar l,r,\bar r,\cdots)+i\,l\, l'
-i\,\bar l\, \bar l'- i\Lambda ^-\,\bar \ID_- r+i \bar \Lambda ^-\,\ID_- \bar r\Big)\nonumber\\
&=&-\int\,d^2 \sigma \,d^2\theta \,d^2  \theta' \, \Big(V(l,\bar l,r,\bar r,\cdots)-
\Lambda ^ +\,\bar \ID_+ l-\bar \Lambda ^+\,\ID_+ \bar l- r\, r'- \bar r\, \bar r'\Big)\nonumber\\
&&+i\int d\tau \,d^2\theta \Big(W(l,\bar l,r,\bar r,\cdots)+i
\Lambda ^ +\,\bar \ID_+ l-i\bar \Lambda ^+\,\ID_+ \bar l- i\,r\, r'+i\,\bar r\,\bar r'\Big)\nonumber\\
&=&-\int\,d^2 \sigma \,d^2\theta \,d^2  \theta' \, \Big(V(l,\bar l,r,\bar r,\cdots)-
l\, l'-\bar l\, \bar l'- r\, r'- \bar r\, \bar r'\Big)\nonumber\\
&&+i\int d\tau \,d^2\theta \Big(W(l,\bar l,r,\bar r,\cdots)+i
\,l\, l'-i\,\bar l\, \bar l'- i\,r\, r'+i\,\bar r\,\bar r'\Big),
\end{eqnarray}
where we introduced the notation $l'=\bar \ID_+\Lambda ^+$, $\bar l'=
\ID_+\bar \Lambda ^+$, $r'=\bar \ID_-\Lambda ^-$, $\bar r'= \ID_-\bar
\Lambda ^-$. Integrating over the unconstrained fields $( l,\bar l
,\Lambda ^-,\bar \Lambda ^-)$, $(\Lambda ^+,\bar \Lambda ^+,r,\bar r)$ or
$(l,\bar l,r ,\bar r )$ resp.~yields three dual formulations of the
model.

\subsubsection{The duality between twisted chiral and twisted complex linear fields}
This duality transformation is fully determined by the following two
equivalent versions of the first order action,
\begin{eqnarray}
 {\cal S}&=&-\int\,d^2 \sigma \,d^2\theta \,d^2  \theta' \, \Big(V(w,\bar w,\cdots)
-\Lambda ^+\bar \ID_+ w-\bar \Lambda ^-\ID_- w-\bar \Lambda ^+\ID_+\bar  w-\Lambda ^-\bar \ID_- \bar w
 \Big)\nonumber\\
 &&+i\,\int\,d\tau  \,d^2\theta \, \Big(W(w,\bar w,\cdots)
+i\,\Lambda ^+\bar \ID_+ w+i\,\bar \Lambda ^-\ID_- w-i\,\bar \Lambda ^+\ID_+\bar  w-i\,\Lambda ^-\bar \ID_- \bar w
 \Big)\nonumber\\
 &=&-\int\,d^2 \sigma \,d^2\theta \,d^2  \theta' \, \Big(V(w,\bar w,\cdots)-w\,x-\bar w\,\bar x\Big)\nonumber\\
&&\qquad +i\,\int\,d\tau  \,d^2\theta \,\Big(W(w,\bar w,\cdots)+i\,w\,x-i\,\bar w\,\bar x\Big),
\end{eqnarray}
where $\Lambda ^\pm$ are unconstrained complex fermionic superfields and
we wrote $x\equiv \bar \ID_+\Lambda ^++\ID_-\bar \Lambda ^-$ and $\bar
x\equiv  \ID_+\bar \Lambda ^++\bar \ID_- \Lambda ^-$. We identify $x$ as
a twisted complex linear superfield defined by the constraints quadratic
in the derivatives: $\bar \ID_+\ID_-x= \ID_+\bar \ID_-\bar x=0$
\cite{Gates:1995du}. Integrating over $\Lambda ^\pm$ and $\bar \Lambda
^\pm$ constrains $w$ and $\bar w$ to be twisted chiral. If on the other
hand we first integrate over the unconstrained fields $w$ and $\bar w$,
we end up with the dual description where the dependence on a twisted
chiral field was exchanged for one on a twisted complex linear
superfield.

\subsubsection{The duality between chiral and complex linear fields}
Starting from the potentials $V(z,\bar z,\cdots)$ and $W(z,\bar
z,\cdots)$, where $z$ is a chiral field, we write a first order action,
\begin{eqnarray}
 {\cal S}&=&-\int\,d^2 \sigma \,d^2\theta \,d^2  \theta' \, \Big(V(z,\bar z,\cdots)
-\Lambda ^+\bar \ID_+ z- \Lambda ^-\bar \ID_- z-\bar \Lambda ^+\ID_+\bar  z-\bar \Lambda ^- \ID_- \bar z
 \Big)\nonumber\\
 &&+i\,\int\,d\tau  \,d^2\theta \, \Big(W(z,\bar z,\cdots)
+i\,\Lambda ^+\bar \ID_+ z-i\, \Lambda ^-\bar \ID_- z-i\,\bar
\Lambda ^+\ID_+\bar  z+i\,\bar \Lambda ^- \ID_- \bar z
 \Big),
\end{eqnarray}
where we now take $z$ and $\bar z$ as unconstrained superfields and
$\Lambda ^\pm$ and $\bar \Lambda ^\pm$ are (unconstrained) Lagrange
multipliers. Varying the Lagrange multipliers gives the original model.
Upon partial integration we can rewrite the first order action as,
\begin{eqnarray}
 {\cal S}&=&-\int\,d^2 \sigma \,d^2\theta \,d^2  \theta' \, \Big(V(z,\bar z,\cdots)-z\,x-\bar z\,\bar x\Big)\nonumber\\
&& +i\,\int\,d\tau  \,d^2\theta \,\Big(W(z,\bar z,\cdots)+i\,z\,
\big(\bar \ID_+\Lambda ^+-\bar \ID_-\Lambda ^-\big)-i\,\bar z\,
\big( \ID_+\bar \Lambda ^+- \ID_-\bar \Lambda ^-\big),
\end{eqnarray}
where $x\equiv\bar \ID_+\Lambda ^++\bar \ID_-\Lambda ^-$ is a complex
linear superfield defined by the constraints $\bar \ID_+\bar
\ID_-x=\ID_+\ID_- \bar x=0$ \cite{Gates:1980az}, \cite{Deo:1985ix}. The
treatment of the boundary term in the action and the boundary conditions
requires special care. We postpone this discussion to a future paper.

\subsection{Dualities with an isometry}
The main idea here is to gauge the isometry and through Lagrange
multipliers enforce the gauge fields to be pure gauge. Integrating over
the Lagrange multipliers brings us back to the original model while
integrating over the gauge fields results in the dual model. The
treatment of the boundary conditions through the duality transformation
requires special care.
\subsubsection{The duality between a pair of chiral and twisted chiral
fields and a semi-chiral multiplet}

The starting point is a bulk potential of the form $V\big(z+\bar z,w+\bar
w,i(z-\bar z-w+\bar w),\cdots\big)$ and a boundary potential
$W\big(z+\bar z,w+\bar w,i(z-\bar z-w+\bar w),\cdots\big)$. This clearly
exhibits the isometry $z\rightarrow z+i\,a$, $w\rightarrow w+i\,a$, with
$a$ an arbitrary real constant\footnote{While this duality transformation
was already found in \cite{Grisaru:1997ep}, the elucidation of the
underlying gauge structure is rather recent
\cite{Lindstrom:2007vc}-\cite{Merrell:2007sr}.}. The first order action
is,
\begin{eqnarray}
 {\cal S}&=&-\int\,d^2 \sigma \,d^2\theta \,d^2  \theta' \, \Big(V\big(Y,\tilde Y,\hat Y,\cdots\big)
 +\Lambda ^+\bar \ID_+\big(Y-\tilde Y-i\,\hat Y\big)
 +\bar \Lambda ^+ \ID_+\big(Y-\tilde Y+i\,\hat Y\big)\nonumber\\
&&-\Lambda ^-\bar \ID_-\big(Y+\tilde Y-i\,\hat Y\big)
 -\bar \Lambda ^- \ID_-\big(Y+\tilde Y+i\,\hat Y\big)\Big)\nonumber\\
 &&+i\int d\tau \,d^2\theta \Big(W\big(Y,\tilde Y,\hat Y,\cdots\big)
-i\,\Lambda ^+\bar \ID_+\big(Y-\tilde Y-i\,\hat Y\big)
 +i\,\bar \Lambda ^+ \ID_+\big(Y-\tilde Y+i\,\hat Y\big)\nonumber\\
&&-i\,\Lambda ^-\bar \ID_-\big(Y+\tilde Y-i\,\hat Y\big)
 +i\,\bar \Lambda ^- \ID_-\big(Y+\tilde Y+i\,\hat Y\big)\Big),
\end{eqnarray}
where $\Lambda ^\pm$ and $\bar \Lambda ^\pm$ are unconstrained complex
fermionic superfields and $Y$, $\tilde Y$ and $\hat Y$ are unconstrained
real bosonic superfields. Integrating over the Lagrange multipliers
$\Lambda ^\pm$ and $\bar \Lambda ^\pm$ returns us to the original model.
Upon partial integration we rewrite the first order action as,
\begin{eqnarray}
 {\cal S}&=&-\int\,d^2 \sigma \,d^2\theta \,d^2  \theta' \, \Big(V\big(Y,\tilde Y,\hat Y,\cdots\big)
 +Y\big(l+\bar l-r-\bar r\big)
-\tilde Y\big(l+\bar l+r+\bar r\big)\nonumber\\
&&-i\,\hat Y\big(l-\bar l-r+\bar r\big)\Big)
 +i\int d\tau \,d^2\theta \Big(W\big(Y,\tilde Y,\hat Y,\cdots\big)
 -i\,Y\big(l-\bar l+r-\bar r\big)\nonumber\\
 &&+i\,
\tilde Y\big(l-\bar l-r+\bar r\big)-
\hat Y\big(l+\bar l+r+\bar r\big)
\Big),\label{1storderCTS}
\end{eqnarray}
where we introduced the semi-chiral multiplet $l=\bar \ID_+\Lambda ^+$,
$\bar l=\ID_+\bar \Lambda ^+$, $r=\bar \ID_-\Lambda ^-$ and $\bar r=
\ID_-\bar \Lambda ^-$. Integrating over $Y$, $\tilde Y$ and $\hat Y$
yields the dual model.

Let us illustrate this with a simple example. Our starting point is a
model on $T^4$ parameterized by a twisted chiral, $w$, and a chiral, $z$,
superfield. We take for the generalized K{\"a}hler potential,
\begin{eqnarray}
 V=-\frac 1 4 \big(z+\bar z-w-\bar w\big)^2+
\frac 1 4 \big(z-\bar z-w+\bar w\big)^2+\big(z+\bar z\big)^2.\label{EXb1}
\end{eqnarray}
We consider a D3-brane whose location is fixed by the Dirichlet boundary
condition,
\begin{eqnarray}
 i\big(z-\bar z-w+\bar w)=i(1-a)\big(z-\bar z\big),\label{EXb2}
\end{eqnarray}
where $a\in\IQ$. Using the methods of section 3 one finds the boundary
potential,
\begin{eqnarray}
 W=\frac i 2 \big(z-\bar z-w+\bar w\big)\big(w+\bar w), \label{EXb4}
\end{eqnarray}
to which we could have added an arbitrary real function of $z$ and $\bar
z$. When dualizing this to a semi-chiral model, we have to distinguish
two cases: $a=1$ and $a\neq 1$.

Consider the case where $a=1$. In that case eq.~(\ref{EXb2}) implies a
Dirichlet boundary condition for the gauge fields: $\hat Y=0$ and the
boundary potential $W$ vanishes. The first order action
eq.~(\ref{1storderCTS}) becomes,
\begin{eqnarray}
 {\cal S}&=&-\int\,d^2 \sigma \,d^2\theta \,d^2  \theta' \, \Big(-\frac 1 4 \,
 \big(Y-\tilde Y\big)^2-\frac 1 4 \,\hat Y{}^2+Y^2
 +Y\big(l+\bar l-r-\bar r\big)
-\tilde Y\big(l+\bar l+r+\bar r\big)\nonumber\\
&&-i\,\hat Y\big(l-\bar l-r+\bar r\big)\Big)
 +i\int d\tau \,d^2\theta \Big(
 -i\,Y\big(l-\bar l+r-\bar r\big)+i\,
\tilde Y\big(l-\bar l-r+\bar r\big)\Big).\label{1storderCTSXX}
\end{eqnarray}
From the bulk equations of motion we get,
\begin{eqnarray}
 Y&=&r+\bar r,\nonumber\\
 \tilde Y&=&-2\big(l+\bar l\big)-\big(r+\bar r\big),\nonumber\\
 \hat Y&=&-2i\big(l-\bar l-r+\bar r\big).\label{EXb3}
\end{eqnarray}
Note that we already had a Dirichlet boundary condition $\hat Y=0$ which
is reproduced by varying $\tilde Y$ in the boundary term in the first
order action eq.~(\ref{1storderCTSXX}). Varying $Y$ in the boundary term
yields a second Dirichlet boundary condition which together with the
first one imply,
\begin{eqnarray}
 l=\bar l,\qquad r=\bar r.\label{EXb5}
\end{eqnarray}
So in the dual model we obtain a generalized lagrangian D2-brane whose
location is specified by eq.~(\ref{EXb5}), the boundary potential
vanishes and the bulk potential is given by,
\begin{eqnarray}
 V_{dual}=\big(l+\bar l+r+\bar r\big)^2- \big(l-\bar l-r+\bar r\big)^2-
 \big(r+\bar r\big)^2.\label{EXb8}
\end{eqnarray}

We now consider the case $a\neq 0$ where for simplicity we choose $a=0$.
Eq.~(\ref{EXb2}) results in the boundary conditions,
\begin{eqnarray}
 \bar \ID\big(\hat Y+iY)=\ID\big(\hat Y-iY)=0,
\end{eqnarray}
which implies that,
\begin{eqnarray}
 Z^1\equiv \hat Y +iY= -2i\,l+2i\,\bar l+3i\,r-i\,\bar r,
\end{eqnarray}
is a boundary chiral field! With this, the first order action
eq.~(\ref{1storderCTS}) becomes,
\begin{eqnarray}
 {\cal S}&=&-\int\,d^2 \sigma \,d^2\theta \,d^2  \theta' \, \Big(-\frac 1 4 \,
 \big(Y-\tilde Y\big)^2-\frac 1 4 \,\hat Y{}^2+Y^2
 +Y\big(l+\bar l-r-\bar r\big)
-\tilde Y\big(l+\bar l+r+\bar r\big)\nonumber\\
&&-i\,\hat Y\big(l-\bar l-r+\bar r\big)\Big)
 +i\int d\tau \,d^2\theta \Big(\frac 1 4 \,\big(Z^1+\bar Z^{\bar 1}\big)\tilde Y+
 i\,\tilde Y\big(l-\bar l-r+\bar r\big)\nonumber\\
 &&-Z^1\big(l+r\big)-\bar Z^{\bar 1}\big(\bar l+\bar r\big)
 \Big).\label{1storderCTSXXX}
\end{eqnarray}
Obviously the bulk equations of motion are again given by
eq.~(\ref{EXb3}). Varying $\tilde Y$ in the boundary term of the first
order action eq.~(\ref{1storderCTSXXX}) gives an expression compatible
with the bulk equations of motion eq.~(\ref{EXb3}). Varying $Z^1$ and
$\bar Z^{\bar 1}$ -- taking into account that they are {\em constrained}
boundary superfields -- gives,
\begin{eqnarray}
 \bar \ID\big(\frac 1 4 \,\tilde Y-l-r\big)=
\ID\big(\frac 1 4 \,\tilde Y-\bar l-\bar r\big)=0,
\end{eqnarray}
implying the existence of a second boundary chiral field $Z^2$,
\begin{eqnarray}
 Z^2\equiv\frac 1 4 \,\tilde Y-l-r= -\frac 3 2 \,l-\frac 1 2 \,\bar l-\frac 5 4 \,r-\frac 1 4\,\bar r.
\end{eqnarray}
So we end up with a maximally coisotropic brane on $T^4$. Labelling rows
and columns as $(l,\bar l,r,\bar r)$, we get for the complex structure
$K$,
\begin{eqnarray}
K=\frac 1 4\, \left(
\begin{array}{cccc} 6 i& -2 i& -i& 3 i \\
2i& -6i&-3i&i \\
-2 i& 6 i& 6 i& -2 i \\
-6 i&2 i&2 i&-6 i
\end{array}\right).
\end{eqnarray}
The bulk potential is given by eq.~(\ref{EXb8}) and the boundary
potential is,
\begin{eqnarray}
 W_{dual}=-i\big(l-\bar l-r+\bar r\big)\big(2(l+\bar l)+(r+\bar
 r)\big).\label{EXb9}
\end{eqnarray}
In terms of the boundary chiral fields this becomes,
\begin{eqnarray}
 W_{dual}=-\frac 1 4 \,\big(Z^1+\bar Z^{\bar 1}\big)\big(
Z^2+\bar Z^{\bar 2}-\frac i 4 \,(Z^1-\bar Z^{\bar 1})
 \big)=-\frac 1 4 \,\big(
Z^1 \bar Z^{\bar 2}+\bar Z^{\bar 1}Z^2
 \big),
\end{eqnarray}
where in the last step we discarded total derivative terms.

We now focus on the inverse transformation. Starting point is a bulk
potential of the form $V\big(l+\bar l,r+\bar r,i(l-\bar l-r+\bar
r),\cdots\big)$ and a boundary potential $W\big(l+\bar l,r+\bar
r,i(l-\bar l-r+\bar r),\cdots\big)$ . The basic relation is given by,
\begin{eqnarray}
 {\cal S}&=&-\int\,d^2 \sigma \,d^2\theta \,d^2  \theta' \, \Big(V\big(Y,\tilde Y,\hat Y,\cdots\big)
+i\,u\bar \ID_+\bar \ID_-\big(Y-\tilde Y-i\hat Y\big)\nonumber\\
&&+i\,\bar u \ID_+ \ID_-\big(Y-\tilde Y+i\hat Y\big)
-i\,v\bar \ID_+ \ID_-\big(Y+\tilde Y-i\hat Y\big)
-i\,\bar v\ID_+\bar \ID_-\big(Y+\tilde Y+i\hat Y\big)
 \Big)\nonumber\\
 &&+i\int d\tau \,d^2\theta \Big(W\big(Y,\tilde Y,\hat Y,\cdots\big)
-\frac 1 2 \bar \ID'u\bar \ID'\big(Y-\tilde Y-i\hat Y\big)
+\frac 1 2  \ID'\bar u \ID'\big(Y-\tilde Y+i\hat Y\big)\nonumber\\
&&-v\,\bar \ID_+ \ID_-\big(Y+\tilde Y-i\hat Y\big)
+\bar v\,\ID_+\bar \ID_-\big(Y+\tilde Y+i\hat Y\big)
 \Big)\nonumber\\
 &=&-\int\,d^2 \sigma \,d^2\theta \,d^2  \theta' \, \Big(V\big(Y,\tilde Y,\hat Y,\cdots\big)
+Y\big(z+\bar z-w-\bar w\big)-\tilde Y\big(z+\bar z+w+\bar w\big)\nonumber\\
&&-i\,\hat Y\big(z-\bar z-w+\bar w\big)\Big)
+i\int d\tau \,d^2\theta \Big(W\big(Y,\tilde Y,\hat Y,\cdots\big)\nonumber\\
&&+i\,\big(Y+\tilde Y\big)\big(w-\bar w\big)+\hat Y\big(w+\bar
w\big) \Big), \label{EXb10}
\end{eqnarray}
where $u,\,v\in\IC$ and $Y,\,\tilde Y,\,\hat Y\in\IR$ are unconstrained
superfields and where we defined $z=i\bar \ID_+\bar \ID_-u$, $\bar z=i
\ID_+ \ID_-\bar u$, $w=i\bar \ID_+ \ID_-v$ and $\bar w=i \ID_+\bar
\ID_-\bar v$. When using this, special attention must be given to the
boundary terms proportional to $\bar \ID'u$ and $ \ID'\bar u$.

Again we will illustrate this with a simple example. Indeed we will
dualize the lagrangian D2- and the coisotropic D4-brane obtained above
back to a D3-brane in terms of a twisted chiral and a chiral field. The
bulk potential we start from is given by eq.~(\ref{EXb8}). For the
D2-brane we consider the Dirichlet boundary conditions eq.~(\ref{EXb5})
and a vanishing boundary potential. The Dirichlet boundary condition
imply $\hat Y=0$ on the boundary. Using the first part of relation
eq.~(\ref{EXb10}), we find the first order action to be,
\begin{eqnarray}
{\cal S}&=&-\int\,d^2 \sigma \,d^2\theta \,d^2  \theta' \, \Big(\big(Y + \tilde Y\big)^2 + \hat Y ^2 - \tilde Y ^2
+i\,u\bar \ID_+\bar \ID_-\big(Y-\tilde Y-i\hat Y\big)\nonumber\\
&&+i\,\bar u \ID_+ \ID_-\big(Y-\tilde Y+i\hat Y\big)
-i\,v\bar \ID_+ \ID_-\big(Y+\tilde Y-i\hat Y\big)
-i\,\bar v\ID_+\bar \ID_-\big(Y+\tilde Y+i\hat Y\big)
 \Big)\nonumber\\
 &&+i\int d\tau \,d^2\theta \Big( -v\,\bar \ID_+ \ID_-\big(Y+\tilde Y-i\hat Y\big)
+\bar v\,\ID_+\bar \ID_-\big(Y+\tilde Y+i\hat Y\big)
\nonumber\\
&&-\frac 1 2 \bar \ID'u \left(\bar \ID'\big(Y-\tilde Y-i\hat Y\big) + \bar \ID\big(Y+\tilde Y \big) \right)
+\frac 1 2  \ID'\bar u \left(\ID'\big(Y-\tilde Y+i\hat Y\big) + \ID\big(Y+\tilde Y \big) \right)
 \Big),\nonumber\\
\end{eqnarray}
where we added two extra terms to the boundary term proportional to $\bar
\ID'u$ and $ \ID'\bar u$ such that the variation of $\bar \ID'u$ and
$\ID'\bar u$ yields expressions compatible with the boundary conditions
and the constraints eq.~(\ref{chicon}). Integrating this action by parts
yields,
\begin{eqnarray}
{\cal S}&=&-\int\,d^2 \sigma \,d^2\theta \,d^2  \theta' \,
\Big(\big(Y + \tilde Y\big)^2 + \hat Y ^2 - \tilde Y ^2
+Y\big(z+\bar z-w-\bar w\big) \nonumber\\
&&\quad \quad-\tilde Y\big(z+\bar z+w+\bar w\big) -i\,\hat Y\big(z-\bar z-w+\bar w\big)\Big)\nonumber \\ && +i\int d\tau
\,d^2\theta \Big( +i\,\big(Y+\tilde Y\big)\big(w-\bar w -z+\bar z\big)
\Big),\nonumber\\ \label{1storderCTSD2D3}
\end{eqnarray}
where the boundary term containing the chiral field $z$, $\bar z$ results
from the additional terms added in the first order action we started
from. We used the boundary condition $\hat Y=0$ as well.

The bulk equations of motion give,
\begin{eqnarray}
Y&=& \frac{1}{2} \big( z+\bar z + w +\bar w \big), \nonumber\\
\tilde Y&=& -\big(z+ \bar z \big), \nonumber\\
\hat Y &=& \frac{i}{2} \big(z-\bar z - w +\bar w \big).
\label{EXb11}
\end{eqnarray}
Inserting these equations of motion back into the bulk part of the
action eq.~(\ref{1storderCTSD2D3}) reproduces the generalized
K{\"a}hler potential eq.~(\ref{EXb1}). The Dirichlet boundary
condition $\hat Y=0$ implies -- using eq.~(\ref{EXb11}) -- the
Dirichlet boundary condition in eq.~(\ref{EXb2}) with $a=1$. So we
do recover the D3-brane discussed previously. Varying either $Y$ or
$\tilde Y$ in eq.~(\ref{1storderCTSD2D3}) yields an expression which
vanishes by virtue of the Dirichlet boundary condition. As a
consequence the boundary term in the dual action vanishes as
expected.

Next, we consider the coisotropic D4-brane constructed above, given by
the Neumann boundary conditions,
\begin{eqnarray}
&&\bar \ID \big(-2 i\, l + 2 i\, \bar l + 3 i\, r - i\, \bar r \big)
= 0,
\quad \ID \big(-2 i\, l + 2 i\, \bar l +  i\, r - 3 i\, \bar r \big) = 0,  \label{Neum1}\\
&&\bar \ID\big(-6 l - 2 \bar l - 5 r - \bar r \big) = 0, \quad \quad
\quad \ID\big(-2 l - 6 \bar l -  r - 5 \bar r \big) = 0,
\label{Neum2}
\end{eqnarray}
and the boundary potential eq.~(\ref{EXb9}). The Neumann boundary
conditions eq.~(\ref{Neum1}) imply that $\tilde Y$ and $\hat Y$ together
form a chiral boundary field,
\begin{eqnarray}
\bar \ID\big(\hat Y - \frac{i}{2} \tilde Y \big) =
\ID\big(\hat Y + \frac{i}{2} \tilde Y \big) = 0.\label{abccba}
\end{eqnarray}
The first order action reads,
\begin{eqnarray}
{\cal S}&=&-\int\,d^2 \sigma \,d^2\theta \,d^2  \theta' \, \Big( \big(Y + \tilde Y\big)^2 + \hat Y ^2 - \tilde Y ^2
+i\,u\bar \ID_+\bar \ID_-\big(Y-\tilde Y-i\hat Y\big)\nonumber\\
&&\qquad+i\,\bar u \ID_+ \ID_-\big(Y-\tilde Y+i\hat Y\big)
-i\,v\bar \ID_+ \ID_-\big(Y+\tilde Y-i\hat Y\big)
-i\,\bar v\ID_+\bar \ID_-\big(Y+\tilde Y+i\hat Y\big)
 \Big)\nonumber\\
 &&+i\int d\tau \,d^2\theta \Big( - \hat Y \left( 2Y +\tilde Y \right)
-\frac 1 2 \bar \ID'u\left(\bar \ID'\big(Y-\tilde Y-i\hat Y\big) - \bar\ID\big(Y+\tilde Y + i\, \hat Y \big) \right) \nonumber \\
&&\qquad+\frac 1 2  \ID'\bar u \left(\ID'\big(Y-\tilde Y+i\hat Y\big) - \ID\big(Y+ \tilde Y - i\, \hat Y \big) \right)
-v\,\bar \ID_+ \ID_-\big(Y+\tilde Y-i\hat Y\big) \nonumber\\
&&\qquad +\bar v\,\ID_+\bar \ID_-\big(Y+\tilde Y+i\hat Y\big)
 \Big),
\end{eqnarray}
where once more we inserted  two additional terms in the boundary term
such that the variation of $\bar \ID'u$ and $ \ID'\bar u$ gives
expressions consistent with the Neumann boundary conditions and the
constraints eq.~(\ref{chicon}). After integrating this action by parts,
we find the following action,
\begin{eqnarray}
{\cal S}&=&-\int\,d^2 \sigma \,d^2\theta \,d^2  \theta' \,
\Big(\big(Y + \tilde Y\big)^2 + \hat Y ^2 - \tilde Y ^2
+Y\big(z+\bar z-w-\bar w\big) \nonumber\\
&&\quad \quad -\tilde Y\big(z+\bar z+w+\bar w\big) -i\,\hat Y\big(z-\bar z-w+\bar w\big)\Big)\nonumber \\ && +i\int d\tau
\,d^2\theta \Big( -\hat Y \left(2Y+\tilde Y\right)  +i\,\big(Y+\tilde Y\big)\big(w-\bar w + z - \bar z \big)+\hat
Y\big(w+\bar w - z - \bar z \big)
\Big).\nonumber\\ \label{1storderCTSD4D3}
\end{eqnarray}
The bulk analysis is similar to the previous case, with equations of
motion given in eqs.~(\ref{EXb11}). Varying the gauge field Y in the
boundary term and imposing the equation of motion for $\hat Y$
yields the Dirichlet boundary condition,
\begin{eqnarray}
i\, \big(w - \bar w \big) = 0, \label{EXb12}
\end{eqnarray}
which is indeed the boundary condition eq.~(\ref{EXb2}) for $a=0$. When
varying $\tilde Y$ and $\hat Y$ in the boundary term, one should take
into account that they are constrained at the boundary (see
eq.~(\ref{abccba})). Doing so correctly, one recovers again the Dirichlet
boundary condition eq.~(\ref{EXb12}).

\subsubsection{The duality between a chiral and a twisted chiral field}
Starting from a potential of the form $V(z+\bar z,\cdots)$ and $W(z+\bar
z,\cdots)$, we write the first order action,
\begin{eqnarray}
 {\cal S}&=&-\int\,d^2 \sigma \,d^2\theta \,d^2  \theta' \, \Big(V\big(Y,\cdots\big)
 -i\,u\,\bar \ID_+\ID_-Y-i\,\bar u \,\ID_+\bar \ID_-Y\Big)\nonumber\\
 &&\qquad+i\,
 \int d\tau \,d^2\theta \Big(W\big(Y,\cdots\big)-u\,\bar \ID_+\ID_-Y+\bar u \,\ID_+\bar \ID_-Y\Big).\label{xxfoa1}
\end{eqnarray}
Integrating over the complex unconstrained Lagrange multipliers $u$ and
$\bar u$ brings us back to the original model. Upon integrating by parts
one gets,
\begin{eqnarray}
 {\cal S}&=&-\int\,d^2 \sigma \,d^2\theta \,d^2  \theta' \, \Big(V\big(Y,\cdots\big)
 -Y\big(w+\bar w\big)\Big)\nonumber\\
 &&\qquad +i\,
 \int d\tau \,d^2\theta \Big(W\big(Y,\cdots\big)+i\,Y\big(w-\bar w\big)\Big),\label{xxfoa2}
\end{eqnarray}
where we introduced the twisted chiral fields $w=i\,\bar \ID_+\ID_-u$ and
$\bar w=i\,\ID_+\bar \ID_-\bar u$. Integrating over the unconstrained
gauge field $Y$ gives us the dual model in terms of a twisted chiral
field $w$.

We illustrate this with a simple example, a two-torus parameterized by a
chiral field with K{\"a}hler potential $V=(z+\bar z)^2/2$. Either D0- or
D2-brane configurations are allowed.

Let us start with a D0-brane. We take the Dirichlet boundary condition
$z=(a+i\,b)/2$ with $a,\,b\in\IR$ and constant. The boundary potential
vanishes. The first order action is given in eq.~(\ref{xxfoa1}) where the
gauge field $Y$ satisfies the boundary condition $Y=a$. Dualizing the
model using eq.~(\ref{xxfoa2}) we obtain the bulk equation of motion
$Y=w+\bar w$ which -- using the boundary condition for $Y$ -- gives us
the boundary condition for the twisted chiral field: $w+\bar w=a$.
Performing the duality transformation gives the potentials,
\begin{eqnarray}
 V_{dual}=-\frac 1 2\,\big(w+\bar w\big)^2,\qquad
 W_{dual}=i\,a\big(w-\bar w\big).
\end{eqnarray}
So we end up with a lagrangian D1-brane whose position is determined by
$w+\bar w=a$.

We now turn to the D2-brane. We still have the bulk potential $V=(z+\bar
z)^2/2$ but we can now allow for a boundary potential as well, which for
simplicity we choose as $F\,(z+\bar z)^2/2$ with $F\in\IR$ and constant.
The boundary conditions are fully Neumann and explicitly given by,
\begin{eqnarray}
 \ID'z=i\,F\,\ID z,\qquad \bar \ID'\bar z=-i\,F\,\bar \ID \bar z.
\end{eqnarray}
Once more our starting point is the first order action eq.~(\ref{xxfoa1})
where the gauge field $Y$ satisfies the boundary conditions
$\ID'Y=i\,F\,\ID Y$ and $\bar \ID'Y=-i\,F\,\bar \ID Y$. Using
eq.~(\ref{xxfoa2}) we obtain the dual model. The bulk equation of motion
gives $Y=w+\bar w$ which combined with the boundary conditions for $Y$
results in the boundary conditions $\ID\big(-i(w-\bar w)-F\,(w+\bar
w)\big) =\bar \ID\big(-i(w-\bar w)-F\,(w+\bar w)\big)=0$ where we used
eq.~(\ref{twichcon}). These equations are equivalent to a single
Dirichlet boundary condition,
\begin{eqnarray}
 -i\big(w-\bar w\big)=F\,\big(w+\bar w\big).\label{diducon}
\end{eqnarray}
The potentials for the dual model are given by,
\begin{eqnarray}
 V_{dual}=-\frac 1 2 \big(w+\bar w\big)^2,\qquad W_{dual}=-\frac 1 2\, F\,\big(w+\bar w\big)^2.
\end{eqnarray}
As was to be expected we find a lagrangian D1-brane whose position is
determined by eq.~(\ref{diducon}).

The inverse transformation starts from potentials of the form $V(w+\bar
w,\cdots)$ and $W(w+\bar w,\cdots)$. One has
\begin{eqnarray}
&&-\int\,d^2 \sigma \,d^2\theta \,d^2  \theta' \, \Big(V\big(\tilde Y,\cdots\big)
 -i\,u\,\bar \ID_+\bar \ID_-\tilde Y-i\,\bar u \,\ID_+\ID_-\tilde Y\Big)\nonumber\\
 &&\qquad+i\,
 \int d\tau \,d^2\theta \Big(W\big(\tilde Y,\cdots\big)+\frac 1 2 \,\bar \ID' u\,\bar \ID' \tilde Y
 -\frac 1 2\, \ID' \bar u \,\ID' \tilde Y\Big)\nonumber\\
 &&=-\int\,d^2 \sigma \,d^2\theta \,d^2  \theta' \, \Big(V\big(\tilde Y,\cdots\big)
 -\tilde Y\big(z+\bar z)\Big)+i\,
 \int d\tau \,d^2\theta \,W\big(\tilde Y,\cdots\big),\label{xxfoa3}
\end{eqnarray}
where we have put $z=i\,\bar \ID_+\bar \ID_-u$ and $\bar
z=i\,\ID_+\ID_-\bar u$.

Here more care is required with the treatment of the boundary term as we
will illustrate with a simple example. Starting point is a lagrangian
D1-brane with K{\"a}hler potential $V=-(w+\bar w)^2/2$ and whose position is
determined by the Dirichlet boundary condition $-i(w-\bar w)=m\,(w+\bar
w)$ with $m\in\IZ$. As a consequence we find a boundary potential
$W=-m\,(w+\bar w)^2/2$. From the boundary condition on the twisted chiral
field we get the boundary conditions for the gauge field $\tilde Y$:
$\ID'\tilde Y=i\,m\,\ID\tilde  Y$ and $\bar \ID'\tilde Y=-i\,m\,\bar
\ID\tilde  Y$. We modify the expression in eq.~(\ref{xxfoa3}) to,
\begin{eqnarray}
 &&{\cal S}=-\int\,d^2 \sigma \,d^2\theta \,d^2  \theta' \, \Big(-\frac 1 2 \,\tilde Y^2
 -i\,u\,\bar \ID_+\bar \ID_-\tilde Y-i\,\bar u \,\ID_+\ID_-\tilde Y\Big)\nonumber\\
 &&+i\,
 \int d\tau \,d^2\theta \Big(-\frac m 2\,\tilde Y^2+\frac 1 2 \,\bar \ID' u\,\big(\bar \ID' \tilde Y+
 i\,m\,\bar \ID\tilde  Y\big)
 -\frac 1 2\, \ID' \bar u \,\big(\ID' \tilde Y-i\,m\,\ID\tilde  Y\big)\Big)
\end{eqnarray}
such that the variation of $\bar \ID' u$ and $ \ID'\bar  u$ in the
boundary term precisely reproduces the boundary conditions for $\tilde Y$.
Upon partial integration, this becomes,
\begin{eqnarray}
&&{\cal S}=-\int\,d^2 \sigma \,d^2\theta \,d^2  \theta' \, \Big(-\frac 1 2 \tilde Y^2
 -\tilde Y\big(z+\bar z)\Big)\nonumber\\
 &&+i\,
 \int d\tau \,d^2\theta \,\Big(-\frac m 2\,\tilde Y^2-m\,\tilde Y\big(z+\bar z\big)\Big),
\end{eqnarray}
where we used $\bar \ID_+\bar \ID_-=-\bar \ID\bar \ID'/2$. Both the bulk
and the boundary variation of $\tilde Y$ yields $\tilde Y=-(z+\bar z)$
which results in the dual potentials,
\begin{eqnarray}
 V_{dual}=\frac 1 2 \,\big(z+\bar z\big)^2,\qquad
 W_{dual}=\frac m 2 \,\big(z+\bar z\big)^2.
\end{eqnarray}
Combining the boundary condition for $\tilde Y$ with the bulk equation of
motion results in the Neumann boundary conditions for the chiral field:
$\ID'z=i\,m\,\ID z$ and $\bar \ID'\bar z=-i\,m\,\bar \ID \bar z$ so that
we end up with a D2-brane.

One can also dualize a lagrangian D1-brane on a two-torus
parameterized by a twisted chiral superfield to a D0-brane. Let us
start from the K{\"a}hler potential $V = - \frac{1}{2} (w+\bar w)^2$
and the Dirichlet boundary condition $w +\bar w = -i\, n \big( w -
\bar w \big)$,  with $ n \in \IZ$, describing the position of the
D1-brane. For this model we can consider two different possible
dualizations, depending on the value of $n$. If $n \neq 0$ we can
dualize the D1-brane to a D2-brane with a worldvolume flux
characterized by the integer n, analogous to the situation described
above. However, if $n = 0$ the D1-brane is dualized to a D0-brane.
The boundary potential vanishes in that case and the first order
action we start from reads,
\begin{eqnarray}
{\cal S} &=& - \int\, d^2 \sigma\, d^2 \theta\, d^2  \theta' \, \Big( -\frac{1}{2} \tilde Y^2
-i\,u\,\bar \ID_+\bar \ID_-\tilde Y-i\,\bar u \,\ID_+\ID_-\tilde Y \Big)\nonumber \\
&& +\, i\, \int\, d\tau\, d^2\theta\, \Big( +\frac 1 2 \,\bar \ID' u\,\bar \ID' \tilde Y
 -\frac 1 2\, \ID' \bar u \,\ID' \tilde Y \Big),
\end{eqnarray}
and $\tilde Y$ satisfies the boundary condition $\tilde Y=0$. When
varying the Lagrange multipliers $u$ and $\bar u$ we recover the original
model parameterized by a twisted chiral superfield. It is however crucial
to notice that the fermionic derivatives of the Lagrange multipliers
$\bar\ID' u$ and $\ID' \bar u$ should satisfy a Dirichlet boundary
condition in order to reproduce the D1-brane with $n=0$. Upon integration
by parts we find,
\begin{eqnarray}
{\cal S} &=& - \int\, d^2 \sigma\, d^2 \theta\, d^2  \theta' \, \Big( -\frac{1}{2} \tilde Y^2
-\tilde Y \big( z + \bar z \big) \Big).
\end{eqnarray}
The bulk equation of motion reads $ \tilde Y = - \big( z + \bar z \big)$,
while the boundary term vanishes completely. The dual potentials are
therefore given by,
\begin{eqnarray}
V_{dual} = \frac 1 2 \, \big( z + \bar z \big)^2,\qquad W_{dual} = 0.
\end{eqnarray}
Since the gauge field $\tilde Y$ satisfies the Dirichlet boundary
condition $\tilde Y = 0$, we conclude that the chiral field $z$ and
its complex conjugate $\bar z$ also satisfy a Dirichlet boundary
condition,
\begin{eqnarray}
z = i\, b, \quad \bar z = -i\, b.
\end{eqnarray}
Moreover, these Dirichlet boundary conditions are fully consistent
with the Dirichlet boundary conditions for the Lagrange mulitpliers
we had to impose in the original model. We thus find a D0-brane
localized in the point $Re(z)=0$ and $Im(z) =  b$, where $b$ is a
free parameter.

\section{Examples}\label{examples}
\subsection{The WZW model on $S^3\times S^1$ and its dual formulation}
We will use the Hopf surface $S^3\times S^1$ -- better known as the
WZW-model on $SU(2)\times U(1)$ -- as a non-trivial example of various
issues discussed in the preceding two sections. We parameterize the Hopf
surface with coordinates $z$ and $w$ where $z, \,w\in (\IC^2\setminus
0)/\Gamma$ where $\Gamma$ is generated by $(z,w)\rightarrow (e^{2 \pi
}\,z,e^{2 \pi }\,w)$. The connection with the group manifold $SU(2)\times
U(1)$ is made explicit when parameterizing a group element as,
\begin{eqnarray}
{\cal G}= \frac{e^{-i\, \ln \sqrt{z \bar z+ w \bar w} }}{\sqrt{z
\bar z+ w \bar w}}\,\left(
\begin{array}{cc} w & \bar z\\
-z & \bar w\end{array}\right).
\end{eqnarray}
A very useful parameterization is in terms of Hopf coordinates $\phi
_1,\, \phi_2,\, \rho \in \IR\,\mbox{mod}\,2 \pi $ and $\psi\in[0, \pi
/2]$ where we put,
\begin{eqnarray}
 z=\cos \psi \,e^{ \rho +i \phi _1},\qquad w=\sin \psi \,e^{ \rho + i\phi
_2}.
\end{eqnarray}

In \cite{Spindel:1988nh} it was shown that any WZW-model which has an
even-dimensional target manifold has $N\geq (2,2)$. In
\cite{Rocek:1991vk} an explicit formulation of the $SU(2)\times U(1)$
model was given in terms of a chiral and a twisted chiral
superfield\footnote{In fact it was also shown that this is the only
WZW-model which can be described without the use of semi-chiral
superfields.}. The chiral superfield $z$ and the twisted chiral
superfield $w$ are precisely identified with the coordinates $z$ and $w$
introduced above. The generalized K{\"a}hler potential was found to be,
\begin{eqnarray}
V(z,\bar z,w,\bar w)=+\int^{z \bar z/w \bar w} \frac{dq}{q}\ln\big(1+q\big)-
\frac 1 2 \big(\ln w\, \bar w\big)^2.\label{WZWpot2}
\end{eqnarray}
which is everywhere well defined except when $w=0$. However -- as noted
in \cite{Sevrin:2008tp} -- we can rewrite the generalized K{\"a}hler
potential as,
\begin{eqnarray}
V(z,\bar z,w,\bar w)=-\int^{w \bar w/z \bar z} \frac{dq}{q}\ln\big(1+q\big)+
\frac 1 2 \big(\ln z\, \bar z\big)^2 -\ln\big( z\bar z\big)\ln\big( w\bar w\big),\label{WZWpot2b}
\end{eqnarray}
where the last term can be removed by a generalized K{\"a}hler
transformation resulting in an expression for the potential well defined
in $w=0$ (but not in $z=0$). The non-vanishing components of the metric
are in these coordinates,
\begin{eqnarray}
 g_{z\bar z}=g_{w\bar w}= \frac{1}{z\bar z+w\bar w}\,,
\end{eqnarray}
and we get for the torsion 3-form,
\begin{eqnarray}
 H_{z\bar zw}=-\frac 1 2 \,\frac{\bar w}{(z\bar z+w\bar w)^2}\,,\qquad
H_{zw\bar w}=-\frac 1 2 \,\frac{\bar z}{(z\bar z+w\bar w)^2}\,,
\end{eqnarray}
and complex conjugates. In \cite{Sevrin:2008tp}, D1- and D3-branes on
$S^3\times S^1$ were explicitely constructed using the above formulation.
Below we will show that D2- and D4 branes exist as well on $S^3\times
S^1$, although they require a semi-chiral parameterization of the Hopf
surface.

By making a different choice for the complex structures on $S^3\times
S^1$ an alternative parameterization in terms of a semi-chiral multiplet
was found in \cite{Sevrin:1996jr}. The generalized K{\"a}hler potential is
now,
\begin{eqnarray}
 V(l,\bar l,r,\bar r)=\ln \frac{l}{\bar r}\,\ln \frac{\bar l}{r}\,-\int^{r\bar r}\frac{dq}{q}\,
 \ln\big(1+q\big).\label{s3s1v1}
\end{eqnarray}
Using this we calculate the metric,
\begin{eqnarray}
 g_{l\bar l}=\frac{1}{l\bar l}\,,\quad g_{r\bar r}=\frac{1}{r\bar r}\,\frac{1}{1+r\bar r}\,,\quad
 g_{lr}=-\frac{1}{lr}\,\frac{1}{1+r\bar r}\,,\quad g_{\bar l\bar r}=-\frac{1}{\bar l\bar r}\,\frac{1}{1+r\bar r}\,,
\end{eqnarray}
and the torsion 3-form,
\begin{eqnarray}
 H_{lr\bar r}=-\frac{1}{l}\,\frac{1}{(1+r\bar r)^2}\,,\qquad
 H_{\bar l r\bar r}=+\frac{1}{\bar l}\,\frac{1}{(1+r\bar r)^2}\,.
\end{eqnarray}
Geometrically the two parametrizations are related by the coordinate
transformation,
\begin{eqnarray}
 l=w,\quad \bar l=\bar w,\quad r=\frac{\bar w}{z}\,,\quad  \bar r=\frac{w}{\bar z}\,.
\end{eqnarray}
One easily verifies that the expressions for the metric and torsion are
indeed equivalent in both coordinate systems. The complex structures in
both formulations are obviously different. In the chiral/twisted chiral
formulation we find that $J_+$ and $J_-$ are diagonal where $J_+$ has
eigenvalue $+i$ on $dz$ and $dw$ while for $J_-$ one finds eigenvalue
$+i$ on $dz$ and eigenvalue $-i$ on $dw$. In the semi-chiral
parameterization we find for $J_+$ and $J_-$,
\begin{eqnarray}
 J_+&=&\left(\begin{array}{cccc}
   +i & 0 & 0 & 0 \\
   0 &-i & 0 & 0 \\
   0 & -2i\,\frac{r}{\bar l} & +i&0\\
   +2i\,\frac{\bar r}{l}&0&0&-i
 \end{array}   \right),\nonumber\\
J_-&=&\left(\begin{array}{cccc}
   i&0&0&-2i\,\frac{l}{\bar r}\,\frac{1}{1+r\bar r} \\
   0& -i& +2i\,\frac{\bar l}{ r}\,\frac{1}{1+r\bar r}&0\\
   0 & 0 & +i & 0 \\
   0 & 0 & 0 & -i
 \end{array}   \right).
\end{eqnarray}
where we labelled the rows and columns in the order $l\bar l r\bar r$.

In \cite{Sevrin:2008tp} D1- and D3-branes were constructed on $S^3\times
S^1$ in the chiral/twisted chiral parameterization. In this section we
will study lagrangian D2-branes and maximally
coisotropic D4-branes on $S^3\times S^1$ in its semi-chiral
parameterization. As the direct construction of such branes is rather
non-trivial we will make use of a duality transformation. Indeed the
semi-chiral model on $S^3\times S^1$ is dual to a model on $T^2\times D$
where $T^2$ is parameterized by a twisted chiral and $D$ (the disk) by a
chiral field. In the dual model it is very easy to construct general D1-
and D3-brane configurations which when dualizing back to $S^3\times S^1$
will give rise to the desired D2- and D4-branes.

We make a coordinate transformation in eq.~(\ref{s3s1v1}) by replacing
$l$ by $e^l$ and $r$ by $e^{-r}$ which gives,
\begin{eqnarray}
 V(l,\bar l,r,\bar r)
 &=&\big(l+\bar r\big)\big(\bar l+r\big)+\int^{r+\bar r}dq\,
 \ln\big(1+e^{-q}\big)\nonumber\\
 &=&\frac 1 4\, \big(l+\bar l+r+\bar r\big)^2-
 \frac 1 4\, \big(l-\bar l-r+\bar r\big)^2+\int^{r+\bar r}dq\,
 \ln\big(1+e^{-q}\big).\label{s3s1v2}
\end{eqnarray}
In these coordinates we get that the two-form $\Omega^{(-)} $ defined in
eq.~(\ref{Omega}) is explicitly given by,
\begin{eqnarray}
 \Omega ^{(-)}_{l\bar l}=\Omega^{(-)} _{lr}=\Omega^{(-)} _{\bar r\bar l}=-i,\quad
 \Omega^{(-)} _{l\bar r}=\Omega ^{(-)}_{\bar lr}=0,\quad
 \Omega^{(-)} _{r\bar r}=\frac{i}{1+e^{-r-\bar r}}.\label{OmegaS3S1}
\end{eqnarray}
The potential eq.~(\ref{s3s1v2}) is readily dualized to,
\begin{eqnarray}
 V_{dual}&=&-\big(z+\bar z-w-\bar w\big)^2+\big(z-\bar z-w+\bar w\big)^2
 -2\,\int^{z+\bar z}dq\,
 \ln\big(e^{-2q}-1\big)\nonumber\\
 &=&-4\,\big(z-w\big)\big(\bar z-\bar w\big)-2\,\int^{z+\bar z}dq\,
 \ln\big(e^{-2q}-1\big).\label{s3s1v3}
\end{eqnarray}
Modulo a generalized K{\"a}hler transformation, one finds that the dual
potential factorizes in a part which describes a disk, $\mbox{Re}\, z\leq
0$ and a part which describes a two torus, $w\simeq w+\pi \,(n_1+i\,n_2)$
with $n_1,\,n_2\in\IZ$\footnote{In \cite{Rocek:1991vk} the $S^3\times
S^1$ model in terms of a chiral and twisted chiral field was shown to be
dual to the model on $D\times T^2$ in terms of chiral fields with the
same singular metric on $D$ as here. Note that superconformal invariance
at the quantum level requires a non-trivial dilaton as well
\cite{Rocek:1991vk}.}. Here it is rather straightforward (see
\cite{Sevrin:2008tp}) to construct globally well defined D-brane
configurations. We have two cases.
\begin{enumerate}
  \item \underline{A D1-brane}\\
  The position of the D1-brane is given by the three Dirichlet
  boundary conditions,
\begin{eqnarray}
 &&-i\,\big(w-\bar w\big)=\frac{m}{n}\,\big(w+\bar w\big),\nonumber\\
 &&z=\frac 1 2 \big(a+i\,b\big),\qquad \bar z=\frac 1 2 \big(a-i\,b\big),\label{diriD1}
\end{eqnarray}
where $m,\,n\in\IZ$, $a,\,b\in\IR$ and constant and $a\leq 0$. In
order to be consistent we need -- besides the bulk potential in
eq.~(\ref{s3s1v3}) -- a boundary potential given by,
\begin{eqnarray}
 W_{dual}=2\left(\frac m n \,a-b\right)\,\big(w+\bar w\big).
\end{eqnarray}
  \item \underline{A D3-brane}\\
  The position of the D3-brane is fixed by the Dirichlet boundary
  condition,
\begin{eqnarray}
 &&-i\,\big(w-\bar w\big)=\frac{m}{n}\,\big(w+\bar w\big)+\alpha \,z+\bar \alpha \,\bar z,\label{diriD3}
\end{eqnarray}
where $m,\,n\in\IZ$ and  $\alpha \in\IC$. Consistency requires the
presence of a boundary potential,
\begin{eqnarray}
 W_{dual}=2\left(\alpha \,z+\bar \alpha \,\bar z+i\big(z-\bar z\big)+\frac m n\,
 \big(z+\bar z\big)\right)\,\big(w+\bar w\big)+g(z+\bar z),
\end{eqnarray}
where $g$ is an arbitrary real function of $z+\bar z$.
\end{enumerate}
We have now all ingredients which will allow us to dualize this to
lagrangian D2-branes and maximally coisotropic D4-branes on the Hopf
surface $S^3\times S^1$.

\subsection{From D1-branes on $T^2\times D$ to D2-branes on $S^3\times S^1$}
The Dirichlet boundary conditions given in eq.~(\ref{diriD1}) imply the
following Dirichlet boundary conditions on the gauge fields,
\begin{eqnarray}
 &&Y=a,\nonumber\\
 &&\hat Y=-b+\frac m n \tilde Y.
\end{eqnarray}
Using this, the first order action eq.~(\ref{1storderCTS}) becomes,
\begin{eqnarray}
 {\cal S}&=&-\int\,d^2 \sigma \,d^2\theta \,d^2  \theta' \, \Big(V\big(Y,\tilde Y,\hat Y\big)_{dual}
 +Y\big(l+\bar l-r-\bar r\big)
-\tilde Y\big(l+\bar l+r+\bar r\big)\nonumber\\
&&-i\,\hat Y\big(l-\bar l-r+\bar r\big)\Big)
 +i\int d\tau \,d^2\theta \Big(\big(2\big(\frac m n \,a-b\big)+
 i\big(l-\bar l-r+\bar r\big)\nonumber\\
 &&-\frac m n \,\big(l+\bar l+r+\bar r\big)\big)\tilde Y+b\, \big(l+\bar l+r+\bar r\big)-i\,a
 \big(l-\bar l+r-\bar r\big)
\Big),\label{1storderD1D2}
\end{eqnarray}
Varying $\tilde Y$ in the boundary term gives a Dirichlet boundary
condition which is compatible with a combination of the boundary
conditions for the gauge fields and the bulk equations of motion of the
gauge fields. Integrating over the gauge fields gives in this way a
D2-brane on $S^3\times S^1$ whose position is given by,
\begin{eqnarray}
 r+\bar r&=&-\ln \big(e^{-2a}-1\big),\nonumber\\
 r-\bar r&=&l-\bar l+i\,\frac m n \,\big(l+\bar l\big)+i\,\big(
2b-\frac m n \,\ln(1-e^{2a} )
 \big). \label{DiriD21}
\end{eqnarray}
The bulk potential is given in eq.~(\ref{s3s1v2}) and the boundary
potential is now,
\begin{eqnarray}
 W=\big(b+\frac m n\,a\big)\big(l+\bar l\big)-2i\,a\big(l-\bar l\big).
\end{eqnarray}
One checks that the Dirichlet boundary conditions in terms of Hopf
coordinates are rephrased as,
\begin{eqnarray}
&& \psi=\arcsin \sqrt{1-e^{2a}}\in\big[0,\frac \pi 2\big],\nonumber\\
&& \phi _1=\frac m n \,\rho+b,\label{gkjgkj}
\end{eqnarray}
where we used $a\neq 0$. This is indeed a generalized lagrangian brane
with respect to the symplectic form given in eq.~(\ref{OmegaS3S1}). It is
gratifying to notice -- see eq.~(\ref{gkjgkj}) -- that also globally
everything works out perfectly.

\subsection{From D3-branes  on $T^2\times D$ to D2-branes on $S^3\times S^1$}
Taking $\mbox{Im}\, \alpha =-1$ allows us to translate the Dirichlet
boundary condition in eq.~(\ref{diriD3}) into a Dirichlet boundary
condition for the gauge fields,
\begin{eqnarray}
 \hat Y=a\,Y+ \frac{m}{n}\,\tilde Y,\label{ddd2}
\end{eqnarray}
where $a\equiv \mbox{Re}\, \alpha $. Using this we rewrite the first
order action eq.~(\ref{1storderCTS}) as,
\begin{eqnarray}
 {\cal S}&=&-\int\,d^2 \sigma \,d^2\theta \,d^2  \theta' \, \Big(V\big(Y,\tilde Y,\hat Y\big)_{dual}
 +Y\big(l+\bar l-r-\bar r\big)
-\tilde Y\big(l+\bar l+r+\bar r\big)\nonumber\\
&&-i\,\hat Y\big(l-\bar l-r+\bar r\big)\Big)
 +i\int d\tau \,d^2\theta \Big(
2\big(a+\frac m n\big)Y\,\tilde Y+g(Y)\nonumber\\
&&-Y\big(
(a+i)l+(a-i)\bar l+(a+i)r+(a-i)\bar r\big)-\nonumber\\
&&\tilde Y\big((\frac m n -i)l+(\frac m n +i)\bar l+(\frac m n +i)r+(\frac m n -i)\bar r
\big) \Big).\label{1storderD3D2}
\end{eqnarray}
Varying $\tilde Y$ in the boundary term in eq.~(\ref{1storderD3D2}) gives
a Dirichlet boundary condition which upon using the bulk equations of
motion is equivalent to eq.~(\ref{ddd2}). Varying $Y$ in the boundary
term in eq.~(\ref{1storderD3D2}) gives a second Dirichlet boundary
condition so that we end up with a D2-brane on $S^ 3\times S^1$.
Explicitly the Dirichlet boundary conditions are,
\begin{eqnarray}
 -i\big(l-\bar l\big)&=&a\big(l+\bar l+r+\bar r\big)-\frac 1 2 \,
 g'\big(-\frac 1 2\,\ln (1+e^{-(r+\bar r)})\big),\nonumber\\
 -i\big(r-\bar r\big)&=&\big(a+\frac m n\big)\big(
l+\bar l+r+\bar r+\ln (1+e^{-(r+\bar r)})\big)\nonumber\\
&&-\frac 1 2 \,
 g'\big(-\frac 1 2\,\ln (1+e^{-(r+\bar r)})\big).
\end{eqnarray}
In Hopf coordinates this gives,
\begin{eqnarray}
 \phi _1&=&\frac m n \,\rho -a\,\ln\big(\cos\psi\big),\nonumber\\
 \phi _2&=&a\,\rho+a\,\ln\big(\cos\psi\big)-\frac 1 4 \,g'\big(\ln(\cos\psi)\big).
\end{eqnarray}
Using the Dirichlet boundary conditions and the equations of motion
we can write the dual boundary potential as,
\begin{eqnarray}
W_{dual} &=& \frac 1 2 \ln\left( 1 + e^{-(r+\bar r)} \right)\big( (a+i\,) l + (a-i\,) \bar l + (a+i\,) r + (a-i\,) \bar r \big) \nonumber \\
&& + g\left( -\frac 1 2 \ln\big( 1 + e^{-(r + \bar r)} \big) \right).
\end{eqnarray}

Neglecting the function $g'(\ln(\cos \psi) )$ (by interpreting it as
a manner to describe fluctuations of the D2-branes) one can easily
check that the pullback of the two-form in eq.~(\ref{OmegaS3S1})
w.r.t. the D2-brane vanishes. Also this D2-brane is a generalized
lagrangian brane w.r.t. the symplectic form in
eq.~(\ref{OmegaS3S1}).

\subsection{From D3-branes  on $T^2\times D$ to D4-branes on $S^3\times S^1$}
For generic values of $\alpha $ we find that the Dirichlet boundary
condition eq.~(\ref{diriD3}) implies,
\begin{eqnarray}
 &&\bar \ID\left(\hat Y +\big(i-\bar \alpha\big)Y-\frac m n \,\tilde Y \right)=0\nonumber\\
  && \ID\left(\hat Y +\big(-i- \alpha\big)Y-\frac m n \,\tilde Y \right)=0,\label{ddd4}
\end{eqnarray}
which implies that $Z^1\equiv\left(\hat Y +(i-\bar \alpha  )Y-\frac m n
\,\tilde Y \right)$ is a boundary chiral field. Note that if $\mbox{Im}\,
\alpha =-1$, the boundary chiral field is real and as a consequence is a
constant which is precisely the case previously studied. For simplicity
we take here $\alpha =0$ and we find that
\begin{eqnarray}
 Z^1=\hat Y+i\, Y -\frac m n \,\tilde Y,
\end{eqnarray}
is a boundary chiral field. Using this we write the first order action
eq.~(\ref{1storderCTS}) as,
\begin{eqnarray}
 {\cal S}&=&-\int\,d^2 \sigma \,d^2\theta \,d^2  \theta' \, \Big(V\big(Y,\tilde Y,\hat Y\big)_{dual}
 +Y\big(l+\bar l-r-\bar r\big)
-\tilde Y\big(l+\bar l+r+\bar r\big)\nonumber\\
&&-i\,\hat Y\big(l-\bar l-r+\bar r\big)\Big)
 +i\int d\tau \,d^2\theta \Big(
\big(Z^ 1+\bar Z^{\bar 1}\big)\,\tilde Y-i\frac m n \,\big(Z^ 1-\bar Z^{\bar 1}\big)\,\tilde Y\nonumber\\
&&+
g\big( -\frac i 2 (Z^ 1-\bar Z^{\bar 1}) \big)-\frac 1 2\,\big(Z^ 1-\bar Z^{\bar 1}\big)
\big( l-\bar l+r-\bar r \big)+i\,\tilde Y\,\big(
l-\bar l-r+\bar r
\big)\nonumber\\
&&-\frac m n \,\tilde Y\big( l+\bar l+r+\bar r\big)-\frac 1 2 \,
\big(Z^ 1+\bar Z^{\bar 1}\big) \big(l+\bar l+r+\bar r\big)
  \Big).\label{1storderD3D4}
  \end{eqnarray}
Varying the unconstrained field $\tilde Y$ in the boundary term yields an
equation fully compatible with the bulk equations of motion. When varying
$Z^1$ and $\bar Z^{\bar 1}$ in the boundary term one needs to take into
account that they are constrained fields. This variation implies the
existence of a second boundary chiral field $Z^ 2$,
\begin{eqnarray}
 Z^2=\tilde Y-\frac 1 2\,\big(l+\bar l+r+\bar r\big)-\frac 1 2\, \big(l-\bar l+r-\bar r\big)-i\,
 \big(\frac 1 2 \,g' (Y)+\frac m n \,\tilde Y\big),
\end{eqnarray}
where a prime denotes a derivative. The variation of $Z^1$ and $\bar
Z^{\bar 1}$ in the boundary term gives $\bar \ID Z^2=\ID\bar Z^{\bar 2} =
0$. Hence, we constructed a (space-filling) coisotropic D4-brane with the
complex structure $K$ w.r.t. the basis $\{l, \bar l , r, \bar r \}$ given
by,
\begin{eqnarray}
K= \frac{i}{2n} \left(\small{\begin{array}{cccc}  2n+\big(n +i\, m
\big)e^{r+\bar r}& -(n-i\,m)e^{r+\bar r} & -(n-i\,m) \frac{e^{2r +
2\bar r}}{1+e^{r+\bar r}}
& \frac{ 2n + \big(n +i\, m \big) e^{r+ \bar r}}{1+e^{-(r+ \bar r)} }\\
(n+i\,m)e^{r+\bar r} & -2n - (n-i\,m)e^{r+\bar r} & - \frac{2n +
\big(n -i\, m \big) e^{r+ \bar r}}{1+e^{-(r+ \bar r)}} & (n+i\,m)
\frac{e^{2r + 2\bar
r}}{1+e^{r+\bar r}}  \\
-(n-i\, m) e^{r+\bar r}& (n-i\, m) (2 + e^{r+ \bar r})  & 2n + (n
-i\, m)e^{r+\bar r}
& -(n +i\, m)e^{r+\bar r}  \\
-(n+i\, m) (2 + e^{r+\bar r})
 & (n-i\,m) e^{r+\bar r}
 &(n-i\,m)e^{r+\bar r}
 &-2n-(n+i\,m)e^{r+\bar r}
\end{array}}\right).\nonumber
\end{eqnarray}
The dual boundary potential reads,
\begin{eqnarray}
W_{dual}&=& \frac 1 2 \big( l + \bar l + r +\bar r\big) \Big[ i\, \big(l-\bar l -r -\bar r\big) -
\frac m n \big( l+\bar l +r +\bar r \big) \Big] \nonumber \\
&& + \frac 1 2 \ln\left( 1 + e^{-(r + \bar r)} \right) \Big[ i\, \big( l-\bar l +r-\bar r \big) -
\frac m n \big( l + \bar l + r + \bar r \big) \Big] \nonumber \\
&& + g\left(- \frac 1 2 \ln\left( 1 + e^{-(r + \bar r)}
\right)\right),
\end{eqnarray}
which can also be written in terms of the boundary chiral fields as
follows, when ignoring total derivative terms,
\begin{eqnarray}
W_{dual}&=& -\frac 1 4 \big(1-i\, \frac m n \big)Z^1 \bar Z^{\bar 2} -
\frac 1 4 \big(1+i\, \frac m n \big) \bar Z^{\bar 1} Z^2 - \frac m 2n Z^1 \bar Z^{\bar 1} \nonumber \\
&&+\frac i 2 \big( Z^1 - \bar Z^{\bar 1} \big) g'
\left( -\frac i 2 \big(  Z^1 - \bar Z^{\bar 1} \big) \right) +
g\left( -\frac i 2 \big(  Z^1 - \bar Z^{\bar 1} \big) \right).
\end{eqnarray}

So we arrive at the conclusion that $S^3\times S^1$ (or the WZW model on
$SU(2)\times U(1)$) allows for D1, D3, D2 and D4 supersymmetric brane
configurations. We need the description of $S^3\times S^1$ in terms of a
twisted chiral and a chiral field if we have D1- or D3-branes
\cite{Sevrin:2008tp}. Lagrangian D2-branes or
maximally coisotropic D4-branes require the semi-chiral description. From
the above it should be clear that duality transformations provide for a
powerful method to construct highly non-trivial supersymmetric D-brane
configurations.

\section{Conclusions and discussion}
The off-shell description of a general $d=2$, $N=(2,2)$ supersymmetric
non-linear $\sigma $-model requires semi-chiral, twisted chiral and
chiral superfields. In the present paper we identified the allowed
boundary conditions for these fields. The cleanest case is where only
semi-chiral and twisted chiral fields are involved. These fields share
the property that they are a priori unconstrained on the boundary. For
these fields two classes of boundary conditions are possible: either we
impose a Dirichlet boundary condition -- which in its turn implies a
Neumann boundary condition as well -- or we require them to be chiral on
the boundary. The result is a straightforward generalization of A-branes
on K{\"a}hler manifolds: the allowed D-brane configurations are either
lagrangian or coisotropic with respect to the symplectic structure
$\Omega ^{(-)}=2g\,(J_+-J_-)^{-1}$. When no semi-chiral superfields are
present, $\Omega ^{(-)}$ reduces to the K{\"a}hler two-form and we recover
the usual lagrangian and coisotropic A-branes on K{\"a}hler manifolds. Once
semi-chiral superfields are present as well, non-K{\"a}hler geometries become
possible, but the notion of lagrangian and coisotropic branes carries
over. An example of this are the lagrangian D2-branes and the maximally
coisotropic D4-branes on $S^3\times S^1$ (which is certainly not a K{\"a}hler
manifold).

The picture gets murkier once chiral fields get involved. Chiral fields
remain chiral -- {\em i.e.} constrained -- on the boundary. When only
chiral fields are present, the situation is still quite simple. The
branes wrap around holomorphic cycles of K{\"a}hler manifold. These are
nothing but the standard B-branes on K{\"a}hler manifolds.

Once all three types of superfields are present we get into a situation
interpolating between the two cases mentioned above. In general the
target manifold is not symplectic anymore, however any bihermitian
manifold is still a Poisson manifold. This allows us to view the
resulting D-brane configurations as generalized coisotropic submanifolds
defined through a foliation of the Poisson manifold by symplectic leaves.

While the precise form of the torsion potential $b$ is gauge
dependent, we found that there is a particular gauge such that
$\Omega _+=-(g-b)J_+$ is a closed two-form. As -- at least with this
definition -- this two-form is not globally defined, it does not
define a symplectic structure. However, when it is globally defined
it allows for an alternative classification of the allowed
supersymmetric D-brane configurations. Consider {\em e.g.} the
four-dimensional case. When described in terms of two chiral
fields, we can have D0-, D2- or D4-branes which are all
symplectic submanifolds with respect to $\Omega _+$. Having one
chiral and one twisted chiral superfield gives a D1-brane which is
isotropic and a D3-brane which is coisotropic. Finally a semi-chiral
multiplet or two twisted chiral fields gives a lagrangian D2-brane
or a maximally coisotropic D4-brane. The latter case is indeed
always lagrangian or coisotropic as $\Omega _+$ coincides with the
symplectic structure $\Omega ^{(-)}$.

An unexpected\footnote{Note however that the present analysis holds only
for models with a constant dilaton and no RR-fluxes.} result of the
present analysis is the fact that supersymmetric D0- and D1-branes are
rather
``rare''. Indeed, the only way to get a D0-brane is by imposing Dirichlet
boundary conditions in all directions. This is only possible if the model
is formulated in terms of chiral superfields only. So supersymmetric
D0-branes are always B-branes on K{\"a}hler manifolds! Similarly, in order to
obtain D1-branes, we need a single twisted chiral and an arbitrary number
of chiral fields. The fact that D0- and D1-branes behave differently from
the other D-branes is somewhat puzzling (note however that such an
unusual behaviour of D0- and D1-branes viz. other Dp-branes was -- though
in a very different context -- already seen before \cite{Craps:2006vx}).

The superspace formulation of these models allows for the study of
T-duality transformations while keeping the $N=2$ supersymmetry manifest.
As usual, the possibility of making a T-duality transformation requires
the existence of an isometry in the target manifold geometry. Having an
isometry which acts on chiral or twisted chiral fields only results in a
T-duality transformation which exchanges chiral and twisted chiral
fields. An isometry which mixes chiral and twisted chiral fields non
trivially yields a T-duality transformation which exchanges a pair
consisting of a twisted chiral and a chiral field for a semi-chiral
multiplet. The inverse transformation exists as well. A consequence of
this is that these duality transformations often simplify the
construction of D-branes. {\em E.g.} coisotropic branes require the
existence of an additional complex structure on (a subspace of) the
worldvolume. As we illustrate in this paper, such branes can often be
obtained through a T-duality transformation from much simpler brane
configurations.

Looking at the case relevant to compactified string theory, we arrive at
the following possible parametrizations of a six dimensional target
manifold. We denote chiral fields by $z$, twisted chiral fields by $w$
and semi-chiral fields by $l$ and $r$.
\begin{description}
  \item[\underline{$z_1$, $z_2$, $z_3$}:] The geometry is necessarily
  K{\"a}hler and one can have D0-, D2-, D4- and D6-branes
  wrapping holomorphic cycles.
  \item[\underline{$z_1$, $z_2$, $w_3$}:] A non-trivial $H$-flux can be present.
  One can have D1-, D3- or D5-branes where the branes wrap
  holomorphic cycles in the chiral directions and a lagrangian
  submanifold in the twisted chiral direction.
  \item[\underline{$z_1$, $w_2$, $w_3$}:] Once more a non-trivial $H$-flux might be
  present. There are D2- or D4-brane configurations which wrap a
  holomorphic cycle in the chiral direction and which are lagrangian
  in the twisted chiral directions. Also D4- and D6-branes can be
  possible where the branes are now maximally coisotropic in the
  twisted chiral directions.
  \item[\underline{$w_1$, $w_2$, $w_3$}:] The geometry is again K{\"a}hler. One either
  has a lagrangian D3-brane or a coisotropic D5-brane.
  \item[\underline{$l$, $r$, $z$}:] A non-trivial $H$-flux can be present. When the
  branes wrap a lagrangian submanifold in the semi-chiral directions
  we can have D2- or D4-branes. When the brane is maximally
  coisotropic in the semi-chiral directions we have D4- or D6-branes.
  \item[\underline{$l$, $r$, $w$}:] Once more a non-trivial $H$-flux can be present.
  We either have a lagrangian D3-brane or a coisotropic D5-brane.
\end{description}
Presently an analysis of supersymmetric branes on various tori described
by any of the superfield combinations given above is being investigated
with applications along the lines of \cite{Font:2006na} in mind.

The whole analysis in this paper was performed at the classical level. In
order to make contact with the ($\alpha '$ corrected) supergravity
equations of motion and their solutions, one needs to study the
superconformal invariance of these models at the quantum level. Having no
boundaries, the one loop $\beta $-function for a general $N=(2,2)$
non-linear $\sigma $-model was calculated and analysed in
\cite{Grisaru:1997pg} and recently shown to be consistent with
supergravity results \cite{Halmagyi:2007ft}. The results in this paper
are perfectly tailored for a systematic study of the one-loop $\beta
$-functions in the presence of D-branes. As argued in
\cite{Nevens:2006ht}, the superspace treatment automatically yields the
stability conditions for the supersymmetric D-branes which would allow to
extend and reinterpret the results of \cite{Kapustin:2003se} in a more
physical context. Work in this direction is now in progress. We would
also like to stress that an economic formulation of $\sigma $-models {\em
with} the dilaton in $N=(2,2)$ or $N=2$ superspace would be most useful
for numerous applications.

Finally a study of D- and F-terms in $N=2$ boundary superspace using the
technology developed in the present paper might be very interesting.
Indeed, supersymmetric D-branes sometimes cease to remain supersymmetric
when a small closed string perturbation is switched on. Another
interesting event is when a D-brane decays into a superposition of
D-branes when crossing a line of marginal stability (for both phenomena
see {\it e.g.} \cite{Brunner:2009mn}). A manifest supersymmetric
formulation might reveal the systematics of this.

\acknowledgments

\bigskip

We thank Matthias Gaberdiel, Jim Gates, Chris Hull, Paul Koerber, Ulf
Lindstr{\"o}m, Martin Ro\v cek and Maxim Zabzine for useful discussions and
suggestions. AS and WS are supported in part by the Belgian Federal
Science Policy Office through the Interuniversity Attraction Pole P6/11,
and in part by the
``FWO-Vlaanderen'' through project G.0428.06. AW is supported in part by
grant 070034022 from the Icelandic Research Fund.

\appendix

\section{Conventions, notations and identities}\label{app conv}
The conventions used in the present paper are essentially the same as
those in \cite{Sevrin:2008tp} and \cite{Sevrin:2007yn}. However we did
modify some of the notations. The torsion which was previously called $T$
is now more conventionally renamed to $H$. Semi-chiral fields were
previously labelled by $r$, $\bar r$, $s$ and $\bar s$ and are now called
$l$, $\bar l$, $r$ and $\bar r$.

We denote the worldsheet coordinates by $ \tau \in\IR$ and $ \sigma \in
\IR$, $ \sigma \geq 0$, and the worldsheet light-cone coordinates are
defined by,
\begin{eqnarray}
\sigma ^\pp= \tau + \sigma ,\qquad \sigma ^== \tau - \sigma .\label{App1}
\end{eqnarray}
The $N=(1,1)$ (real) fermionic coordinates are denoted by $ \theta ^+$ and $ \theta ^-$ and the
corresponding derivatives satisfy,
\begin{eqnarray}
D_+^2= - \frac{i}{2}\, \partial _\pp \,,\qquad D_-^2=- \frac{i}{2}\, \partial _= \,,
\qquad \{D_+,D_-\}=0.\label{App2}
\end{eqnarray}
The $N=(1,1)$ integration measure is explicitely given by,
\begin{eqnarray}
\int d^ 2 \sigma \,d^2 \theta =\int d^2 \sigma \,D_+D_-.
\end{eqnarray}
Passing from $N=(1,1)$ to $ N=(2,2)$ superspace requires
the introduction of two more real fermionic coordinates $ \hat \theta ^+$ and $ \hat \theta ^-$
where the corresponding fermionic derivatives satisfy,
\begin{eqnarray}
\hat D_+^2= - \frac{i}{2} \,\partial _\pp \,,\qquad \hat D_-^2=- \frac{i}{2} \,\partial _= \,,
\end{eqnarray}
and again all other -- except for (\ref{App2}) -- (anti-)commutators do vanish.
The $N=(2,2)$ integration measure is,
\begin{eqnarray}
\int d^2 \sigma \,d^2 \theta \, d^2 \hat \theta =
\int d^2 \sigma \,D_+D_-\, \hat D_+ \hat D_-.
\end{eqnarray}
Quite often a complex basis is used,
\begin{eqnarray}
\ID_\pm\equiv \hat D_\pm+i\, D_\pm,\qquad
\bar \ID_\pm\equiv\hat D_\pm-i\,D_\pm,
\end{eqnarray}
which satisfy,
\begin{eqnarray}
\{\ID_+,\bar \ID_+\}= -2i\, \partial _\pp\,,\qquad
\{\ID_-,\bar \ID_-\}= -2i\, \partial _=,
\end{eqnarray}
and all other anti-commutators do vanish.

When dealing with boundaries in $N=(2,2)$ superspace, we introduce
various derivatives as linear combinations of the previous ones. We
summarize their definitions together with the non-vanishing
anti-commutation relations. We have,
\begin{eqnarray}
&&D\equiv D_++D_-,\qquad \hat D\equiv \hat D_++ \hat D_-, \nonumber\\
&& D'\equiv D_+-D_-,\qquad \hat D'\equiv \hat D_+- \hat D_-,
\end{eqnarray}
with,
\begin{eqnarray}
&&D^2=\hat D^2=D'{}^2=\hat D'{}^2=- \frac{i}{2} \partial _ \tau , \nonumber\\
&&\{D,D'\}=\{\hat D, \hat D'\}=-i \partial _ \sigma .
\end{eqnarray}
In addition we also use,
\begin{eqnarray}
&&\ID\equiv \ID_++\ID_-=\hat D+i\,D,\qquad \ID'\equiv \ID_+-\ID_-=\hat D'+i\,D', \nonumber\\
&& \bar \ID\equiv \bar \ID_++ \bar \ID_-=\hat D-i\,D,\qquad \bar \ID'\equiv \bar \ID_+- \bar
\ID_-=\hat D'-i\,D'.
\end{eqnarray}
They satisfy,
\begin{eqnarray}
&&\{\ID, \bar \ID\}=\{\ID', \bar \ID'\}= -2i\, \partial _ \tau ,\, \nonumber\\
&&\{\ID, \bar \ID'\}=\{\ID', \bar \ID\}=-2i\, \partial _ \sigma  \,.
\end{eqnarray}
The integration measure we use when boundaries are present is defined by,
\begin{eqnarray}
 \int d^2\sigma \,d^2\theta \,d^2\theta '\equiv \int d^2\sigma \,D\hat D D'\hat D' \,,
\end{eqnarray}
and on the boundary we take,
\begin{eqnarray}
 \int d\tau \,d^2\theta \equiv\int d\tau \,D\hat D.
\end{eqnarray}
When integrating by parts one finds that the following relations are most
useful,
\begin{eqnarray}
&& \int d^2\sigma \,d^2\theta \,d^2\theta '\,\ID_\pm=\mp\int d\tau \,d^2\theta\,\ID_\pm=
-\frac 1 2 \,\int d\tau \,d^2\theta\,\ID' \,,\nonumber\\
&& \int d^2\sigma \,d^2\theta \,d^2\theta '\,\bar \ID_\pm=\pm\int d\tau \,d^2\theta\,\bar \ID_\pm=
+\frac 1 2 \,\int d\tau \,d^2\theta\,\bar \ID'\,.
\end{eqnarray}

\section{$N=1$ non-linear $\sigma $-models}\label{app nis1}
While a comprehensive review of the $N=1$ non-linear $\sigma $-model in
the presence of boundaries can be found in \cite{Sevrin:2007yn}, we
summarize here -- in order to be self contained -- its most relevant
properties.

In the absence of boundaries a non-linear $ \sigma $-model (with
$N\leq(1,1)$) on some $d$-dimensional target manifold $ {\cal M}$ is
characterized by a metric $g_{ab}(X)$ and a closed 3-form $H_{abc}(X)$
where $X^a$ are local coordinates on ${\cal M}$ and
$a,b,c,...\in\{1,\cdots ,d\}$, we also use a locally defined 2-form
potential $b_{ab}(X)=-b_{ba}(X)$ for the torsion: $H_{abc}=- (3/2)
\partial _{[a}b_{bc]}$. We introduce a boundary at $ \sigma =0$
( $ \sigma \geq 0$ ) and $ \theta ^+= \theta ^-$ which breaks the
invariance under translations in both the $ \sigma $ and the $  \theta'
\equiv \theta ^+- \theta ^-$ direction thus reducing the $N=(1,1)$
supersymmetry to an $N=1$ supersymmetry. The action,
\begin{eqnarray}
{\cal S}=-4\int d^2 \sigma \, d \theta
\,D'\left(D_+X^aD_-X^b\left(g_{ab}+b_{ab}\right)\right)+2i\,\int d \tau \, d \theta\,A_a(X)\,DX^a,\label{an1}
\end{eqnarray}
is manifestly invariant under the $N=1$ supersymmetry and differs from
the usual action in the absence of boundary terms by a total derivative
term \cite{Lindstrom:2002mc}, \cite{Koerber:2003ef}. We can drop the
boundary term provided we replace $b$ in the bulk term by $\cal F$,
\begin{eqnarray}
 b_{ab}\rightarrow{\cal F}_{ab}=b_{ab}+F_{ab},\label{frfr1}
\end{eqnarray}
with,
\begin{eqnarray}
 F_{ab}=\partial _aA_b- \partial _b A_a.\label{frfr2}
\end{eqnarray}
Dimensionally, one could as well add a non-standard boundary term to the
action,
\begin{eqnarray}
 {\cal S}_{\hat b}=2i\,\int d \tau \, d \theta\,\hat A_a(X)\,D'X^a. \label{Eva1}
\end{eqnarray}
A priori such a term is problematic, however through appropriate Neumann
boundary conditions it can be reduced to the standard boundary term. This
is precisely the situation we encounter when dealing with twisted chiral
and semi-chiral superfields.

Varying the action eq.~(\ref{an1}) yields a boundary term,
\begin{eqnarray}
\delta {\cal S}\big| _{boundary}= -2i\,\int d \tau d \theta \,\delta X^a\left(
g_{ab} \,D'X ^b-\, {\cal F}_{ab}\,DX^b\right),\label{var1}
\end{eqnarray}
which will only vanish upon imposing suitable boundary conditions. In
order to do so we start by imposing a set of Dirichlet boundary
conditions,
\begin{eqnarray}
 Y^{\hat A}(X)=0,\qquad \hat A\in\{1,\cdots , d-p\}.\label{appb1}
\end{eqnarray}
We denote the remainder of the coordinates -- the world volume
coordinates of the brane -- by,
\begin{eqnarray}
 \sigma ^A(X),\qquad A\in\{1,\cdots, p\}.
\end{eqnarray}
In order to make the boundary term in the variation vanish, we need to
impose in addition to the Dirichlet boundary conditions
eq.~(\ref{appb1}), $p$ Neumann boundary conditions,
\begin{eqnarray}
\frac{\partial X^c}{\partial \sigma ^A}\,g_{cb}D'X^b=
\frac{\partial X^c}{\partial \sigma ^A}\,{\cal F}_{cd}\frac{\partial X^d}{\partial \sigma ^B}\,D\sigma ^B.\label{nis1NEU}
\end{eqnarray}
We end up with a D$p$-brane whose position is determined by
eq.~(\ref{appb1}), with a possibly non-trivial $U(1)$ bundle with field strength ${\cal F}$ on it.

\section{Some geometry} \label{app:geo}

\subsection{Generalized complex geometry}

In this section, we review some aspects of generalized complex geometry (GCG) that are useful for understanding
some discussions in the main text, section \ref{sec:GCG} in particular. For a much more detailed discussion,
see \cite{Gualtieri:2003dx}.

To get started, let us recall some better known structures. An almost complex structure on a manifold ${\cal M}$
is a linear map $J: T \rightarrow T$ (where $T$ is the tangent bundle of ${\cal M}$)\footnote{In order to be correct,
we should be speaking of smooth sections ${\cal C}^\infty (T)$ of $T$. We will however be a bit sloppy here and use the
same notation for a bundle and the space of its sections.}
, which satisfies $J^2 = -1$. For our purposes this should be contrasted with the notion of a pre-symplectic
structure on ${\cal M}$, which is simply a non-degenerate two-form $\Omega$ on ${\cal M}$. More abstractly,
this means that a pre-symplectic structure is an isomorphism $\Omega: T \rightarrow T^\ast$ (where $T^\ast$ is
the dual of $T$, the cotangent bundle of ${\cal M}$), satisfying $\Omega^\ast = - \Omega$.

Both notions can be naturally combined once we look at structures on the direct sum $T \oplus T^\ast$, leading to the
notion of a generalized complex structure (GCS). As usual, it is useful to have a bilinear form at one's disposal.
The natural symmetric pairing on $T \oplus T^\ast$ is given by,
\begin{eqnarray}
\langle X + \xi, Y + \eta \rangle = \frac 1 2 \left( \eta(X) + \xi(Y) \right), \qquad X+\xi,\, Y+\eta \in T \oplus T^\ast \label{np}
\end{eqnarray}
Using this bilinear form, an almost GCS is a linear map ${\cal J}:  T \oplus T^\ast \rightarrow T \oplus T^\ast$,
satisfying ${\cal J}^2 = -1$, which preserves the natural pairing,
$\langle {\cal J} W, {\cal J} Z \rangle = \langle W, Z \rangle$ for all $W, Z \in T \oplus T^\ast$.
Using the defining relation for the dual map $\langle W, {\cal J} Z \rangle = \langle {\cal J}^\ast W, Z\rangle$,
the latter condition is nothing but ${\cal J}^\ast = - {\cal J}$.

The next step is to introduce an appropriate notion of integrability. To this end one defines the Courant bracket,
\begin{eqnarray}
[X+\xi,Y+\eta] = [X,Y] + {\cal L}_X \eta - {\cal L}_Y \xi - \frac 1 2 d (\eta(X) - \xi(Y)).
\end{eqnarray}
Here the first term is the usual Lie bracket on $T$ and ${\cal L}_X$ is the Lie derivative corresponding to $X$.
This clearly reduces to the Lie bracket when projecting to $T$. One of the main useful properties of the Courant
bracket is its covariance with respect to b-transforms. A b-transform is a symmetry of the natural pairing eq.~(\ref{np}),
\begin{eqnarray}
e^b \left( \begin{array}{c} X \\ \xi \end{array} \right) \equiv
             \left(\begin{array}{cc}
                     1 & 0 \\
                     b & 1 \end{array} \right)
             \left(\begin{array}{c} X \\ \xi \end{array} \right) = \left(\begin{array}{c} X \\ \xi + \iota_X b \end{array} \right) ,
\end{eqnarray}
where $b$ is a locally defined two-form and $\iota_X b$ is the inner product, $\iota_X b(Y) = b(X,Y)$ for all vector fields $Y$. It is then not hard to show that,
\begin{eqnarray}
[e^b (W), e^b (Z)] = e^b [W,Z] , \quad \mbox{if } db = 0. \label{Binv}
\end{eqnarray}

Analogously to the case of an almost complex structure, given an almost GCS ${\cal J}$ we can consider its $+i$-eigenbundle $L$, namely ${\cal J} W = +i W$, for all $W \in L$. A GCS is then an almost GCS for which its $+i$-eigenbundle $L$ is involutive with respect to the Courant bracket. Symbolically we will write this as $[L,L] \subset L$. In this case we say that the almost GCS is integrable. Note that eq.~(\ref{Binv}) implies that if ${\cal J}$ is integrable with $+i$-eigenbundle $L$, then $e^b {\cal J} e^{-b}$ is integrable with $+i$-eigenbundle $e^b L$ as long as $db = 0$.

In the presence of a non-zero three-form $H$ one can twist the Courant bracket by $H$,
\begin{eqnarray}
[X+\xi,Y+\eta]_H = [X+\xi,Y+\eta] + \iota_X \iota_Y H,
\end{eqnarray}
where $\iota_X \iota_Y H(Z)=H(Y,X,Z)$. With this definition,
eq.~(\ref{Binv}) becomes
\begin{eqnarray}
[e^b (W), e^b (Z)]_H = e^b [W,Z]_{H -db}.
\end{eqnarray}
This shows that this is still only a symmetry of the twisted bracket if $db=0$. On the other hand it shows that performing a b-transform with $db \neq 0$ changes the twisting. An almost GCS which is integrable with respect to an $H$-twisted Courant bracket will be called an $H$-twisted GCS. If $L \subset T \oplus T^\ast$ is involutive with respect to $[,]_H$ then $e^b L$ is involutive with respect to $[,]_{H+db}$. In other words, if ${\cal J}$ is $H$-twisted, then $e^{b} {\cal J} e^{-b}$ is $(H+db)$-twisted.

A pair $({\cal J}_1, {\cal J}_2)$ of commuting GCSs, such that ${\cal G} = -{\cal J}_1{\cal J}_2$ defines a
positive definite metric on $T \oplus T^\ast$, is called a generalized K{\"a}hler structure (GKS). When
both ${\cal J}_1$ and ${\cal J}_2$ are $H$-twisted, the resulting GKS is also called $H$-twisted. As was shown in \cite{Gualtieri:2003dx}, a twisted GKS is equivalent to a bihermitian structure. Given the bihermitian data $(g, H, J_+, J_-)$, the corresponding $H$-twisted GKS $({\cal J}_+, {\cal J}_-)$ is, up to a b-transform,
\begin{eqnarray}
{\cal J}_{\pm} = \frac 1 2
                                        \left( \begin{array}{cc}
                                        J_+ \pm J_- & \omega^{-1}_+ \mp \omega^{-1}_- \\
                                        -(\omega_+ \mp \omega_-) & -(J^t_+ \pm J^t_-)
                                        \end{array}\right),
                                        \label{gcs2}
\end{eqnarray}
where $\omega_\pm = -g J_\pm$ are two-forms,\footnote{In this section we use a more abstract notation, viewing tensors as maps between the appropriate sets. For instance $g J_\pm$ corresponds to $g_{ac} J_{\pm b}^c$ in the rest of the text (apart from section \ref{sec:GCG}). A good check for the validity of expressions is thus that lower indices should always be contracted with upper indices when recovering the index structure. \label{ftn:ind}} because $g$ is hermitian with respect to both $J_\pm$.

\subsubsection{Example: K{\"a}hler structure}

As an illustration of the definition of a GKS and in preparation of the discussion in the next section, let us look at the simplest example of a GKS -- a K{\"a}hler structure. A K{\"a}hler structure $(g,J, \Omega)$ is a Riemannian metric $g$, a complex structure $J$ and a symplectic structure $\Omega$ (i.e. a pre-symplectic structure satisfying $d\Omega = 0$), with the compatibility condition $\Omega = -gJ$. This last condition is usually phrased as $g$ being hermitian with respect to $J$. Now, a complex structure $J$ and a symplectic structure $\Omega$ correspond to the GCSs ${\cal J}_J$ and ${\cal J}_\Omega$, respectively, where
\begin{eqnarray}
{\cal J}_J =
                                        \left( \begin{array}{cc}
                                        J & 0 \\
                                        0 & -J^t
                                        \end{array}\right),
                         \quad
{\cal J}_\Omega =
                                        \left( \begin{array}{cc}
                                        0 & \Omega^{-1} \\
                                        -\Omega & 0
                                        \end{array}\right). \label{cands}
\end{eqnarray}
Courant integrability of ${\cal J}_J$ is equivalent with the integrability of the complex structure $J$, while Courant integrability of ${\cal J}_\Omega$ can be written as $d\Omega = 0$, indeed the integrability condition required for a symplectic structure.
For a K{\"a}hler manifold -- so given a Riemannian metric expressible as $g = \Omega J$ -- it is easily seen that ${\cal J}_J$ and ${\cal J}_\Omega$ commute and their product leads to a positive metric on $T\oplus T^\ast$,
\begin{eqnarray}
{\cal G} = -{\cal J}_J {\cal J}_\Omega = - {\cal J}_\Omega {\cal J}_J =
                                        \left( \begin{array}{cc}
                                        0 & g^{-1} \\
                                        g & 0
                                        \end{array}\right).
\end{eqnarray}
In other words a K{\"a}hler manifold is an example of a generalized K{\"a}hler manifold. Note that taking $J_+ = J_- = J$ in \rref{gcs2} results in the K{\"a}hler structure $({\cal J}_+, {\cal J}_-) = ({\cal J}_J, {\cal J}_\Omega)$. In our conventions this corresponds to a local description entirely in terms of chiral fields. The mirror description in terms of only twisted chiral fields by sending $J_- \rightarrow -J_-$ results in the K{\"a}hler structure $({\cal J}_+, {\cal J}_-) = ({\cal J}_\Omega, {\cal J}_J)$ where indeed complex and symplectic structure data are interchanged. More generally, on defines mirror symmetry to act locally by interchanging ${\cal J}_+$ and ${\cal J}_-$.

\subsection{Generalized complex submanifolds} \label{app:sub}

We now want to define the appropriate notion of generalized submanifold
of a generalized complex manifold. Again a more in-depth discussion
can be found in \cite{Gualtieri:2003dx}. Consider a manifold ${\cal M}$
and a closed three-form $H$ living on it. With the application to
D-branes in mind, one defines a generalized submanifold $({\cal N}, {\cal
F})$ of the manifold $({\cal M}, H)$ as a submanifold ${\cal N}$ of
${\cal M}$ along with a two-form ${\cal F}$ living on ${\cal N}$ such
that $d{\cal F} = H\vert_{\cal N}$.\footnote{In the absence of $H$, this
reduces to the existence of a closed two-form on ${\cal N}$, which is the
magnetic field strength.} One then defines the generalized tangent bundle
of ${\cal N}$ to be
\begin{eqnarray}
\tau_{\cal N}^{\;\cal F} = \{X + \xi \in T_{\cal N} \oplus T^\ast_{\cal M}
\big\vert_{\cal N}\; : \; \xi \big\vert_{\cal N} = i_X{\cal F}\}, \label{gtb}
\end{eqnarray}
where from now on we denote the tangent bundle of a manifold ${\cal M}$ by $T_{\cal M}$ to avoid confusion between the tangent bundle of the total space and that of the submanifold. We use the notation $T^\ast_{\cal M}
\big\vert_{\cal N}$ to denote the restriction of the cotangent bundle of ${\cal M}$ to the submanifold ${\cal N}$, i.e of all vector fields tangent to ${\cal M}$, only those that ``start at'' a point in ${\cal N}$ are sections of this restricted bundle.
Finally, a generalized complex submanifold of a generalized complex manifold
$({\cal M}, {\cal J}, H)$, where ${\cal J}$ is an $H$-twisted GCS, is a
submanifold $({\cal N}, {\cal F})$ of $({\cal M}, H)$ which is stable under ${\cal J}$,
\begin{eqnarray}
{\cal J} (\tau_{\cal N}^{\;\cal F}) \subset \tau_{\cal N}^{\;\cal F}. \label{stab}
\end{eqnarray}
This mimics (and generalizes) the definition of a holomorphic submanifold ${\cal N}$ of
a complex manifold ${\cal M}$ with complex structure $J$, where $T_{\cal N}$ is required to be stable under $J$. Notice that this definition of a generalized submanifold and tangent bundle is consistent with changing of the twisting. Indeed, since $e^b({\cal N}, {\cal F}) = ({\cal M}, {\cal F}' = {\cal F} + b)$, we find $d{\cal F}' = H + db$ on ${\cal N}$. On the other hand, $e^{b} {\cal J} e^{-b}$ is $(H+db)$-twisted, so that it is indeed $H+db$, and not just $H$, which enters the definition eq.~(\ref{gtb}) of the generalized tangent bundle.

Let us get a feeling for this definition and its usefulness by examining the two limiting cases. The more general case is developed to some extend in section \ref{sec:GCG}.

\subsubsection{Example 1: complex manifolds}

Consider a complex manifold $({\cal M}, J)$. We can examine what it means for a submanifold to be a generalized complex submanifold with respect to ${\cal J}_J$. Eq.~\rref{stab} implies
\begin{itemize}
\item $J(T_{\cal N}) \subset T_{\cal N}$, i.e. ${\cal N}$ is a complex submanifold of ${\cal M}$.
\item $J^t {\cal F} + {\cal F} J = 0$ on ${\cal N}$, i.e. ${\cal F}$ is of type (1,1) on ${\cal N}$.
\end{itemize}
Note that this conclusion works for any complex manifold and arbitrary $H$. In the case of a K{\"a}hler manifold (which also implies  that $H=0$), this however shows that a B brane $({\cal N}, F)$ is a generalized complex submanifold with respect to ${\cal J}_J$ of the K{\"a}hler manifold ${\cal M}$.

\subsubsection{Example 2: symplectic manifolds} \label{sympl}

Since some aspects of symplectic geometry might be less familiar, we start by reviewing these briefly.
As stated before, a symplectic form $\Omega $ is a closed, non-degenerate two-form. A
manifold endowed with a symplectic form is called a symplectic manifold.
A symplectic manifold ${\cal M}$ has several types of submanifolds. A
submanifold ${\cal N}$ is called symplectic, isotropic, coisotropic or
lagrangian resp.~if its tangent space $T_{{\cal N}}$ is a symplectic,
isotropic, coisotropic or lagrangian subspace resp.~of the tangent space
$T_{{\cal M}}$ of the manifold ${\cal M}$.

Given a symplectic vector space $M$, {\em i.e.}~an even dimensional
($d=2k$, $k\in\IN$) vector space equipped with a non-degenerate,
skew-symmetric, bilinear form $\Omega $. Consider a subspace $N$ of $M$
and define its symplectic complement $N^\bot$ by,
\begin{eqnarray}
 N^\bot=\{m\in M\vert \Omega \big(m,n\big)=0,\,\forall n\in N\}. \label{sc}
\end{eqnarray}
We distinguish four cases:
\begin{description}
  \item[Symplectic subspace:] $N$ is a symplectic subspace of $M$ if ${ N}^\bot \cap N =\emptyset$. Note that e.g. for a holomorphic submanifold ${\cal N}$ of a K{\"a}hler manifold ${\cal M}$, $T_{{\cal N}}$ is a symplectic subspace of $T_{{\cal M}}$.
  \item[Isotropic subspace:] ${ N}$ is an isotropic subspace of ${
  M}$  if ${N} \subseteq{N}^\bot$. This is true if and only if
 $\Omega$ restricts to zero on $N$ and we get $\dim (N)\leq k$. Every
 one-dimensional subspace is isotropic.
  \item[Coisotropic subspace:] ${ N}$ is a coisotropic subspace of ${
  M}$ if ${ N}^\bot \subseteq{ N}$. In other words, $N$ is
 coisotropic if and only if $N^\bot$ is isotropic. Equivalently, $N$
 is coisotropic if and only if $\Omega$ descends to a non-degenerate
 form on the quotient space $N/N^\bot$. We get $\dim (N)\geq k$ and
 any codimension one subspace is always coisotropic.
  \item[Lagrangian subspace:] ${ N}$ is a lagrangian subspace of ${
  M}$ if it is simultaneously isotropic and coisotropic, {\em i.e.}
  if ${ N}^\bot ={ N}$. This implies that, because of the
non-degeneracy of $\Omega$, a lagrangian subspace is $k$-dimensional.
Obviously $\Omega$ vanishes on a lagrangian subspace.
\end{description}
We are now ready to analyze the conditions for a generalized complex submanifold of a symplectic manifold ${\cal M}$. For this we consider the stability of the generalized tangent bundle under ${\cal J}_\Omega$, as in eq.~\rref{stab} . This results in the following conditions:
\begin{itemize}
\item $\Omega^{-1} (\Ann T_{\cal N}) \subset T_{\cal N}$, where
\begin{eqnarray}
\Ann T_{\cal N} = \{ \xi \in T_{\cal M}^\ast \, \vert \, \xi (X) = 0, \forall X \in T_{\cal N}  \}. \label{ann}
\end{eqnarray}
It is easily shown that $\Omega^{-1}(\Ann T_{\cal N}) = T_{\cal N}^\bot$, so that this is equivalent to $T_{\cal N}^\bot \subset T_{\cal N}$, i.e. ${\cal N}$ is a coisotropic submanifold of ${\cal M}$.
In other words $\Omega$ is non-degenerate
on $T_{\cal N}/T_{\cal N}^\bot$.
\item $\Omega^{-1} (\iota_X {\cal F}) = X_T \in T_{\cal N}$ for all $X \in T_{\cal N}$. This implies that ${\cal F} (X,Y) = \Omega (X_T, Y)$, for all $X, Y \in T_{\cal N}$. This in turn implies that $\iota_Y {\cal F} = 0$ for all $Y \in T_{\cal N}^\bot$. In other words, ${\cal F}$ descends to a form on $T_{\cal N}/T_{\cal N}^\bot$.
\item $(\Omega + {\cal F} \Omega^{-1} {\cal F})(T_{\cal N}) \subset \Ann T_{\cal N}$, or $(1 + K^2)(T_{\cal N}) \subset T_{\cal N}^\bot$, where $K = \Omega^{-1} {\cal F}$, so that $K$ is a complex structure on $T_{\cal N}/T_{\cal N}^\bot$. This in turn implies that ${\cal F}$ is non-degenerate on $T_{\cal N}/T_{\cal N}^\bot$ and both $\Omega$ and ${\cal F}$ are (2,0)+ (0,2) forms with respect to $K$.
\end{itemize}
When $T_{\cal N}^\bot = T_{\cal N}$, the submanifold is lagrangian and ${\cal F} = 0$
on ${\cal N}$. These conditions are precisely those for a A branes on symplectic manifolds. In particular, they coincide with the conditions for coisotropic branes first proposed in \cite{Kapustin:2001ij}. The fact that coisotropic branes on symplectic manifolds are generalized complex submanifolds with respect to the symplectic structure was first established in \cite{Gualtieri:2003dx}.

Summarizing, a brane $({\cal N}, {\cal F})$ is coisotropic if ${\cal N}$ is a coisotropic submanifold and ${\cal F}$ is zero on $T_{\cal N}^\bot$ but non-degenerate on $T_{\cal N}/T_{\cal N}^\bot$ so that $\Omega^{-1} {\cal F}$ is a complex structure on $T_{\cal N}/T_{\cal N}^\bot$.

\subsection{Poisson structures} \label{app:pois}

A Poisson manifold $({\cal M}, \Pi)$ is a manifold ${\cal M}$ endowed with a Poisson structure $\Pi$. A Poisson structure is an antisymmetric bivector $\Pi$ such that the associated Poisson bracket
\begin{eqnarray}
\{f,g\} \equiv \Pi (df, dg) = \Pi^{ab} \partial_a f \partial_b g, \label{poisson}
\end{eqnarray}
for smooth functions $f$ and $g$ on ${\cal M}$ obeys the Poisson algebra, i.e. it is a Lie algebra that acts as a derivation on the algebra of smooth functions on ${\cal M}$,
\begin{eqnarray}
\{f,gh\} = \{f,g\} h + g\{f,h\}.
\end{eqnarray}
All required conditions follow automatically from the definition (\ref{poisson}), except for the Jacobi identity. The latter is equivalent to the set of conditions
\begin{eqnarray}
\Pi^{d\,[a} \partial_d \Pi^{ bc ]} = 0 \label{jacobi}
\end{eqnarray}
on the antisymmetric bivector $\Pi$. So in short, a Poisson structure is an antisymmetric bivector which satisfies (\ref{jacobi}). See \cite{Vaisman} for more details. When $\Pi$ is invertible, this condition translates to $d \Omega = 0$, for $\Omega = \Pi^{-1}$. This implies that an invertible Poisson structure yields a symplectic structure.

An interesting property of Poisson manifolds is that they are foliated by symplectic leaves. The construction is very roughly as follows. A Hamiltonian vector field is a vector field $X_f$ associated with some function $f$ for which
\begin{eqnarray}
X_f(g) = \{f,g\}, \quad \mbox{for any function } g.
\end{eqnarray}
In components, this implies
\begin{eqnarray}
X_f^a = \Pi^{ba} \partial_b f.
\end{eqnarray}
We call $S_x$ the subspace of $T_{\cal M}$ spanned by these Hamiltonian vector fields at a point $x$ of ${\cal M}$.  If we regard $\Pi$ as a map from $T^\ast_{\cal M}$ to $T_{\cal M}$, i.e. $\Pi(df) = X_f$, we see that the dimension of $S_x$ is the rank of the map $\Pi$. A point $x$ is called regular when the rank of $\Pi$ is constant in a neighborhood of $x$. We implicitly only consider regular points in this text. Now, one can show \cite{Vaisman} that the subspaces $S_x$ define a (generalized) integrable distribution, and the Poisson structure induces a symplectic structure on the leaves $S$. This symplectic structure is essentially the inverse of the restriction of $\Pi$ to $S$.

The notion of a coisotropic submanifold carries over to Poisson manifolds in the following way. A submanifold ${\cal N}$ of a Poisson manifold $({\cal M}, \Pi)$ is called coisotropic if
\begin{eqnarray}
\Pi (\Ann T_{\cal N}) \subset T_{\cal N}, \label{pcois}
\end{eqnarray}
where the annihilator $\Ann T_{\cal N}$ was defined in eq.~(\ref{ann}). Equivalently, for any two functions $f$ and $g$ which vanish on a coisotropic submanifold ${\cal N}$, their Poisson bracket $\{ f, g\}$ also vanishes on ${\cal N}$ \cite{Vaisman}. It is clear from eq.~(\ref{gcsm}) that all generalised complex submanifolds of generalized K{\"a}hler manifolds are coisotropic in this general sense.

If $\Pi$ is invertible eq.~\rref{pcois} reduces to the coisotropy condition on a symplectic manifold of the previous section since $\Omega^{-1} (\Ann T_{\cal N}) = T_{\cal N}^\bot$, where $T_{\cal N}^\bot$ denotes the symplectic complement with respect to $\Omega$ as before.

This characterization of a coisotropic submanifold by the symplectic complement of the tangent space has a natural generalization to the Poisson case \cite{Vaisman}. Indeed, it is not hard to see that in general $\Pi (\Ann T_{\cal N})$ for some submanifold ${\cal N}$ is the symplectic complement of $T_{\cal N} \cap S_x$ in $S_x$ with respect to the induced symplectic structure on $S$. Eq.~\rref{pcois} thus becomes
\begin{eqnarray}
(T_{\cal N} \cap S_x)^\bot \subset T_{\cal N}. \label{pcois2}
\end{eqnarray}
This obviously reduces to the standard definition on symplectic
manifolds, where the foliation comprises only one leaf, namely $S={\cal
M}$.

\section{Auxiliary fields and boundary conditions}

As noted in section \ref{boundaries} the fields $\ID'
l^{\tilde \alpha }$, $\bar \ID'l^{\bar{ \tilde \alpha} }$, $
\ID'r^{\tilde \mu  }$ and $\bar \ID'r^{\bar{ \tilde \mu }  }$
should be treated as auxiliary fields. In an
$N=1$ superspace formulation these auxiliary fields are essential
for the extended supersymmetry-algebra to close off-shell in the directions along
which the two complex structures $J_+$ and $J_-$ do not commute.
The expressions for the auxiliary fields in terms of the $N=2$ superfields
can be found by working out the $\ID'$ and $\bar \ID'$ derivatives
in the action
eq.~(\ref{bsfa}) and varying the resulting action with respect to $\ID'
l^{\tilde \alpha }$, $\bar \ID'l^{\bar{ \tilde \alpha} }$, $
\ID'r^{\tilde \mu  }$ and $\bar \ID'r^{\bar{ \tilde \mu }  }$.
Performing this set of manipulations yields the following relations,
\begin{eqnarray}
\mathcal{N}\, \bar\ID' \bar{\IX}  = && - \mathcal{M}_1\, \bar\ID \IX
- \mathcal{M}_2\, \bar\ID\bar \IX  - \mathcal{M}_3\, \bar\ID'
\bar{\IY} \nonumber \\
&& - \mathcal{M}_4\, \bar\ID \bar{\IY} - \mathcal{M}_5\,
\bar\ID' \IY - \mathcal{M}_6\, \ID \IY, \label{aux1}
\end{eqnarray}
and,
\begin{eqnarray}
\ID' \IX^{T}\, \mathcal{N}^{\dagger} = && - \ID \bar{\IX}^{T}\,
\mathcal{M}_1^{\dagger} - \ID {\IX}^{T}\, \mathcal{M}_2^{\dagger} -
\ID'\IY^{T}\, \mathcal{M}_3^{\dagger}\nonumber \\
&& - \ID \IY^{T}\, \mathcal{M}_4^{\dagger} -
\ID' \bar{\IY}^{T}\, \mathcal{M}_5^{\dagger} - \ID \bar{\IY}^{T}\,
\mathcal{M}_6^{\dagger} \label{aux2}
\end{eqnarray}
where we introduced $\IX^{T} \equiv \left( l^{\tilde \beta},  r^{\tilde \mu} \right)$ and $\IY^{T} \equiv \left( z^{\beta}, w^{\nu} \right) $ and
\begin{eqnarray}
\mathcal{N} &\equiv& \left(\begin{array}{cc}
 V_{\tilde \alpha \bar{\tilde \beta}} & V_{\tilde \alpha \bar{\tilde \nu}}
 \\
 V_{\tilde \mu \bar{\tilde \beta}} & V_{\tilde \mu \bar{\tilde \nu}}
\end{array}  \right), \,
\mathcal{M}_1 \equiv \left(\begin{array}{cc} 0 & 2V_{\tilde \alpha \tilde \nu}\\ -2V_{\tilde \mu \tilde \beta} & 0\end{array}\right), \, \mathcal{M}_2 \equiv \left(\begin{array}{cc} V_{\tilde \alpha \bar{\tilde \beta}} & V_{\tilde \alpha \bar{\tilde \nu}}\\ -V_{\tilde \mu\bar{\tilde\beta}} & -V_{\tilde \mu \bar{\tilde \nu}} \end{array}\right), \,
\mathcal{M}_3 \equiv \left(\begin{array}{cc} V_{\tilde \alpha \bar \beta} & V_{\tilde \alpha \bar \nu}\\ V_{\tilde \mu \bar \beta} & 0 \end{array}\right), \nonumber \\
\mathcal{M}_4 &\equiv& \left(\begin{array}{cc} V_{\tilde \alpha \bar \beta } & V_{\tilde \alpha \bar \nu}\\ -V_{\tilde \mu \bar \beta} & 0 \end{array}\right), \,
\mathcal{M}_5 \equiv \left(\begin{array}{cc} V_{\tilde \alpha \beta} & 0\\ V_{\tilde \mu \beta} & V_{\tilde \mu \nu} \end{array}\right), \, \mathcal{M}_6 \equiv \left(\begin{array}{cc} V_{\tilde \alpha \beta} & 0\\ -V_{\tilde \mu \beta} & -V_{\tilde \mu \nu} \end{array}\right).
\end{eqnarray}
Using the $N=2$ superfield constraints eqs.~(\ref{semicon}), (\ref{twichcon}) and (\ref{chicon}) these relations can be written more elegantly as,
\begin{eqnarray}
\bar\ID' V_{\tilde\alpha} &=& - \bar\ID V_{\tilde\alpha},\nonumber \\
\bar\ID' V_{\tilde\mu} &=& + \bar\ID V_{\tilde\mu}, \label{auxil1}
\end{eqnarray}
and,
\begin{eqnarray}
\ID' V_{\bar {\tilde \alpha} } &=& - \ID V_{\bar{\tilde\alpha}}, \nonumber \\
\ID' V_{\bar {\tilde\mu} } &=& + \ID V_{\bar{\tilde\mu}}. \label{auxil2}
\end{eqnarray}

In the second part of this section we will discuss how the relations for the
auxiliary fields eqs.~(\ref{auxil1}) and (\ref{auxil2}) arise, when a chiral/twisted
chiral pair is dualized to a semi-chiral multiplet. While the Dirichlet boundary
conditions in the original model are dualized to the Dirichlet or/and Neumann
boundary conditions in the dual model, the Neumann boundary conditions
from the original model result in the expressions for the auxiliary fields after
dualization. This connection between the original Neumann boundary conditions and the
expressions for the auxiliary fields in the dual model thus forms an additional consistency check for
the dualization. Let us clarify these statements with the examples constructed in section
\ref{examples}.

Starting with the Dirichlet boundary conditions eq.~(\ref{diriD1}) for the D1-brane,
we can deduce from the associated Neumann boundary condition for the twisted chiral superfield that
the gauge field $\tilde Y$ should satisfy the relations,
\begin{eqnarray}
\ID' \tilde Y = i\, \frac{m}{n} \ID \tilde Y, \quad
\bar \ID' \tilde Y = - i\, \frac{m}{n} \bar\ID \tilde Y. \label{dualD1N}
\end{eqnarray}
Using the equations of motion,
\begin{eqnarray}
\hat Y &=& -\frac{i}{2} \big(l -\bar l - r+ \bar r \big),\nonumber \\
\tilde Y &=& Y - \frac 1 2 \big(l+\bar l + r + \bar r \big), \nonumber \\
Y &=& -\frac 1 2 \ln\left(e^{-(r+\bar r) } + 1\right),\label{eomdual}
\end{eqnarray}
and imposing the Dirichlet boundary conditions of the dual D2-brane
eqs.~(\ref{DiriD21}) enables us to write eqs.~(\ref{dualD1N}) as,
\begin{eqnarray}
\ID' \left( l + \bar l + r + \bar r + \ln\big(1+e^{-r-\bar r}\big) \right) &=& - \ID\left( l - \bar l - r - \bar r  \right), \nonumber\\
\bar\ID' \left( l + \bar l + r + \bar r + \ln\big(1+e^{-r-\bar r}\big) \right)&=& +  \bar\ID\left( l - \bar l - r - \bar r  \right).
\end{eqnarray}
These relations can also be obtained from eqs.~(\ref{auxil2})
and (\ref{auxil1}) respectively, after taking a linear combination
and imposing the dual Dirichlet boundary conditions.

When dualizing the D3-brane given in eq.~(\ref{diriD3}) we need to
distinguish between two different cases: $\alpha = a - i$ and
$\alpha \neq a -i$. In the first case the D3-brane is dualized
to a D2-brane, in the latter case to a D4-brane. Focusing first
on the lagrangian D2-brane, we can deduce from the associated
Neumann boundary condition for eq.~(\ref{diriD3}) that the
gauge fields $\tilde Y$ and $Y$ should satisfy the following
expressions at the boundary,
\begin{eqnarray}
\ID' \tilde Y &=& + i\, \frac{m}{n}\, \ID \tilde Y + i\, a\, \ID Y + \ID Y, \nonumber \\
\bar \ID' \tilde Y &=& -i\, \frac{m}{n}\, \bar \ID \tilde Y - i\, a\, \bar \ID Y + \bar\ID Y.
\end{eqnarray}
Imposing the Dirichlet boundary condition eq.~(\ref{ddd2}) and
implementing the equations of motion eq.~(\ref{eomdual}), we
find the following relations,
\begin{eqnarray}
\ID' \left( l + \bar l + r + \bar r + \ln\big(1+e^{-r-\bar r}\big)\right) &=& - \ID \left( l - \bar l - r + \bar r - \ln\big(1+e^{-r-\bar r}\big)\right), \nonumber\\
\bar \ID' \left( l + \bar l + r + \bar r + \ln\big(1+e^{-r-\bar r}\big)\right) &=& + \bar \ID \left( l - \bar l - r + \bar r - \ln\big(1+e^{-r-\bar r}\big)\right), \label{osher}
\end{eqnarray}
which are just linear combinations of the expressions in
eqs.~(\ref{auxil2}) and (\ref{auxil1}) respectively. The Neumann
boundary conditions for the chiral superfield on the other hand can be properly dualized
to the first expression in eqs.~(\ref{auxil1}) and (\ref{auxil2}).

If we choose $\alpha = 0$, the D3-brane is dualized to a coisotropic D4-brane,
and the associated Neumann boundary condition for eq.~(\ref{diriD3}) then
yields the same relations for $\tilde Y$ as in eq.~(\ref{dualD1N}).
However, we need to impose the relations in eq.~(\ref{ddd4}) for this situation,
after which we implement the equations of motion eq.~(\ref{eomdual}). These
manipulations lead to the same expressions as in eq.~(\ref{osher}), and
thus reproduce the expressions for the auxiliary fields. One can also properly dualize the
Neumann boundary conditions for the chiral superfield to the first expression given
in eqs.~(\ref{auxil1}) and (\ref{auxil2}) respectively.


\begin{thebibliography}{99}
\bibitem{Kachru:2003aw}
  S.~Kachru, R.~Kallosh, A.~Linde and S.~P.~Trivedi,
{\em De Sitter vacua in string theory},
  \prd{68}{2003}{046005},
{\tt  [arXiv:hep-th/0301240]}.
\bibitem{Berkovits:2000fe}
  N.~Berkovits,
{\em Super-Poincare covariant quantization of the superstring},
 \jhep{0004}{2000}{018},
{\tt  [arXiv:hep-th/0001035]}.
\bibitem{Alvarez-Gaume:1981hm}
  L.~Alvarez-Gaume and D.~Z.~Freedman,
{\em Geometrical Structure And Ultraviolet Finiteness In The
Supersymmetric Sigma Model}, \cmp{80}{1981}{443}.
\bibitem{Gates:1984nk}
  S.~J.~Gates, C.~M.~Hull and M.~Ro\v cek,
{\em Twisted Multiplets And New Supersymmetric Nonlinear Sigma Models},
\npb{248}{1984}{157}.
\bibitem{Curtright:1984dz}
  T.~L.~Curtright and C.~K.~Zachos,
{\em Geometry, Topology And Supersymmetry In Nonlinear Models},
\prl{53}{1984}{1799}.
\bibitem{Howe:1985pm}
  P.~S.~Howe and G.~Sierra,
{\em Two-Dimensional Supersymmetric Nonlinear Sigma Models With Torsion},
\plb{148}{1984}{451}.
\bibitem{Lindstrom:2005zr}
  U.~Lindstr{\"o}m, M.~Ro\v cek, R.~von Unge and M.~Zabzine,
{\em Generalized K{\"a}hler manifolds and off-shell supersymmetry},
\cmp{269}{2007}{833}, \hepth{0512164}.
\bibitem{Buscher:1987uw}
  T.~Buscher, U.~Lindstr{\"o}m and M.~Ro\v cek,
{\em New Supersymmetric Sigma Models With Wess-Zumino Terms},
\plb{202}{1988}{94}.
\bibitem{Ivanov:1994ec}
  I.~T.~Ivanov, B.~Kim and M.~Ro\v cek,
{\em Complex structures, duality and WZW models in extended superspace},
\plb{343}{1995}{133}, \hepth{9406063}.
\bibitem{Sevrin:1996jr}
  A.~Sevrin and J.~Troost,
{\em Off-shell formulation of N = 2 non-linear sigma-models},
\npb{492}{1997}{623}, \hepth{9610102}.
\bibitem{Bogaerts:1999jc}
  J.~Bogaerts, A.~Sevrin, S.~van der Loo and S.~Van Gils,
{\em Properties of semi-chiral superfields}, \npb{562}{1999}{277},
\hepth{9905141}.
\bibitem{MS}
J.~Maes and A.~Sevrin, {\em A note on N = (2,2) superfields in two
dimensions}, \plb{642}{2006}{535}, \hepth{0607119}.
\bibitem{Ooguri:1996ck}
  H.~Ooguri, Y.~Oz and Z.~Yin,
{\em D-branes on Calabi-Yau spaces and their mirrors},
\npb{477}{1996}{407}, \hepth{9606112}.
\bibitem{Hanany:1997vm}
  A.~Hanany and K.~Hori,
{\em Branes and N = 2 theories in two dimensions}, \npb{513}{1998}{119},
\hepth{9707192}.
\bibitem{Hori:2000ck}
  K.~Hori, A.~Iqbal and C.~Vafa,
{\em D-branes and mirror symmetry}, \hepth{0005247}.
\bibitem{Hori:2000ic}
  K.~Hori,
{\em Linear models of supersymmetric D-branes}, \hepth{0012179}.
\bibitem{Albertsson:2001dv}
  C.~Albertsson, U.~Lindstr{\"o}m and M.~Zabzine,
{\em N = 1 supersymmetric sigma model with boundaries. I},
\cmp{233}{2003}{403}, \hepth{0111161}.
\bibitem{Albertsson:2002qc}
  C.~Albertsson, U.~Lindstr{\"o}m and M.~Zabzine,
{\em N = 1 supersymmetric sigma model with boundaries. II},
\npb{678}{2004}{295}, \hepth{0202069}.
\bibitem{Lindstrom:2002jb}
  U.~Lindstr{\"o}m and M.~Zabzine,
{\em N = 2 boundary conditions for non-linear sigma models and
Landau-Ginzburg
 models},
\jhep{0302}{2003}{006}, \hepth{0209098}.
\bibitem{Lindstrom:2002mc}
  U.~Lindstr{\"o}m, M.~R\v ocek and P.~van Nieuwenhuizen,
{\em Consistent boundary conditions for open strings},
\npb{662}{2003}{147}, \hepth{0211266}.
\bibitem{Koerber:2003ef}
P.~Koerber, S.~Nevens and A.~Sevrin, {\em Supersymmetric non-linear
sigma-models with boundaries revisited}, \jhep{11}{2003}{066},
\hepth{0309229}.
\bibitem{Sevrin:2007yn}
  A.~Sevrin, W.~Staessens and A.~Wijns,
{\em The world-sheet description of A and B branes revisited},
\jhep{0711}{2007}{061},
 {\tt arXiv:0709.3733 [hep-th]}.
\bibitem{Sevrin:2008tp}
  A.~Sevrin, W.~Staessens and A.~Wijns,
{\em An N=2 worldsheet approach to D-branes in bihermitian geometries: I.
Chiral and twisted chiral fields},
  \jhep{0810}{2008}{108}, {\tt [arXiv:0809.3659 [hep-th]]}.
\bibitem{Sevrin:2008qx}
  A.~Sevrin, W.~Staessens and A.~Wijns,
{\em N = 2 world-sheet approach to D-branes on generalized Kaehler
geometries: I. General formalism}, \forp{57}{2009}{684}, {\tt
arXiv:0810.5355 [hep-th]}.
\bibitem{Hull:2008vw}
  C.~M.~Hull, U.~Lindstrom, M.~Rocek, R.~von Unge and M.~Zabzine,
  {\em Generalized Kahler geometry and gerbes},
{\tt  arXiv:0811.3615 [hep-th]}.
\bibitem{Kapustin:2001ij}
  A.~Kapustin and D.~Orlov,
  {\em Remarks on A-branes, mirror symmetry, and the Fukaya category},
  \jgp{48}{2003}{84},
  {\tt hep-th/0109098}.
\bibitem{Nevens:2006ht}
  S.~Nevens, A.~Sevrin, W.~Troost and A.~Wijns,
{\em Derivative corrections to the Born-Infeld action through
beta-function
  calculations in N = 2 boundary superspace},
 \jhep{0608}{2006}{086}, \hepth{0606255}.
\bibitem{Hitchin:2004ut}
  N.~Hitchin,
  {\em Generalized Calabi-Yau manifolds}, {\em
  Quart.\ J.\ Math.\ Oxford Ser.\ } {\bf 54}, 281 (2003)
  {\tt [arXiv:math/0209099]}.
\bibitem{Gualtieri:2003dx}
  M.~Gualtieri,
{\em Generalized complex geometry}, Oxford University DPhil thesis, 2003,
{\tt arXiv:math/0401221};
M.~Gualtieri, {\em Generalized complex geometry}, {\tt
arXiv:math/0703298}.
\bibitem{Zabzine:2004dp}
  M.~Zabzine,
  {\em Geometry of D-branes for general N = (2,2) sigma models},
  \lmp{70}{211}{2004},
  {\tt [arXiv:hep-th/0405240]}.
\bibitem{Lyakhovich:2002kc}
  S.~Lyakhovich and M.~Zabzine,
  {\em Poisson geometry of sigma models with extended supersymmetry},
  \plb{548}{2002}{243},
  {\tt arXiv:hep-th/0210043}.
\bibitem{Gualtieri:2007}
  M.~Gualtieri,
  {\em Branes on Poisson varieties},
  {\tt  arXiv:0710.2719v1 [math.DG]}.
\bibitem{Kapustin:2005vs}
  A.~Kapustin,
  {\em A-branes and noncommutative geometry},
  {\tt arXiv:hep-th/0502212}.
\bibitem{Grisaru:1997ep}
  M.~T.~Grisaru, M.~Massar, A.~Sevrin and J.~Troost,
{\em Some aspects of N = (2,2), D = 2 supersymmetry},
  \forp{47}{1999}{301},
  \hepth{9801080}.
\bibitem{Gates:1995du}
  S.~J.~Gates, M.~T.~Grisaru and M.~E.~Wehlau,
{\em A Study of General 2D, N=2 Matter Coupled to Supergravity in
Superspace},
  \npb{460}{1996}{579},
  [arXiv:hep-th/9509021].
\bibitem{Gates:1980az}
  S.~J.~Gates and W.~Siegel,
{\em Variant Superfield Representations},
  \npb{187}{1981}{389}.
\bibitem{Deo:1985ix}
  B.~B.~Deo and S.~J.~Gates,
{\em Comments On Nonminimal N=1 Scalar Multiplets},
  \npb{254}{1985}{187}.
\bibitem{Lindstrom:2007vc}
  U.~Lindstr{\"o}m, M.~Ro\v cek, I.~Ryb, R.~von Unge and M.~Zabzine,
{\em New N = (2, 2) vector multiplets},
  \jhep{0708}{2007}{008},
{\tt  [arXiv:0705.3201 [hep-th]]}.
\bibitem{Gates:2007ve}
  S.~J.~J.~Gates and W.~Merrell,
{\em D=2 N=(2,2) Semi Chiral Vector Multiplet},
  \jhep{0710}{2007}{035},
{\tt   [arXiv:0705.3207 [hep-th]]}.
\bibitem{Lindstrom:2007sq}
  U.~Lindstr{\"o}m, M.~Ro\v cek, I.~Ryb, R.~von Unge and M.~Zabzine,
{\em T-duality and Generalized Kahler Geometry}, \jhep{0802}{056}{2008},
\hepth{0707.1696}.
\bibitem{Merrell:2007sr}
  W.~Merrell and D.~Vaman,
{\em T-duality, quotients and generalized K{\"a}hler geometry},
\plb{665}{401}{2008}, \hepth{0707.1697}.
\bibitem{Spindel:1988nh}
  Ph.~Spindel, A.~Sevrin, W.~Troost and A.~Van Proeyen, {\em Complex
  structures on parallelized group manifolds and supersymmetric sigma models},
  \plb{206}{1988}{71} and
{\em Extended Supersymmetric Sigma Models on Group Manifolds. 1. The Complex
Structures}, \npb{308}{1988}{662}.
\bibitem{Rocek:1991vk}
  M.~Ro\v cek, K.~Schoutens and A.~Sevrin,
{\em Off-shell WZW models in extended superspace}, \plb{265}{1991}{303};
 M.~Rocek, C.~H.~Ahn, K.~Schoutens and A.~Sevrin,
{\em Superspace WZW models and black holes},
\hepth{9110035}.
\bibitem{Font:2006na}
  A.~Font, L.~E.~Ibanez and F.~Marchesano,
{\em Coisotropic D8-branes and model-building}, \jhep{0609}{2006}{080},
\hepth{0607219}.
\bibitem{Grisaru:1997pg}
  M.~T.~Grisaru, M.~Massar, A.~Sevrin and J.~Troost,
{\em The quantum geometry of N = (2,2) non-linear sigma-models},
  \plb{412}{1997}{53},
{\tt  [arXiv:hep-th/9706218]}.
\bibitem{Halmagyi:2007ft}
  N.~Halmagyi and A.~Tomasiello,
{\em Generalized Kaehler Potentials from Supergravity},
\cmp{291}{2009}{1}, {\tt
  arXiv:0708.1032 [hep-th]}.
\bibitem{Kapustin:2003se}
A.~Kapustin and Y.~Li, {\em Stability conditions for topological
D-branes: A worldsheet approach}, \hepth{0311101}.
\bibitem{Vaisman}
I.~Vaisman, {\em Lectures on the Geometry of Poisson Manifolds}, Birkh{\"a}user Verlag, Basel, 1994.
\bibitem{Craps:2006vx}
  B.~Craps, O.~Evnin and S.~Nakamura,
  {\em Local recoil of extended solitons: A string theory example},
  \jhep{0701}{2007}{050},
  {\tt [arXiv:hep-th/0608123]}.
\bibitem{Brunner:2009mn}
  I.~Brunner, M.~R.~Gaberdiel, S.~Hohenegger and C.~A.~Keller,
{\em Obstructions and lines of marginal stability from the world-sheet},
{\tt arXiv:0902.3177 [hep-th]}.
\end{thebibliography}
\end{document}